\pdfoutput=1
\documentclass[11pt]{article}
\usepackage[T2A]{fontenc}
\usepackage[utf8]{inputenc}

\usepackage{jheppub}
\usepackage{float}
\usepackage{yfonts}
\usepackage{mathrsfs}

\usepackage{romannum}

\usepackage{customprelude}

\usepackage{pgfornament}

\usepackage{tikz-cd}

\newcommand{\nR}{\text{Я}}
\newcommand{\Ux}{U(1)_x}
\newcommand{\Uy}{U(1)_y}
\newcommand{\Uz}{U(1)_z}
\newcommand{\UsuR}{U(1)_{R}^{\cN=2}}

\newcommand{\QsuR}{Q_{R}^{\cN=2}}
\newcommand{\Ur}{U(1)_r^{\cN=2}}

\newcommand{\su}{\mathfrak{su}}

\protected\def\[#1\]{\begin{equation}\begin{split}#1\end{split}\end{equation}}

\newcommand{\shortitle}{Deconfining \alt{$\cN=2$}{N=2} SCFTs}

\title{\centering \shortitle
 \\
       \vspace{0.4cm}
       \includegraphics[height=1cm]{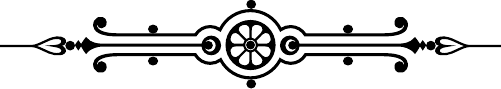}\\
       \vspace{0.2cm}or the Art of Brane Bending}

\hypersetup{pdfinfo={
		Title={\shortitle}
}}

\newcommand{\durham}{$^\sharp$}
\newcommand{\amherst}{$^\flat$}
\newcommand{\Rma}{}
\newcommand{\Rmb}{\prime}
\newcommand{\RmaP}[2]{{#1}^{#2}}

\author{Iñaki García Etxebarria,\durham}
\author{Ben Heidenreich,\amherst}
\author{Matteo Lotito,\amherst}
\author{Ajit Kumar Sorout\amherst}

\affiliation{\durham Department of Mathematical Sciences,\\
	Durham University, Durham, DH1 3LE, United Kingdom}
\affiliation{\amherst Amherst Center for Fundamental Interactions, \\ Department of Physics,
	University of Massachusetts, Amherst, MA 01003 USA}

\emailAdd{inaki.garcia-etxebarria@durham.ac.uk}
\emailAdd{bheidenreich@umass.edu}
\emailAdd{ajitkumar@umass.edu}
\emailAdd{mlotito@umass.edu}

\abstract{We introduce a systematic approach to constructing $\cN=1$
  Lagrangians for a class of interacting $\cN=2$ SCFTs. We analyse in
  detail the simplest case of the construction, arising from placing
  branes at an orientifolded $\bC^2/\bZ_2$ singularity. In this way we
  obtain Lagrangian descriptions for all the $R_{2,k}$ theories. The
  rank one theories in this class are the $E_6$ Minahan-Nemeschansky
  theory and the $C_2\times U(1)$ Argyres-Wittig theory. The
  Lagrangians that arise from our brane construction manifestly
  exhibit either the entire expected flavour symmetry group of the
  SCFT (for even $k$) or a full-rank subgroup thereof (for odd $k$),
  so we can compute the full superconformal index of the $\cN=2$
  SCFTs, and also systematically identify the Higgsings associated to
  partial closing of punctures.
}

\begin{document}

\makeatletter
\let\old@fpheader\@fpheader
\renewcommand{\@fpheader}{\old@fpheader\hfill
ACFI-T21-13}
\makeatother

\maketitle

\newpage

\section{Introduction}

During the last several years our understanding of four-dimensional
$\cN=2$ superconformal field theories (SCFTs) has been greatly
enhanced, particularly in the case in which these theories can be
constructed by compactifying the six-dimensional $(2,0)$ theory of
type $\fg$ on a punctured Riemann surface $\Sigma$
\cite{Gaiotto:2009we}. These SCFTs often come in continuously
connected families, forming a ``conformal manifold''. The geometry of
this conformal manifold can be understood as arising from the geometry
of the Riemann surface $\Sigma$. Of particularly interest are subloci
of this manifold where $\Sigma$ degenerates. In this case we sometimes
have a weakly coupled Lagrangian description of the SCFT, but more
often we end up with a collection of basic building blocks, given by
three-punctured spheres, connected by a weak gauging of subgroups of
their flavour groups.\footnote{At least in most cases. In some rare
  cases one needs to refine this picture, see \cite{Chacaltana:2012ch}
  for the detailed analysis of one such instance.}  The different
degeneration limits of $\Sigma$ are known as the different duality
frames for the theory, and we say that the theories arising in these
degeneration limits are dual to each other. We refer the reader to the
nice reviews \cite{Tachikawa:2013kta,Tachikawa:2015bga} for a detailed
explanation of these facts.

A particularly important case is the one where all three punctures are
``full'' punctures. The resulting SCFTs are known as the $T[\fg]$
theories. Except in the $\fg=A_1$ case, the $T[\fg]$ theories are
intrinsically strongly coupled $\cN=2$ SCFTs. Other types of punctures
can be obtained by Higgsing fields in the $T[\fg]$ theory, an
operation known in the literature as ``(partial) closing'' of
punctures. In addition, for our discussion it is important to
introduce ``twisted'' punctures. Such punctures have the property that
there is a monodromy around them acting as an outer automorphism of
the $(2,0)$ theory. In configurations with three punctures we have the
possibility of twisting two of the punctures. The resulting twisted
theories were analysed in detail in
\cite{Tachikawa:2009rb,Tachikawa:2010vg,Chacaltana:2012ch} and play an
important role below.

Remarkably, much of the knowledge gained in the last few years about
these $\cN=2$ SCFTs has been obtained without requiring any knowledge
of a Lagrangian description, or perhaps more accurately, despite the
fact that no Lagrangian is known.  This situation has recently started
to change. An important development was the construction --- initiated
by Maruyoshi and Song
\cite{Maruyoshi:2016tqk,Maruyoshi:2016aim,Agarwal:2016pjo} and then
further extended by a number of authors
\cite{Agarwal:2017roi,Benvenuti:2017bpg,Giacomelli:2017ckh,Carta:2020plx}
--- of $\cN=1$ preserving renormalization group flows connecting
$\cN=2$ SCFTs. If the starting theory is Lagrangian, then this
provides a Lagrangian theory in the universality class of the $\cN=2$
SCFTs at the end of the flow.

A second approach constructs $\cN=1$
Lagrangians by probing the $\cN=1$ preserving conformal manifold of
the $\cN=2$ SCFTs. These deformations of the $\cN=2$ theory typically
break the symmetry group of the target SCFT to a lower rank subgroup,
but the resulting Lagrangians are still useful. There are some
powerful constraints that any Lagrangian theory in the same conformal
manifold should obey, and exploration of such constraints have led to
the construction of Lagrangians for a number of important $\cN=2$ (and
even $\cN=3$) SCFTs \cite{Razamat:2019vfd,Razamat:2020gcc,Razamat:2020pra,Zafrir:2020epd}.

In this paper we introduce a third class of constructions, which take
advantage of a number of recent results in the context of duality for
$\cN=1$ SCFTs
\cite{Garcia-Etxebarria:2015hua,Garcia-Etxebarria:2016bpb}. We will
review these results in detail below.\footnote{See also
  \cite{GarciaEtxebarria:2012qx,Garcia-Etxebarria:2013tba} for earlier
  work on the class of string configurations we study in this paper,
  and
  \cite{Bianchi:2020fuk,Antinucci:2020yki,Antinucci:2021edv,Amariti:2021lhk}
  for recent work on $\cN=1$ dualities arising from orientifolded
  toric singularities.} These results apply to the class of $\cN=1$
SCFTs that arise from isolated orientifolds of D3 branes probing
isolated toric singularities. This is a large class of theories, and
all have non-trivial conformal manifolds: the value of the string
coupling provides an exactly marginal parameter, and in some
exceptional low-rank\footnote{By rank we mean the number of mobile D3
  branes probing the singularity, as measured by the $F_5$ flux at
  infinity.} cases some additional marginal deformations might
exist. As in
\cite{Garcia-Etxebarria:2015hua,Garcia-Etxebarria:2016bpb}, we will
focus on the directions in the conformal manifold associated to the
ambient string coupling, which persist for arbitrary rank (for
sufficiently high rank the conformal manifold for this class of
theories has complex dimension one, and is parametrised by the ambient
string coupling). The physics at the cusps of the conformal manifold
is very reminiscent of that appearing in $\cN=2$ class-$\cS$ theories:
there is a class of isolated $\cN=1$ SCFTs, denoted by $TO_k$ in
\cite{Garcia-Etxebarria:2016bpb}, and the generic duality frame can be
described in terms of a set of $TO_k$ theories coupled via weak
gauging of diagonal subgroups of their flavour symmetry groups.

Rather surprisingly, the situation regarding Lagrangian descriptions
is much more developed in the $\cN=1$ case than in the $\cN=2$ case:
there are known Lagrangians for \emph{all} $TO_k$ SCFTs. These
Lagrangians are obtained by a rather natural operation in the string
theory description, which we call \emph{deconfinement}, as it
generalizes the familiar notion of deconfinement of free antisymmetric
tensors in the context of Seiberg duality
\cite{Seiberg:1994pq,Intriligator:1995ne,Berkooz:1995km,Pouliot:1995me}. We
will review this operation below. Additionally, in some cases
providing a Lagrangian description of the full theory, not only the
$TO_k$ sectors, requires \emph{brane bending}: a (natural) deformation
of overlapping branes in the system so that gauge couplings of some
branes in the brane tiling become finite. An example of brane bending
is given in figure~\ref{sfig:F03-III-tiling} below. Using these two
operations one can provide $\cN=1$ Lagrangians for \emph{any} $\cN=1$
SCFT in the class described above.

These developments raise a very natural question: what are the
analogues of brane bending and deconfinement in the $\cN=2$ setting?
The answer is not straightforward, as $\cN=2$ theories do not fall in
the class of theories analysed in
\cite{Garcia-Etxebarria:2015hua,Garcia-Etxebarria:2016bpb}, and in
fact we will not provide a complete answer in this paper. However, we make
a small step in this direction by using our knowledge of the $\cN=1$
setting to derive Lagrangians, in a fairly systematic way, for a
number of interesting $\cN=2$ SCFTs.

\medskip

The basic argument goes as follows. Recall that the dualities in
\cite{Garcia-Etxebarria:2015hua,Garcia-Etxebarria:2016bpb} describe
what happens to the field theory as we crank up the ambient string
coupling, and switch descriptions to a new weakly coupled duality
frame. The effect in the field theory is a deformation by an exactly
marginal operator, moving us from one cusp in the conformal manifold
to another.

We will combine this operation with partial resolution of the
singularity. In particular, we will study partial resolutions of the
form\footnote{Here $\rho$ is the ``blow-down'' map from the partially
resolved space to the more singular one.}$^{,}$\footnote{We
emphasise that this is a special case of a much more general
construction. However, even the special case $n=1$ (which we focus on in the present paper) will yield very interesting and nontrivial results.}
\begin{equation}
\rho\colon (\bC^2/\bZ_{2n} \times \bC) + (\bC^2/\bZ_{2n} \times \bC) \to Y^{2n,0} \,,
\end{equation}
where the left hand side denotes a Calabi-Yau threefold that is
smooth except at two points, each locally of the form
$\bC^2/\bZ_{2n}\times \bC$. The toric description of the partial
resolution is (in the $n=2$ case)
\[
\includegraphics[width=0.3\textwidth]{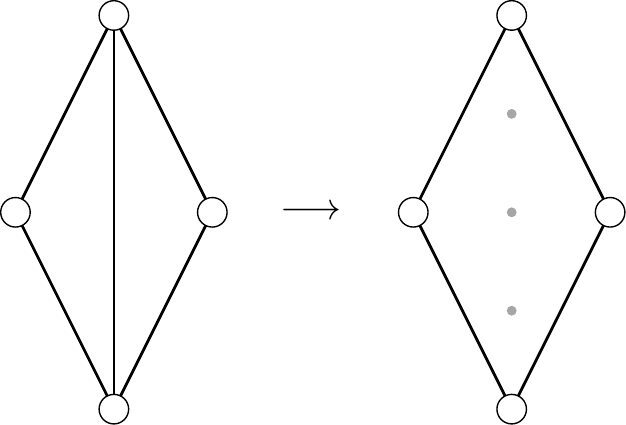}
\]

We are interested in what happens to the IR fixed
point of the theory on branes probing the singularity under this operation. In order to
encode this we introduce the notation $\cT^\phi[X]$, to represent the
CFT describing the IR fixed point of the theory of D3 branes probing
the singular point in $X$, with additional data, such as possible
orientifold planes and fluxes denoted generically by $\phi$. In this
notation, there is an induced map on the space of 4d CFTs, coming from
turning on a baryonic vev (encoding the size of the exceptional cycle
in the partial resolution) and integrating out massive modes
\begin{equation}
\rho^*(\cT^{\phi}[Y^{2n,0}]) = \cT^{\alpha}[\bC^2/\bZ_{2n}\times\bC] + \cT^\beta[\bC^2/\bZ_{2n}\times \bC]
\end{equation}
where now addition on the right hand side means that we have two
decoupled SCFTs.

From the point of view of the string construction it is natural to
expect that the effect of this operation in the field theory commutes
with the duality, in the sense that the following diagram is
commutative:
\begin{equation}
\label{eq:duality-square}
\begin{tikzcd}
\cT^\phi[Y^{2n,0}] \arrow{r}{\rho^*} \arrow[swap]{d}{S} & \cT^\alpha[\bC^2/\bZ_{2n}\times\bC] + \cT^\beta[\bC^2/\bZ_{2n}\times \bC] \arrow{d}{S}\\
\cT^{S(\phi)}[Y^{2n,0}] \arrow{r}{\rho^*} & \cT^{S(\alpha)}[\bC^2/\bZ_{2n}\times\bC] + \cT^{S(\beta)}[\bC^2/\bZ_{2n}\times \bC]
\end{tikzcd}
\end{equation}
where we have denoted by $S$ the action on the background data induced
by taking the string coupling to infinity, so that we move to a
different cusp in the conformal manifold. The assumption
that~\eqref{eq:duality-square} is commutative allows us to understand
the behaviour of S-duality on $\cT^{\alpha}[\bC^2/\bZ_{2n}\times \bC]$
from the behavior of S-duality on $\cT^{\phi}[Y^{2n,0}]$, which was
understood in
\cite{Garcia-Etxebarria:2015hua,Garcia-Etxebarria:2016bpb}, and the
behaviour of $\rho^*$, which is reasonably well understood in the case
of ordinary dimer models --- see for instance
\cite{GarciaEtxebarria:2006aq} for a systematic approach.

For the sake of presentation, it will be convenient to introduce
forgetful maps $\rho^*_L$ and $\rho^*_R$ that focus on each of the
resulting SCFTs as follows
\begin{equation}
\rho^*_L(\cT^{\phi}[Y^{2n}]) = \cT^{\alpha}[\bC^2/\bZ_{2n}\times\bC]
\end{equation}
and similarly $\rho_R^*=\rho^* - \rho^*_L$. It is natural to
conjecture that if~\eqref{eq:duality-square} is commutative then the
reduced version
\begin{equation}
\label{eq:reduced-duality-square}
\begin{tikzcd}
\cT^\phi[Y^{2n,0}] \arrow{r}{\rho^*_L} \arrow[swap]{d}{S} & \cT^\alpha[\bC^2/\bZ_{2n}\times\bC] \arrow{d}{S}\\
\cT^{S(\phi)}[Y^{2n,0}] \arrow{r}{\rho^*_L} & \cT^{S(\alpha)}[\bC^2/\bZ_{2n}\times\bC]
\end{tikzcd}
\end{equation}
is also commutative, and similarly for $\rho^*_R$. This will be our
fundamental assumption in the rest of the paper.

Choosing $\alpha$, $\beta$ and $\phi$ judiciously we can arrange for
$\cT^{\alpha}[\bC^2/\bZ_{2n}\times \bC]$ to preserve $\cN=2$
supersymmetry, thereby connecting the results of
\cite{Garcia-Etxebarria:2015hua,Garcia-Etxebarria:2016bpb} to the
large literature on $\cN=2$ dualities beginning with
\cite{Gaiotto:2009we}.

In this paper we initiate this program, by focusing exclusively on the
simplest case, $n=1$. The resulting singularity is known as the
complex Calabi-Yau cone over $\bF_0=\bP_1\times\bP_1$, or the real
Calabi-Yau cone over $Y^{2,0}$, and can alternatively be described
as a $\bZ_2$ orbifold of the conifold. By following the logic above,
we will be able to systematically construct Lagrangians for all the
$R_{2,k}$ theories, with $k$ even or odd. An important case that we
will study in detail is $k=3$, where $R_{2,3}$ is the rank one $E_6$
Minahan-Nemeschansky theory~\cite{Minahan:1996fg}, also known as the $T[A_2]$ or more simply
$T_3$ theory in the context of class $\cS$. It is also interesting to
consider $k=2$, which engineers the Argyres-Wittig theory with global
symmetry algebra $\fsp(4)\oplus \fu(1)$.

\medskip

We have organised this paper as follows. We start in
\S\ref{sec:review} by providing a short review of the main results
that we will need
from~\cite{Garcia-Etxebarria:2015hua,Garcia-Etxebarria:2016bpb}. Our
main results are presented in \S\ref{sec:Lagrangians}, where we derive
$\cN=1$ Lagrangians for the $R_{2,k}$ $\cN=2$ SCFTs using the strategy
sketched above, and perform some standard checks. We proceed to
further test these Lagrangian descriptions in a number of ways: the
Higgs branch structure of the theories is studied in
\S\ref{sec:Higgs}, the Coulomb branch in \S\ref{sec:Coulomb}, and the
result of turning on mass deformations is described in
\S\ref{sec:mass}. In all cases we find perfect agreement with the
expectations from previous $\cN=2$ results, whenever these exist. We
finish by listing some conclusions and further directions in
\S\ref{sec:conclusions}. The appendix collects explicit expressions
for the $R_{2,2k}$ superconformal indices for $k=1,2,3$.

\section{Review}

\label{sec:review}

\subsection{Deconfinement for isolated \alt{$\cN=1$}{N=1} orientifold SCFTs}

We will now briefly review some of the results in
\cite{GarciaEtxebarria:2012qx,Garcia-Etxebarria:2013tba,Garcia-Etxebarria:2015hua,Garcia-Etxebarria:2016bpb},
placing particular emphasis on those that are particularly important
for our discussion. We will not include derivations of the results, we
refer the interested reader to the original works for proofs.

\begin{figure}
\includegraphics[width=\textwidth]{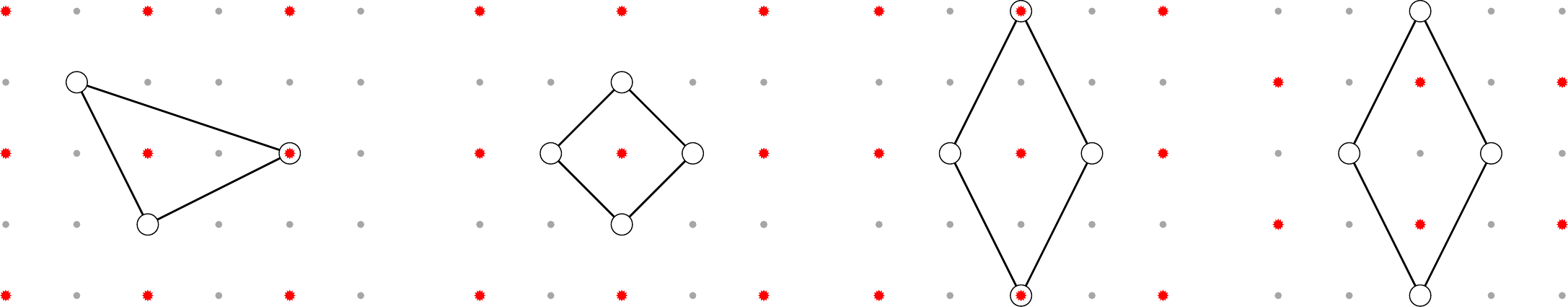}
\caption{Examples of toric diagrams with choice of even sublattice. The first and third graphs describe cases where some vertices of the toric diagam lie within the chosen even sublattice. Such examples will not be considered in the present work.}
\label{fig:toricexamples}
\end{figure}

Consider a toric\footnote{We refer the reader unfamiliar with toric
geometry to the excellent book \cite{Cox:2011tv}. A briefer
introduction summarizing all the ideas that we will need can be
found in \cite{Garcia-Etxebarria:2016bpb}.} Calabi-Yau threefold
$X$, which we assume to be a cone with an isolated singularity at the
origin. All toric Calabi-Yau threefolds can be described by providing
a two-dimensional lattice polytope known as the \emph{toric diagram};
those having isolated singularities have the additional property that
 the edges of the toric diagram do not hit any intermediate lattice points. We
will consider IIB string theory on this background, in the presence
of an orientifold action preserving the toric nature of the
space. Such orientifold actions were classified in
\cite{Garcia-Etxebarria:2016bpb}, they can be characterized by a
choice of even sublattice for the toric $\bZ^2$ lattice. For
simplicity, we will restrict to cases where the orientifold action
leaves only the isolated singularity at the origin fixed. In terms of
the toric diagram this requires that none of the external vertices of
the toric diagram are contained in the even sublattice defining
the orientifold action, see figure~\ref{fig:toricexamples} for examples.

\begin{figure}
\centering
\begin{subfigure}{0.4\textwidth}
\centering
\includegraphics[width=\textwidth]{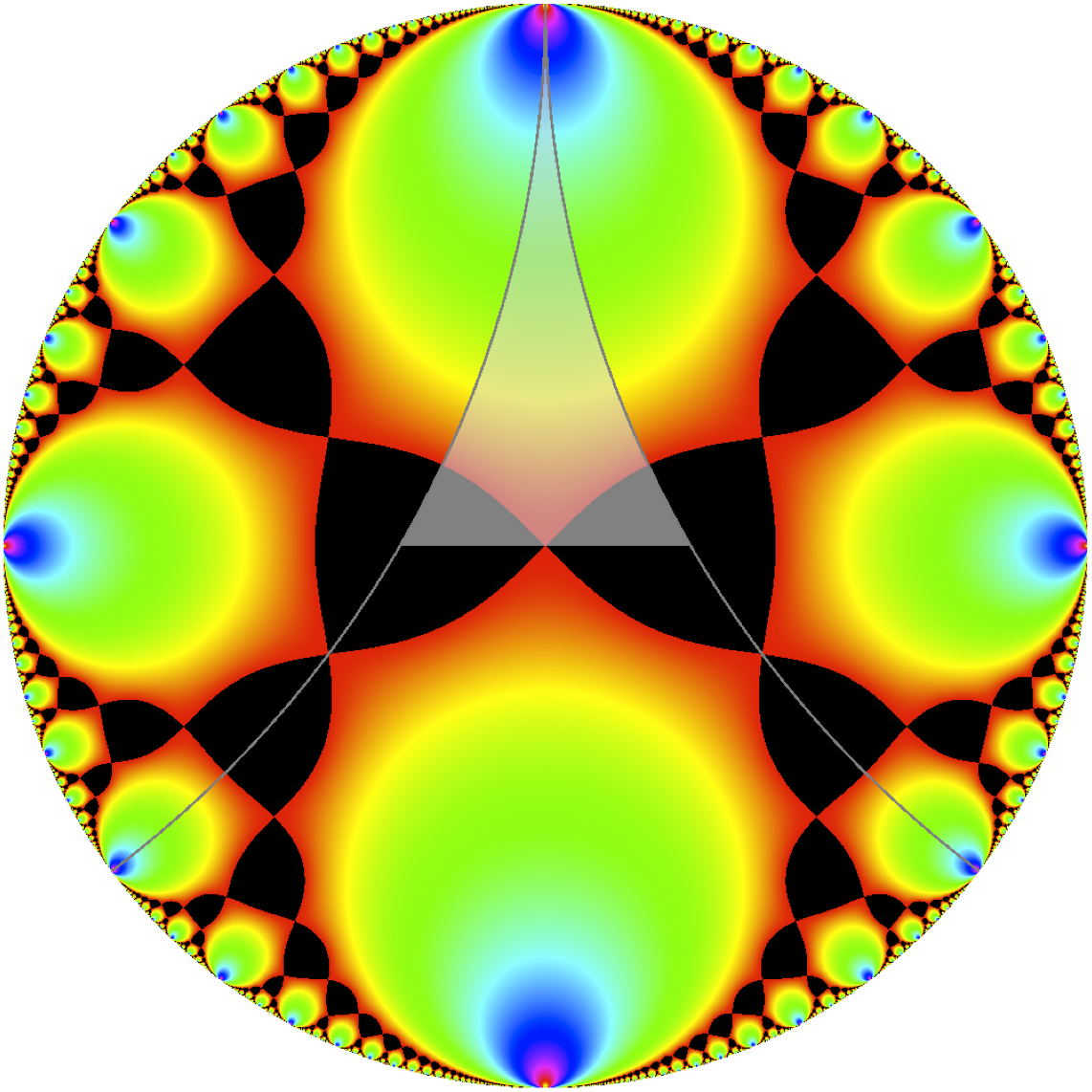}
\caption{Conformal manifold.}
\label{sfig:SL2Zgraphic}
\end{subfigure}
\hfill
\begin{subfigure}{0.4\textwidth}
\centering
\includegraphics[width=\textwidth]{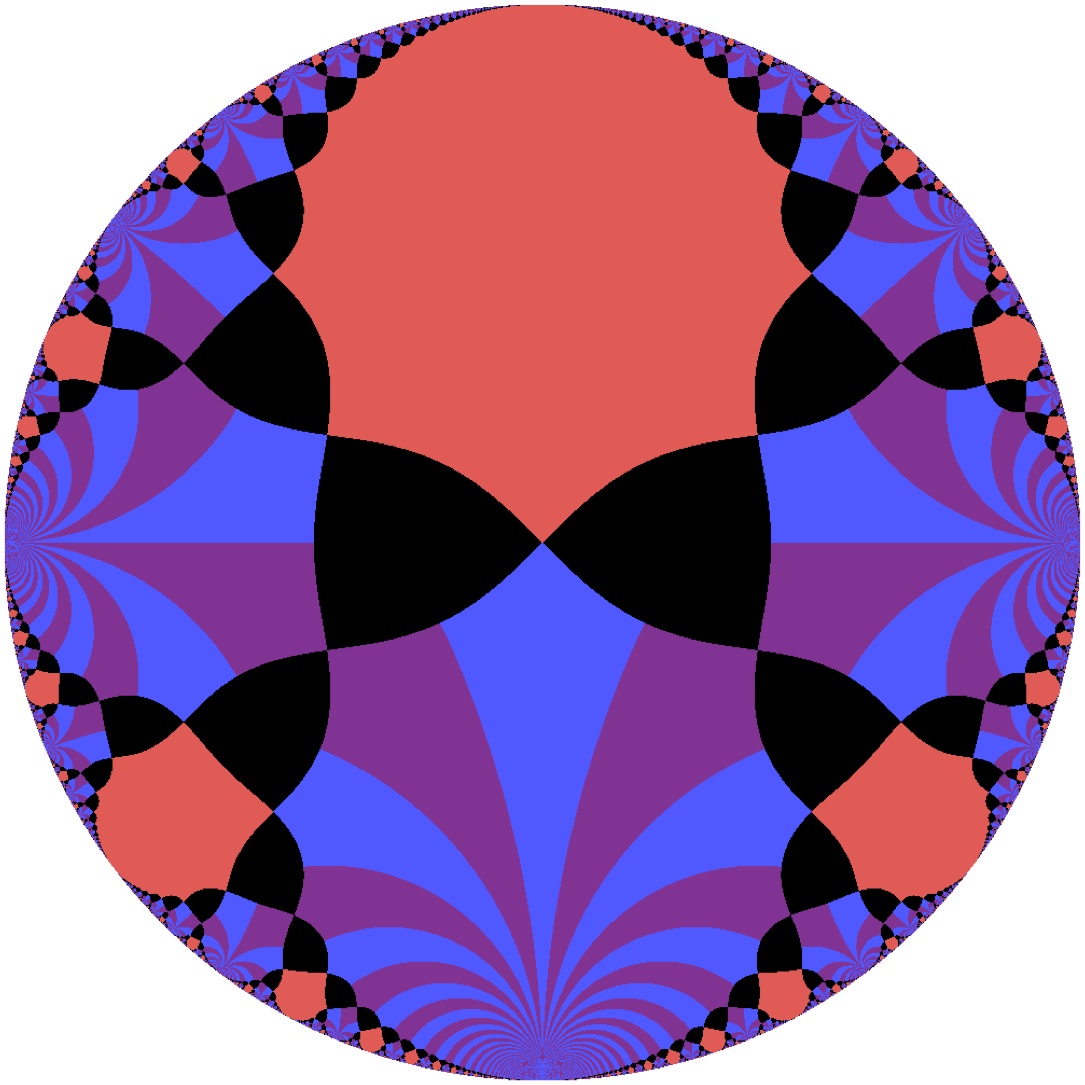}
\caption{Duality phases for the O3 theory.}
\label{sfig:phasediagram}
\end{subfigure}
\caption{\subref{sfig:SL2Zgraphic}~The upper half place, conformally
mapped to a disk. We have lightly shaded a copy of the fundamental
region for $SL(2,\bZ)$. Every other point in the disk can be
mapped to this region by a $SL(2,\bZ)$ transformation. The
blue/purple regions correspond to regions where weakly coupled
descriptions exist in some duality
frame. \subref{sfig:phasediagram} Partition of the conformal
manifold of the $\cN=4$ $\fso(2N+1)$ theory into regions where at
weak coupling the valid description is in terms of a $\fso(2N+1)$
algebra (pink) or $\fsp(2N)$ (blue/purple). \\ \null\hfill {\footnotesize [Figures reproduced from~\cite{GarciaEtxebarria:2012qx}.]}}
\end{figure}

We now place $N$ D3 branes at the singular point of the geometry. For
the moment we do not include an orientifold involution. This leads to
an interacting $\cN=1$ SCFT in four dimensions with a marginal
deformation which we can identify with the value of the IIB
axio-dilaton.\footnote{It is possible to have additional marginal
deformations, but these will play no role in our discussion.} The
axio-dilaton $\tau=C_0+i/g_s$ of IIB takes values in the upper
half-plane, but values related by modular transformations $g\in SL(2,\bZ)$ define the same physical theory:
\[
\tau \to g(\tau) = \frac{a\tau + b}{c\tau + d} \,, \qquad g = \begin{pmatrix}
a & b\\
c & d
\end{pmatrix} \,, \qquad a,b,c,d\in\bZ\,, \quad a d - b c = 1 \,.
\]
Because both the geometric background and
the D3 branes map to themselves under $SL(2,\bZ)$, this implies that
the $\cN=1$ SCFTs on $N$ D3 branes with coupling $\tau$ is equivalent
to that same theory with coupling $g(\tau)$. We have a weakly coupled
Lagrangian description of the $\cN=1$ SCFT at those points related by
an $SL(2,\bZ)$ transformation to $\tau=i\infty$. We will refer to such
points in the conformal manifold as ``cusps''. This situation is illustrated in figure~\ref{sfig:SL2Zgraphic}, in which we map the upper half plane to a disk, shading the cusps in
purple.

This discussion needs to be modified once we introduce orientifold
planes. While the D3s, and by extension the $F_5$ flux created by the
D3 branes, are $SL(2,\bZ)$ invariant, the $H_3$ and $F_3$ fluxes
sourced by the orientifold transform as a $SL(2,\bZ)$
doublet. Introducing the notation $\cF_3\df (F_3,H_3)$, we have:
\[
g(\cF_3) = \begin{pmatrix}
a & b\\
c & d
\end{pmatrix} \begin{pmatrix}
F_3 \\ H_3
\end{pmatrix} = \begin{pmatrix}
aF_3 + bH_3\\
cF_3 + dH_3
\end{pmatrix}\, .
\]
So we have additional data to keep track in determining the physics at
the cusps: the precise CFT at the singularity depends on the pair
$(\tau, \cF_3)$ subject to the relation
\[
\label{eq:flux-manifold}
(\tau, \cF_3) \sim (g(\tau), g(\cF_3))\, .
\]
Note in particular that the physics at different cusps can depend
quite strongly on the value of $\cF_3$. As a simple example, consider
the case of $N$ D3 branes probing an O3 plane in flat space. As
discussed in \cite{Witten:1998xy} in this case $\cF_3$ takes values in
$\bZ_2\oplus \bZ_2$ (we will explain and expand on this statement
momentarily), so there are four possibilities for $\cF_3$. Assume that
on a given cusp, in a duality frame where $g_s\ll 1$, we have a flux
$\cF_3$. Then it was argued in \cite{Witten:1998xy} that the dynamics
at this cusp is described by weakly coupled $\cN=4$ SYM with gauge
algebra $\fg$, with\footnote{The cases $\cF_3=(0,1)$ and $\cF_3=(1,1)$
are related by $\tau\to\tau+1$, which induces a shift of the
$\theta$ angle in the field theory, so they give rise to the same
perturbative behaviour.}
\[
\begin{array}{c|c}
\cF_3 & \fg\\
\hline
(0,0) & \fso(2N)\\
(1,0) & \fso(2N+1)\\
(0,1) & \fsp(2N)\\
(1,1) & \fsp(2N)
\end{array}
\]
Consider for instance the case that in one duality frame we have
$g_s\ll 1$ and $\cF_3=(1,0)$, and let us take $C_0=0$ for simplicity
of exposition. This corresponds to a $\cN=4$ SYM theory with gauge
algebra $\fso(2N+1)$ and coupling
$g_{\fso(2N+1)}^2=g_s$ (in general the mapping is $C_0+i/g_s=\frac{\theta}{2\pi} + i/g_{\text{YM}}^2$). By~\eqref{eq:flux-manifold}, and from the
dictionary between flux and gauge algebra in \cite{Witten:1998xy},
this is equivalent to having an $\cN=4$ theory with gauge algebra
$\fsp(2N)$ with gauge coupling $g_{\fsp(2N)}^2=g_s^{-1}$. This is in
agreement with the standard prediction from Montonen-Olive duality
\cite{Goddard:1976qe,Montonen:1977sn}.

An alternative way of thinking about this duality is that if we fix
the flux at one cusp, then every other cusp is decorated with a flux
assignment, which gives rise to potentially different perturbative
descriptions at different cusps. In the example above, this means that
we can interpolate between the weakly coupled $\fso(2N+1)$ and
$\fsp(2N)$ $\cN=4$ theories by moving in the conformal manifold. This
is illustrated in figure~\ref{sfig:phasediagram}, where we have split
the conformal manifold according to whether the weakly coupled
description valid in each region has algebra $\fso(2N+1)$ or
$\fsp(2N)$. We will refer to the different decorated cusps of the
conformal manifold as the \emph{duality phases} of the theory, and we
will say that two such phases are related by an $SL(2,\bZ)$
transformation $g$ if the corresponding couplings are related by $g$.

\subsection{Phases of \alt{$\cC_\bC(\bF_0)$}{Cc(F0)}}
\label{sec:duality-phases}

The picture that we have discussed so far generalises in a beautiful
way to a certain class of $\cN=1$ SCFTs, namely those arising from D3
branes probing an isolated toric singularity in the presence on a
toric orientifold whose only fixed point is at the singularity of the
geometry.

Consider a toric Calabi-Yau cone $X_6$, with base $X_5$. We assume
that the only singularity of $X_6$ is at the base of the cone. We will
quotient by an orientifold action $(-1)^{F_L}\Omega \sigma$, where
$\sigma\colon X_6\to X_6$ leaves only the origin invariant. This
implies that it acts on $X_5$ freely. We additionally demand that
$\sigma$ preserves the toric structure of $X_6$, or in other words
that $X_6/\sigma$ is still toric (although it will not be Calabi-Yau
any more). Our working example will be the Calabi-Yau cone over
$\bF_0=\bP^1\times\bP^1$, which we denote by $\cC_\bC(\bF_0)$. The
toric diagram of this geometry is
\[
\label{eq:F0-toric-diagram}
\includegraphics[height=2cm]{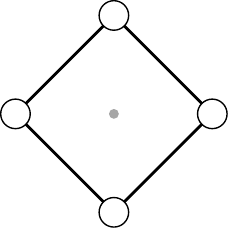}
\]
This Calabi-Yau threefold is a real cone over $(S^3\times S^2)/\bZ_2$.

In order to classify the SCFTs at the cusps, we first need to classify
the possible orientifolds that we can put on the singularity: this
involves fixing the geometric actions $\sigma$ and the choices of
discrete $\cF_3$ flux. We then need to be able to describe the SCFT at
the singularity for any such choices of $\sigma$ and $\cF_3$. This
program was completed in \cite{Garcia-Etxebarria:2016bpb}, building on
previous work in
\cite{GarciaEtxebarria:2012qx,Garcia-Etxebarria:2013tba,Garcia-Etxebarria:2015hua}. In
what follows we will review the results in
\cite{Garcia-Etxebarria:2016bpb}, we refer the reader to that paper
for derivations and a more detailed discussion.

\begin{figure}
\centering
\begin{subfigure}{0.3\textwidth}
\includegraphics[width=\textwidth]{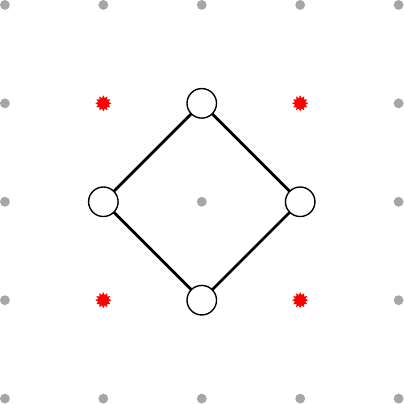}
\caption{O3 involution}
\label{sfig:F0-O3}
\end{subfigure}
\hspace{2cm}
\begin{subfigure}{0.3\textwidth}
\includegraphics[width=\textwidth]{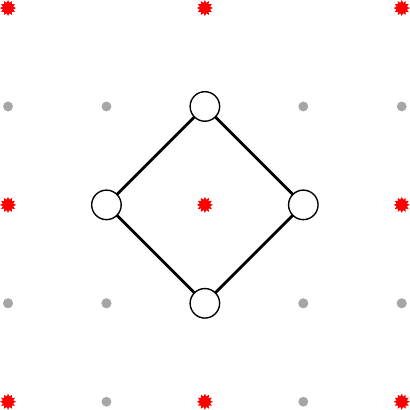}
\caption{O7 involution}
\label{sfig:F0-O7}
\end{subfigure}
\caption{Two choices of even sublattices for $\bZ^2$. As explained
in \cite{Garcia-Etxebarria:2016bpb}, these correspond to the two
choices of toric involutions of $\cC_\bC(\bF_0)$ with compact
orientifold locus, up to equivalences. We have named the two
involutions according to the kind of orientifold planes that arise
when blowing up the singularity.}
\label{fig:F0-sigmas}
\end{figure}

As a first step, the choices of $\sigma$ keeping $X_6/\sigma$ toric
can be classified by choosing an even sublattice of the $\bZ^2$
lattice on which the toric diagram is defined. There are four choices
for the origin of such an even sublattice, although in some cases ---
such as the $\bF_0$ case of interest to us --- various choices might
be related by symmetries of the toric diagram, leading to equivalent
physics. The resulting orientifold action will have a compact fixed
locus iff none of the external vertices of the toric diagram is
contained in the chosen even sublattice. Consider for instance the
case of $X_6=\cC_\bC(\bF_0)$. In this case there are two inequivalent
choices of $\sigma$ keeping the orientifold locus isolated, shown in figure \ref{fig:F0-sigmas}. It is convenient to refer to the
resulting involutions according to the dimension of the fixed locus in the fully
resolved geometry. In one case we obtain an O7 plane wrapping the
exceptional $\bF_0$, while in the other case we obtain four O3
planes. Accordingly, we refer to the two involutions as ``O7'' and
``O3'', respectively. The involution that leads to $\cN=2$ theories
after partial resolution is the O3 one, so henceforth we will focus
exclusively on this one.

The next step is to classify the $\cF_3$ fluxes that one can turn on
the orientifolded geometry. More precisely, we want to classify the
flux as measured at infinity. The manifold at infinity, ignoring the
4d spacetime part, has topology $X_5$, and $\cF_3$ transforms as a
doublet of $SL(2,\bZ)$, so the flux at infinity is classified by
elements of the cohomology with local coefficients\footnote{We will
  use cohomology to classify fluxes. A more precise characterisation,
  at least in the perturbative setting of interest to us, would
  involve K-theory \cite{Moore:1999gb}. See
  \cite{Bergman:2001rp,Garcia-Compean:2002dui,Loaiza-Brito:2004ajy}
  for work classifying orientifolds from this
  perspective. Alternatively, we could work in F-theory, and classify
  those fluxes that preserve Poincaré invariance in 4d, see for
  instance \cite{Denef:2008wq}.}  $H^3(X_5;(\bZ\oplus\bZ)_\rho)$, with
$\rho$ the action of $SL(2,\bZ)$ on the coefficient system. We refer
the reader to \cite{AT} for the definition of cohomology with local
coefficients, and \cite{Witten:1998xy,Aharony:2016kai} for
applications of this formalism in the perturbative and
non-perturbative settings. In our case
$\rho=(-1)^{F_L}\Omega=\begin{psmallmatrix}-1 &
  0\\0&-1\end{psmallmatrix}$, so elements of $\cF_3$ are classified by
$H^3(X_5;\tbZ\oplus\tbZ)=H^3(X_5;\tbZ)\oplus H^3(X_5;\tbZ)$. Each
summand is the group of degree-three cohomology classes with local
$\bZ$ coefficients \cite{AT}, twisted on the non-trivial $\bZ_2$ cycle
of $X_5$. A more down-to-earth summary of all this, at least in
our specific case, is to say that due to the orientifold action the
NSNS and RR fluxes pick up a minus sign as we go around the
non-contractible cycle in $X_5/\sigma$.

For an isolated toric singularity with $n$ external vertices, and an
isolated toric orientifold action, we have
\cite{Garcia-Etxebarria:2016bpb}
\[
H^3(X_5;\tbZ) = \underbrace{\bZ_2\oplus\ldots\oplus\bZ_2}_{n-3\text{ times}}\, .
\]
For $X_6=\cC_\bC(\bF_0)$ we have $n=4$ external vertices in the toric
diagram~\eqref{eq:F0-toric-diagram}, so
\[
H^3(X_5;\tbZ) = \bZ_2\oplus\bZ_2
\]
which implies that there are 16 possible values for the flux $\cF_3$
classifying the orientifold type at the singularity (four for $F_3$
times four for $H_3$), and therefore 16 possible different flux
decorations for a given cusp. Some of these flux choices are related
by shifts of $C_0$, and therefore lead to the same perturbative
description.

\begin{figure}
\centering
\begin{subfigure}[b]{0.325\textwidth}
\centering
\includegraphics[height=4cm]{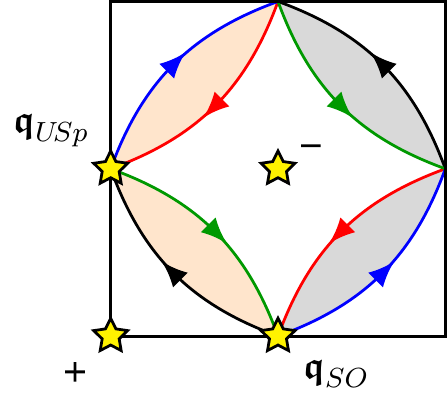}
\caption{Phase \I}
\label{sfig:F03-I-tiling}
\end{subfigure}
\begin{subfigure}[b]{0.325\textwidth}
\centering
\includegraphics[height=3.5cm]{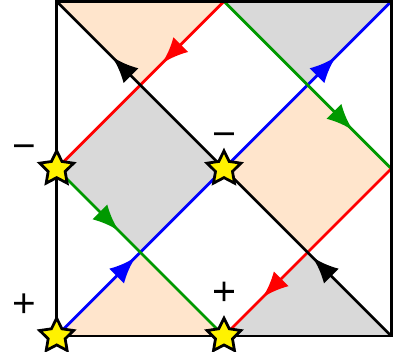}
\vspace{0.5cm}
\caption{Phases \II\ and \tII}
\label{sfig:F03-II-tiling}
\end{subfigure}
\begin{subfigure}[b]{0.325\textwidth}
\centering
\includegraphics[height=4cm]{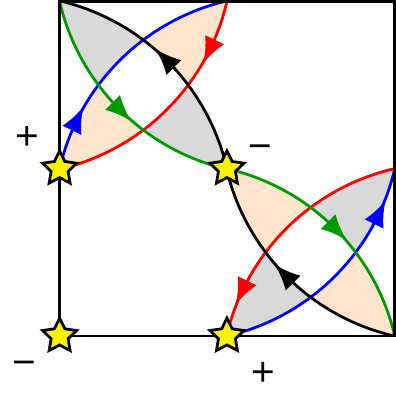}
\caption{Phase \III}
\label{sfig:F03-III-tiling}
\end{subfigure}
\caption{Theories arising at the cusps of the conformal manifold for
the $\cC_\bC(\bF_0)$ theory modded out by the O3 action, for
different choices of flux. Due to the high degree of symmetry of
$\cC_\bC(\bF_0)$ some of the flux choices lead to isomorphic
choices, which we have denoted as phases \II\ and \tII\ in this
figure.\\ \null\hfill {\footnotesize [Figures reproduced from~\cite{Garcia-Etxebarria:2016bpb}.]}}
\label{fig:F03-tiling}
\end{figure}

We will give the detailed dictionary between $\cF_3$ flux and field
theory below, but for the moment let us simply state that there are
three types of theories appearing at the cusps, which we will call
phases \I, \II\ and \III. (There is also a phase \tII\ related to
phase \II. The $\cF_3$ fluxes giving rise to these two phases are
related by symmetries of the singularity, so phases \II\ and \tII\ are
isomorphic as field theories. We will for the most part ignore \tII\
in what follows.) These three phases are best understood in the
language of brane tilings, as in figure~\ref{fig:F03-tiling}. We write
the perhaps more familiar quiver description of these theories below.

\subsection{Brane tiling constructions}

Let us briefly describe how to obtain (and interpret) the brane
tilings in figure~\ref{fig:F03-tiling}, referring the reader to
\cite{Franco:2005rj,Feng:2005gw,Franco:2007ii} for the original works
on brane tilings and their orientifolds, to \cite{Yamazaki:2008bt} for
a review and \cite{Garcia-Etxebarria:2016bpb} for a more detailed
analysis in the particular case that concerns us here. A brane tiling
is a tiling of the torus by D5 branes (the white regions) and bound
states of one D5 with $\pm 1$ NS5 branes (the grey and orange
regions). These two kinds of regions are separated by 1-cycles, which
encode where NS5 branes end on the tiling. The winding numbers of
these cycles on the $T^2$ reproduce the slopes of the external legs in
the $(p,q)$-web diagram for the toric singularity. For our $\bF_0$
example, this means that we have four NS5 branes ending on 1-cycles on
the $T^2$, with winding numbers $(1, 1)$, $(1,-1)$, $(-1,1)$, and
$(-1,-1)$. In the absence of orientifolds, regions of the $T^2$
covered by D5 branes lead to $SU$ gauge factors, intersections of two
NS5 branes lead to bifundamental matter between the D5s touching the
intersection, and D5-NS5 bound states lead to superpotential
couplings.

We are interested in orientifolded configurations, where the
orientifold leaves four points fixed on the $T^2$. These points are O5
planes intersecting the torus, and the sign annotation of the tiling
indicates the type of orientifold plane we are dealing with. The
orientifold projection acts on the gauge theory in a natural way: if
two $SU$ factors are exchanged by the orientifold action then a
diagonal combination survives, while if a $SU$ face is invariant then
it is projected down to $SO$ or $\Sp$ depending on the orientifold
sign. Similarly for matter content: two bifundamental multiplets
exchanged by the orientifold action lead to a single chiral multiplet
in the orientifolded theory, while a bifundamental mapped to itself
leads to a symmetric or antisymmetric representation, depending on the
sign of the orientifold action.

In order to obtain a configuration consistent under the orientifold
action, it must be the case that the NS5 branes map to themselves
under the orientifold action (up to orientation). This implies that
each NS5 brane should pass through two fixed points. Up to
isomorphism, and after some ``brane bending'', this leads to the three
phases in figure~\ref{fig:F03-tiling}. The bending of branes accounts
for the fact that if we draw the NS5s using straight lines on the
$T^2$ then in phases \I\ and \III\ we would have overlapping NS5
branes, which leads to strongly coupled physics. In phase \III\
bending the branes in a way compatible with the orientifold projection
is a simple way of obtaining a Lagrangian theory in the same
universality class as the theory being engineered by the string
construction. Phase \I\ has the further peculiarity that, even after
bending the overlapping NS5 branes so they do not overlap any longer,
we still have four NS5 branes intersecting at a point. This can also
be resolved by a more advanced form of brane bending (which we refer
to as ``deconfinement''), described below.

\begin{figure}
\begin{subfigure}[t]{0.35\textwidth}
\includegraphics[width=\textwidth]{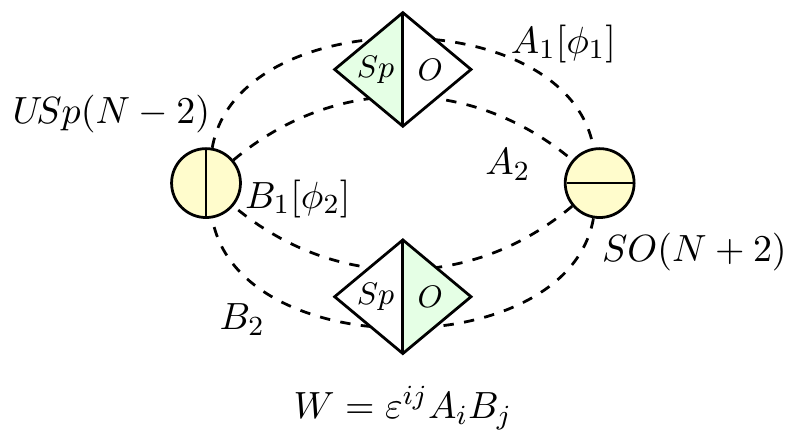}
\caption{Phase \I}
\label{sfig:F03-I-quiver}
\end{subfigure}
\hfill
\begin{subfigure}[t]{0.36\textwidth}
\includegraphics[width=\textwidth]{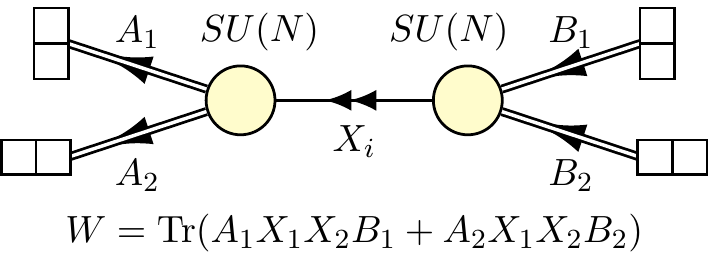}
\caption{Phases \II\ and \tII}
\label{sfig:F03-II-quiver}
\end{subfigure}
\hfill
\begin{subfigure}[t]{0.27\textwidth}
\includegraphics[width=\textwidth]{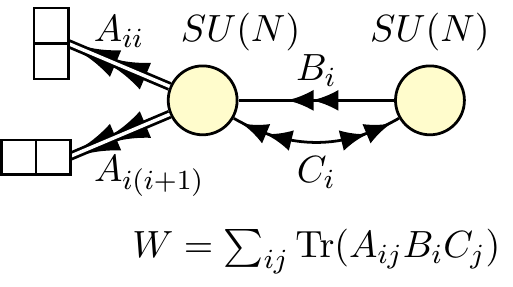}
\caption{Phase \III}
\label{sfig:F03-III-quiver}
\end{subfigure}
\caption{Quiver gauge theories arising from the brane tilings in
figure~\ref{fig:F03-tiling}.
\\ \null\hfill {\footnotesize [Figures reproduced from~\cite{Garcia-Etxebarria:2016bpb}.]}}
\label{fig:F0-O3-phases}
\end{figure}

Applying these rules, we read off the quiver theory associated to each brane
tiling, with the results shown in figure~\ref{fig:F0-O3-phases}. Let us start
with phase \II, which is the most conventional one. The quiver is
shown in figure~\ref{sfig:F03-II-quiver}, and more explicitly we have
a field content
\begin{subequations}
\begin{gather}
\label{eq:F03-II-charges}
\begin{array}{c|cc|cccc}
& SU(N) & SU(N) & U(1)_B & U(1)_X & U(1)_Y & U(1)_R\\
\hline
X_1 & \fund & \ov\fund & \frac{1}{N} & 1 & 0 & \frac{1}{2}\\
X_2 & \fund & \ov\fund & \frac{1}{N} & -1 & 0 & \frac{1}{2}\\
B_1 & \ov\asymm & {\bf 1} & -\frac{1}{N} & 0 & -1-\frac{2}{N} & \frac{1}{2}\\
B_2 & \ov\symm & {\bf 1} & -\frac{1}{N} & 0 & 1-\frac{2}{N} & \frac{1}{2}\\
A_1 & {\bf 1} & \asymm & -\frac{1}{N} & 0 & 1+\frac{2}{N} & \frac{1}{2}\\
A_2 & {\bf 1} & \symm & -\frac{1}{N} & 0 & -1+\frac{2}{N} & \frac{1}{2}
\end{array}\\
\intertext{and superpotential}
W = \Tr(A_1X_1X_2B_1 + A_2X_1X_2B_2)\, .
\end{gather}
\end{subequations}
We emphasize that the Lagrangian just described is \emph{not} conformal. Instead, we are interested in the infrared superconformal fixed point to which this Lagrangian flows.

In fact, the Lagrangian is not even asymptotically free, due to the nonrenormalizable superpotential. There are two ways to view this deficiency: on the one hand, as described in~\cite{GarciaEtxebarria:2012qx} we can view the Lagrangian as an effective field theory with a UV cutoff set by the nonrenormalizable couplings and an IR dynamical scale. By an analysis similar to~\cite{GarciaEtxebarria:2012qx} using the methods of~\cite{Leigh:1995ep}, one can easily check that one combination of the couplings (related to $g_s$) is neutral under all the spurious flavor symmetries, and thus not renormalized at any order, suggesting that there is a fixed line in the infrared. Moreover, this ``exactly dimensionless'' coupling sets the hierarchy between the UV cutoff and the IR dynamical scale, with this hierarchy becoming exponentially large for $g_s \ll 1$. Near this ``cusp'', our ignorance of the (stringy) UV physics becomes unimportant, and the infrared SCFT is accurately described by the IR dynamics of the effective field theory.

\begin{figure}\centering
\includegraphics[width=.7\textwidth]{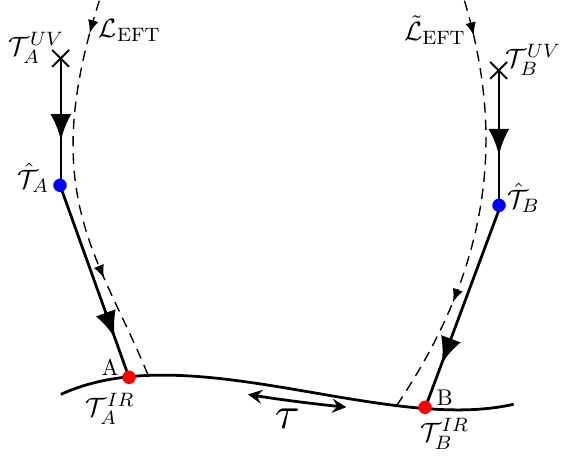}
\caption{A schematic picture of S-duality in $\mathcal{N}=1$ theories. Beginning with a free UV theory $\mathcal{T}^{UV}_A$, we turn on a sequence of relevant operators, producing a sequence of RG flows through intermediate CFTs $\hat{\mathcal{T}}_A$ until we reach an IR CFT $\mathcal{T}^{IR}_A$ with one or more exactly marginal operators. These operators parameterize a fixed line on which $\mathcal{T}^{IR}_A$ is a special point (typically with enhanced global symmetries) that we call a ``cusp''. The fixed line can be reached directly by turning on all these operators simultaneously in the UV, but as some are dangerously irrelevant, the resulting EFT flow $\mathcal{L}_{\text{EFT}}$ is typically not UV complete (except when the cusp is a free theory). S-duality occurs when multiple cusps lie on the same fixed line, or when non-trivial paths along the fixed line return to the same cusp (self-duality).}
\label{fig:rgflow}
\end{figure}

An alternate and perhaps more sophisticated viewpoint is described in~\cite{Garcia-Etxebarria:2016bpb}. We view the effective Lagrangian as a recipe for reaching the desired infrared fixed line via a series of flows beginning at a free UV fixed point, as follows. Starting with the free theory obtained by turning off all the couplings, we produce a series of flows by turning the couplings on one by one, only selecting relevant (or marginally relevant) couplings at each step of the process. After each flow, the dimensions of the remaining couplings will change, but there is always a relevant operator until the last step,\footnote{One way to see this is to note that the exactly dimensionless coupling combination referenced above involves all the individual couplings, and vanishes when any of them vanish. This combination remains neutral under the spurious flavor symmetries at each step, and thus remains exactly dimensionless (in the absence of accidental symmetries). Therefore, it is the product of couplings whose dimensions add to zero, and so at least one of these couplings is relevant (positive dimension) or else they are all exactly marginal.} when the only coupling remaining to be switched on is exactly marginal (see \cite{Green:2010da}). The fixed point so reached is the ``cusp'' itself, whereas turning on this last coupling (now parameterizing an exactly marginal operator) moves us out along the fixed line away from the cusp, see figure \ref{fig:rgflow}. 

Phase \III\ is slightly more subtle, in that the string construction
involves overlapping NS5 branes, give rise to strongly coupled sectors. Thus, there is no perturbative effective field theory description even arbitrarily close to the cusp. However, this is easily avoided  by some straightforward brane bending
which leads to the Lagrangian description in
figure~\ref{sfig:F03-III-quiver}, with field content
\begin{subequations}
\begin{gather}
\label{eqref:F03-III-charges}
\begin{array}{c|cc|cccc}
& SU(N) & SU(N) & U(1)_B & U(1)_X & U(1)_Y & U(1)_R\\
\hline
B_1 & \fund & \ov\fund & \frac{1}{N} & 1 & 0 & \frac{1}{2} \\
B_2 & \fund & \ov\fund & \frac{1}{N} & -1 & 0 & \frac{1}{2} \\
C_1 & \ov\fund & \ov\fund & -\frac{1}{N} & 0 & -1 & \frac{1}{2} \\
C_2 & \ov\fund & \ov\fund & -\frac{1}{N} & 0 & 1 & \frac{1}{2} \\
A_{11} & {\bf 1} & \asymm & 0 & -1 & 1 & 1 \\
A_{22} & {\bf 1} & \asymm & 0 & 1 & -1 & 1 \\
A_{12} & {\bf 1} & \symm & 0 & -1 & -1 & 1 \\
A_{21} & {\bf 1} & \symm & 0 & 1 & 1 & 1
\end{array}\\
\intertext{and superpotential}
\label{eq:W-phase-III}
W = \sum_{ij} \Tr(A_{ij}B_iC_j)\, .
\end{gather}
\end{subequations}
Note that, due to the brane bending, this Lagrangian \emph{does not} describe the correct stringy physics at any finite energy scale. However, we expect it to flow to the same infrared fixed point as the correct string theory, i.e., the deformations introduced by the brane bending are irrelevant in the infrared. This is guaranteed if the brane bending preserves all the flavor symmetries and if the infrared theory lacks flavor-singlet relevant operators. Indeed, the relevant, supersymmetry-preserving deformations of an $\mathcal{N}=1$ SCFTs are superpotential operators carrying $U(1)_R$ charge $2/3 \le r < 2$, hence the associated couplings are charged and the deformation breaks some of the flavor symmetries of the IR CFT. Thus, assuming no accidental abelian symmetries appear in the infrared (which can mix with $U(1)_R$), none of these relevant deformations can induced by the brane bending, and we should reach the same infrared fixed point as the strongly coupled theory we started with.

These expectations are supported by very non-trivial checks, detailed in \cite{Garcia-Etxebarria:2016bpb}, due to the S-dualities relating different cusps on the conformal manifold, which are dynamically very nontrivial, but which follow a pattern that can be predicted by a simple analysis of torsion fluxes in the AdS dual, generalizing~\cite{Witten:1998xy}.

\subsection{Deconfinement and quad CFTs} \label{subsec:quadCFTs}

Phase \I\ is much less familiar, even though (as argued in
\cite{Garcia-Etxebarria:2016bpb}) phases of this type turn out to be far more generic than the simpler cases considered above. As in phase \III, the effective field theory arising from
the string theory configuration does not admit a weakly coupled
description at any energy scale, even arbitrarily close to the
cusp. Rather, the physics is given by a weak gauging of diagonal
subgroups of the symmetry group of strongly interacting SCFTs
(heuristically, a ``gluing of SCFTs'' by weakly coupled sectors), very
reminiscent of what happens in $\cN=2$ cases
\cite{Argyres:2007cn,Gaiotto:2009we}. In fact, we will see momentarily
that resolutions of the singularity connect the SCFTs that arise in
the $\cN=1$ case with those appearing in the $\cN=2$ case, as one
might have guessed. Crucially, all of the $\cN=1$ SCFTs arising at the
cusps can be obtained via a series of flows from a Lagrangian description, i.e., there are known non-conformal $\cN=1$ Lagrangian
theories in the same universality class as the $\cN=1$ SCFTs arising
at the cusps in the string configuration.  This is the key point that makes
many of the results in this paper possible.

\begin{figure}
\centering
\includegraphics[width=0.8\textwidth]{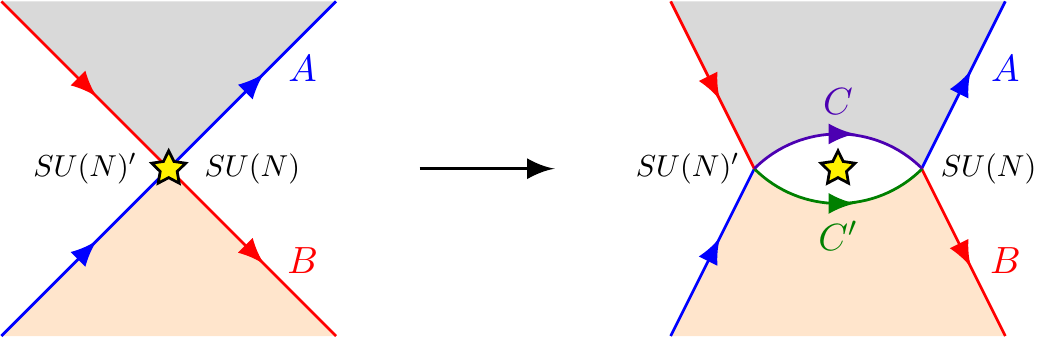}
\caption{Antisymmetric tensor deconfinement as seen from the brane
tiling point of view. On the left hand side we have two NS5
branes, \textcolor{blue}{$A$} and \textcolor{red}{$B$},
intersecting on top of an orientifold plane (the star), giving
rise to an antisymmetric representation of the $SU(N)$ flavour
symmetry group, which here comes from the white wedge on the
left/right (we denote by $SU(N)'$ its orientifold image). The
deconfined description
\cite{Intriligator:1995ne,Berkooz:1995km,Pouliot:1995me} arises
from recombining the NS5 branes in a way that avoids the
orientifold plane. 5-brane charge conservation requires the
appearance of a D5 wrapping the disk bounded by the two new NS5
branes \textcolor{red!30!blue}{$C$} and
\textcolor{green!50!black}{$C'$}, leading to an extra (confining)
$\Sp$ factor.\\ \null\hfill {\footnotesize [Figure reproduced from~\cite{Garcia-Etxebarria:2015hua} with modifications.]}}
\label{fig:deconfinementdimer}
\end{figure}

The following construction explains why these theories exist. Recall
that an intersection of two NS5 branes on top of an orientifold plane
leads to two-index representations of the flavour $SU(N)$ symmetry
group, as sketched on the left half of
figure~\ref{fig:deconfinementdimer}. We will only need to discuss the
case of two-index antisymmetric representations. The strongly coupled
SCFTs mentioned above are precisely those arising from $2k>2$ NS5
branes intersecting atop an orientifold fixed point in the brane
tiling. (Heuristically, the isolated SCFTs appearing at the cusps
are interacting generalisations of the free antisymmetric chiral
multiplet.)

\begin{figure}
\centering
\includegraphics[width=0.7\textwidth]{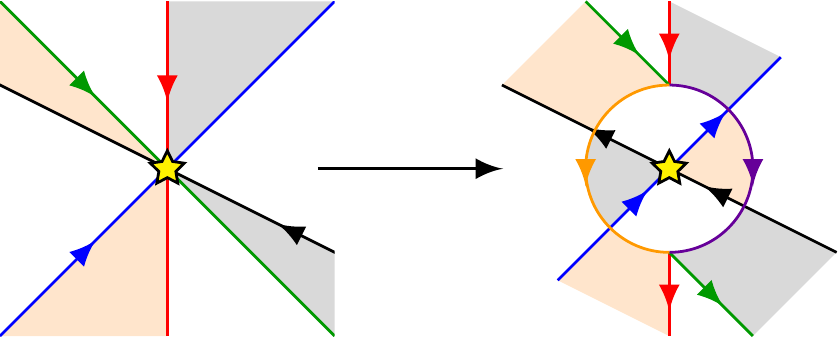}
\caption{Deconfinement in the case of four NS5 brane intersecting on
top of an orientifold point. Recombining two NS5s branes so that
they avoid the intersection gives us a Lagrangian theory in the
same universality class as the original theory. In this case we
have kept two branes intersecting over the orientifold fixed
point, leading to an antisymmetric chiral multiplet. It is
possible to deconfine this multiplet if desired.\\ \null\hfill {\footnotesize [Figure reproduced from~\cite{Garcia-Etxebarria:2015hua} with modifications.]}}
\label{fig:quaddeconfinement}
\end{figure}

In order to give Lagrangian descriptions of these SCFTs, we will
reformulate the old idea of \emph{deconfinement}
\cite{Intriligator:1995ne,Berkooz:1995km,Pouliot:1995me} in the brane
context. In these papers, the authors constructed confining $\cN=1$
theories that lead in the IR to a free chiral $\cN=1$ multiplet in the
two-index antisymmetric representation of a $SU(N)$ flavour group. In
the context of brane tilings, these deconfined descriptions for the
antisymmetric can be understood as coming from a ``bending'' of the
brane system: while there is no motion in moduli space that moves the
branes out of the fixed point, we can (at a cost in energy) recombine
the brane system in a way that avoids the NS5 branes passing through
the orientifold fixed point, as shown in
figure~\ref{fig:deconfinementdimer}. A careful analysis of the
resulting brane system
\cite{Garcia-Etxebarria:2015hua,Garcia-Etxebarria:2016bpb} shows that
it reproduces the deconfined description provided by
\cite{Berkooz:1995km,Pouliot:1995me}.\footnote{The deconfinement
  bubble, when seen from the point of view of the brane tiling, is one
  of the ``geometrically inconsistent'' tilings of
  \cite{Broomhead:2008an,Hanany:2006nm}. We hasten to emphasize that
  thanks to the orientifold projection, and despite the name
  ``inconsistent'' (which we will avoid), the configurations that we
  construct via this method are perfectly cromulent.} But crucially,
the same brane bending operation can be applied in the $k>1$ cases. We
show the $k=2$ example in figure~\ref{fig:quaddeconfinement}. The
Lagrangian theories that we write, although somewhat fearsome when
written in quiver form, can be read off straightforwardly from this
deconfined brane description.

The specific $\cN=1$ SCFTs arising in the $\cC_\bC(\bF_0)$ case
involve four NS5s intersecting over a fixed point. The resulting
theories were denoted $\quadSO^\phi(M)$ and $\quadSp^\Phi(M)$ in
\cite{Garcia-Etxebarria:2015hua,Garcia-Etxebarria:2016bpb}. We will
now briefly review their properties to the extent needed for this
paper. We refer to the original works for a more in-depth discussion.

\begin{figure}
\centering
\begin{subfigure}[b]{0.42\textwidth}
\centering
\includegraphics[height=6cm]{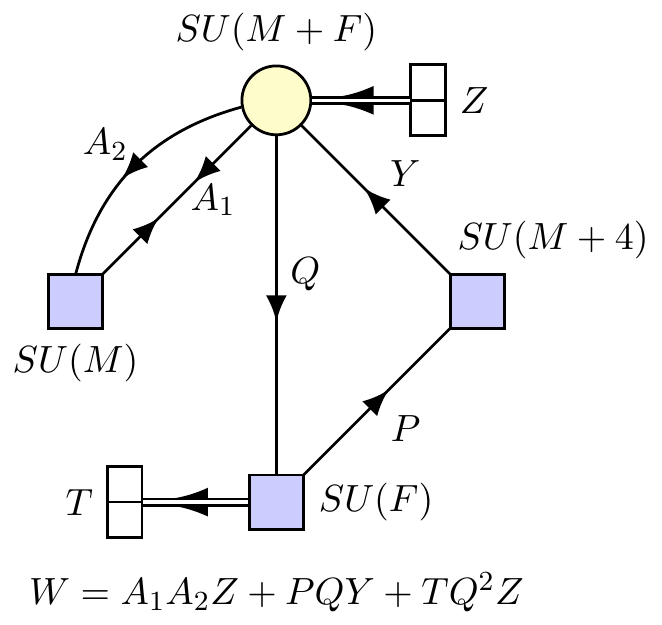}
\caption{Quiver and superpotential for \LSp{A}.}
\label{sfig:qSp-A-quiver}
\end{subfigure}
\hfill
\begin{subfigure}[b]{0.55\textwidth}
\centering
\includegraphics[height=6cm]{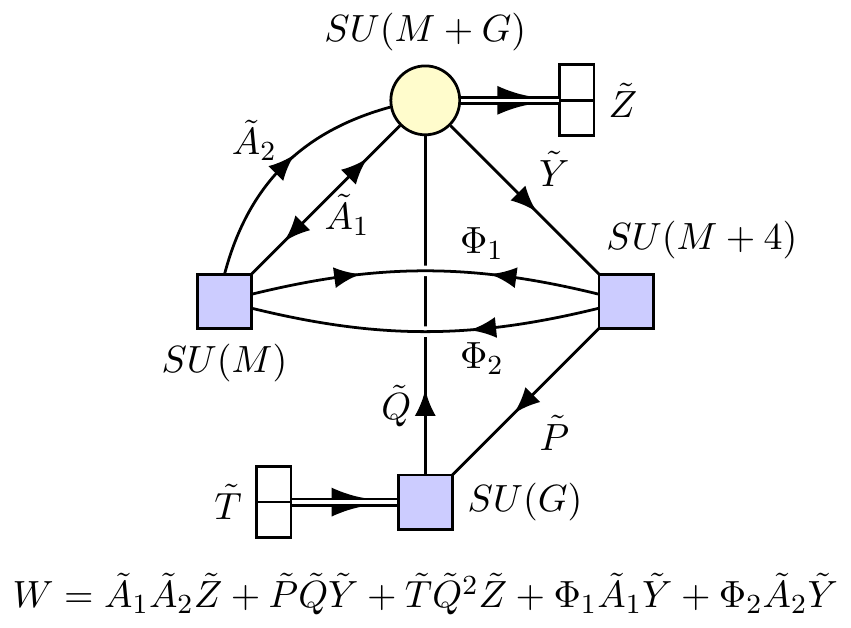}
\caption{Quiver and superpotential for \LSp{B}.}
\label{sfig:qSp-B-quiver}
\end{subfigure}
\caption{Two Lagrangian descriptions of $\quadSp^\phi(M)$, with
$\phi=(-1)^F=(-1)^{G+ M}$.\\ \null\hfill {\footnotesize [Figures reproduced from~\cite{Garcia-Etxebarria:2016bpb} with modifications.]}}
\label{fig:qSp-deconfinement}
\end{figure}

\paragraph{The $\quadSp$ theories.} This family of theories is
parametrised by a positive integer $M$ and a parity $\phi=\pm 1$. We
denote an element of this family by $\quadSp^\phi(M)$. The symmetry
group of the SCFTs is
$\Sp(2M)\times SU(M+4)\times U(1)^2 \times U(1)_R$. Crucially, there
are known Lagrangian theories in the $\quadSp^\phi(M)$ universality
class. We denote these Lagrangian theories by $\LSp{A}(M,F)$ and
$\LSp{B}(M,G)$, where $F$ and $G$ are positive integers such that
$\phi=(-1)^F=(-1)^{G+M}$ and shifting $F$ or $G$ by even numbers leads to theories in the same universality class. The quiver and charge table for $\LSp{A}(M,F)$ are shown in figure~\ref{sfig:qSp-A-quiver} and table~\ref{table:QSp[A]-charges}, respectively.
Note that although we only show $SU(M)\times U(1)_X \subset \Sp(2M)$ explicitly,\footnote{To be precise, the global form of this subgroup is $U(M) = \frac{SU(M)\times U(1)_X}{\mathbb{Z}_M}$. In what follows, we will not track the global form of the global symmetry group for simplicity.}
 it is easy to see that the full symmetry group is indeed $\Sp(2M)$. Our reason
for writing $SU(M) \times U(1)_X$ is that this is the subgroup that is readily
apparent in the brane tiling construction.

\begin{table}
\begin{equation*}
\label{eq:qSpA-charges}
\begin{array}{c|c|ccccccc}
& \SU(M+F) & \SU(M) & SU(M+4) & SU(F)\! & \U(1)_B & \!\!\!U(1)_X\!\!\! & U(1)_Y & U(1)_R\\
\hline
A_1 & \fund & \fund & \singlet & \singlet & -\frac{1}{M+F}  & 1 & \frac{M+4}{2(M+F)} & 1- \frac{M+4}{4(M+F)}\\
A_2 & \fund & \ov\fund & \singlet & \singlet & -\frac{1}{M+F}  & -1 & \frac{M+4}{2(M+F)} & 1- \frac{M+4}{4(M+F)} \\
Y & \ov\fund & \singlet & \fund & \singlet & \frac{1}{M+F} & 0 & 1-\frac{M+4}{2(M+F)} & \frac{M+4}{4(M+F)} \\
Z & \ov\asymm & \singlet & \singlet & \singlet & \frac{2}{M+F} & 0 & -\frac{M+4}{M+F} & \frac{M+4}{2(M+F)} \\
P & \singlet & \singlet & \ov\fund & \fund & \frac{1}{F} & 0 & -1+\frac{M+4}{2F} & 2+\frac{M-4}{4 F} \\
Q & \fund & \singlet & \singlet & \ov\fund & \!\!\!\!-\frac{1}{M+F} \!-\!\frac{1}{F} & 0 & \!\!\!-\!\frac{M+4}{2F}\!+\! \frac{M+4}{2(M+F)} & \,-\frac{M-4}{4F}\!-\! \frac{M+4}{4(M+F)} \\
T & \singlet & \singlet & \singlet & \asymm & \frac{2}{F} & 0 & \frac{M+4}{F} & 2+\frac{M-4}{2F}
\end{array} 
\end{equation*}
\caption{The charge table for \LSp{A}. The $a$-maximized $R$-charge is $U(1)_R^{\text{sc}} = U(1)_R + y_M U(1)_Y + \bigl( \frac{M+4}{4} y_M^2 - \frac{1}{3} \bigr) U(1)_B$ where $y_M$ is the middle root of $9(M+4)y_M^3 + 9 M y_M^2 - 3(3M+4) y_M - M = 0$, 
varying between $y_0 = 0$ and $y_\infty \simeq -0.1018$.}
\label{table:QSp[A]-charges}
\end{table}

A peculiarity of this deconfined description is that there is a field
$Q$ with negative $a$-maximized $R$-charge.\footnote{Specifically, $Q$ has negative $a$-maximized $R$-charge for $M\ge 6$ when $F=1$, $M \ge 4$ when $F=2$, $M \ge 3$ for $F=3,4,5$ and $M\ge 2$ for larger $F$.} This is a more extreme case of the common phenomenon
(already appearing in the conifold theory \cite{Klebanov:1998hh}, for
instance) in which fields in a Lagrangian have $R$-charges below the
unitarity bound. As is well known, this is a not a problem in the
familiar cases: it is perfectly fine for $Q$ to have negative
$R$-charge, since it is not gauge invariant. It is only operators in
the SCFT that need to have $R$-charges above the unitarity bound, but
these operators are built out of gauge invariant combinations of
fundamental fields, and in many cases (such as the conifold) it is
easy to see that the gauge invariant operators do have $R$-charges
above the unitarity bound.

\begin{figure}
\centering
\includegraphics[width=\textwidth]{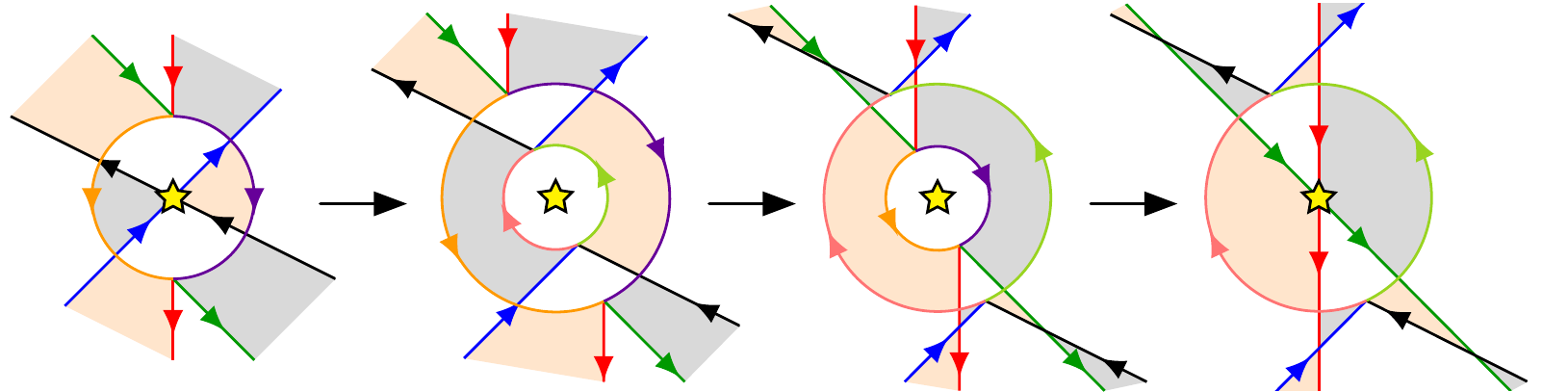}
\caption{``Deconfinement duality'' as seen from the brane tiling, which relates $\LSp{A}(M,F)$ to $\LSp{B}(M,G)$ with the constraint $\phi = (-1)^F = (-1)^{G+M}$.\\ \null\hfill {\footnotesize [Figure reproduced from~\cite{Garcia-Etxebarria:2015hua} with modifications.]}\label{fig:deconfinementduality}}
\end{figure}

In the theories at hand there is a second
phenomenon at play: as explained in
\cite{Garcia-Etxebarria:2015hua,Garcia-Etxebarria:2016bpb}, the $SU(F)$ symmetry is ``trivial,'' i.e., nothing is charged under this symmetry in the infrared. This can be shown by deconfining the antisymmetric tensor field $Z$ in the $\LSp{A}(M,F)$ Lagrangian, then switching to the Seiberg dual description of the $SU(M+F)$ gauge group and reconfining the deconfined $\Sp$ gauge group (which happens to have the right number of flavors to be s-confining), as illustrated in brane tiling description in figure~\ref{fig:deconfinementduality}. This results in the $\LSp{B}(M,G)$ Lagrangian, shown in figure~\ref{sfig:qSp-B-quiver} and table~\ref{table:QSp[B]-charges}, where the $SU(G)$ flavor symmetry was introduced upon deconfinement and the $SU(F)$ becomes manifestly trivial upon reconfinement. The same process can be run in reverse, which brings us back to the $\LSp{A}(M,F')$ Lagrangian, now with an a priori different $F' \ne F$. The only constraint is that $F + G + M$ is even (since $\Sp(n)$ is defined for even $n$), so the parity $\phi = (-1)^F = (-1)^{G+M} = (-1)^{F'}$ remains unchanged. Since different UV descriptions of the same infrared fixed point have different $SU(F)$ or $SU(G)$ symmetries, we conclude that these symmetries must be trivial.

\begin{table}
\begin{equation*}
\label{eq:qSpB-charges}
\begin{array}{c|c|ccccccc}
& \SU(M+G) & \SU(M) & SU(M+4) & SU(G)\! & \U(1)_B & \!\!\!U(1)_X\!\!\! & U(1)_Y & U(1)_R\\
\hline
\tilde A_1 & \ov\fund & \ov\fund & \singlet & \singlet & -\frac{1}{M+G}  & -1 & -\frac{M+4}{2(M+G)} & 1- \frac{M+4}{4(M+G)}\\
\tilde A_2 & \ov\fund & \fund & \singlet & \singlet & -\frac{1}{M+G}  & 1 & -\frac{M+4}{2(M+G)} & 1- \frac{M+4}{4(M+G)} \\
\tilde Y & \fund & \singlet & \ov\fund & \singlet & \frac{1}{M+G} & 0 & -1+\frac{M+4}{2(M+G)} & \frac{M+4}{4(M+G)} \\
\tilde Z & \asymm & \singlet & \singlet & \singlet & \frac{2}{M+G} & 0 & \frac{M+4}{M+G} & \frac{M+4}{2(M+G)} \\
\tilde P & \singlet & \singlet & \fund & \ov\fund & \frac{1}{G} & 0 & 1-\frac{M+4}{2G} & 2+\frac{M-4}{4 G} \\
\tilde Q & \ov\fund & \singlet & \singlet & \fund & \!\!\!\!-\frac{1}{M+G} \!-\!\frac{1}{G} & 0 & \!\!\!\frac{M+4}{2G}\!-\! \frac{M+4}{2(M+G)} & \,-\frac{M-4}{4G}\!-\! \frac{M+4}{4(M+G)} \\
\tilde T & \singlet & \singlet & \singlet & \ov\asymm & \frac{2}{G} & 0 & -\frac{M+4}{G} & 2+\frac{M-4}{2G} \\
\Phi_1 & \singlet & \fund & \fund & \singlet & 0 & 1 & 1 & 1 \\
\Phi_2 & \singlet & \ov\fund & \fund & \singlet & 0 & -1 & 1 & 1 \\
\end{array} 
\end{equation*}
\caption{The charge table for \LSp{B}. Note that this can be obtained from table~\ref{table:QSp[A]-charges} by charge conjugating $U(1)_X$, $U(1)_Y$ and the non-abelian groups, replacing $F$ with $G$ and adding the mesons $\Phi_1 = A_1 Y$ and $\Phi_2 = A_2 Y$.}
\label{table:QSp[B]-charges}
\end{table}

Triviality implies that gauge invariant operators charged under $SU(F)$ or $SU(G)$ must disappear in the infrared. This removes many operators that would otherwise violate the unitarity bound. More generally, to the extent that we have been able to check, \emph{all} operators appearing to violate the unitarity bound are lifted in the infrared, whether by this mechanism or due to other quantum effects. In many cases, this can be seen by choosing a convenient dual description in which the quantum effects in question become obvious, tree-level properties. Alternately, one can express the SCI in terms of a different R-symmetry (not the $a$-maximized, superconformal one) under which all the fundamental chiral superfields have charge $0 < r' < 2$. Computing the SCI order-by-order in this alternate basis, it is straightfoward to check that all the problematic operators cancel from the index, up to the order computed.

Looking ahead, this
subtlety will affect our Lagrangian description of the $R_{2,2n+1}$
theories with $n>1$. So while this is a complication one should keep
in mind in these cases (particularly when expanding the SCI), we
believe that it is a purely technical one. In practice we often deal
with this subtlety by computing the index in a modified basis, as described above,
which is sufficient to check many dualities in great detail.

\begin{figure}
\centering
\begin{subfigure}{0.45\textwidth}
\centering
\includegraphics[width=\textwidth]{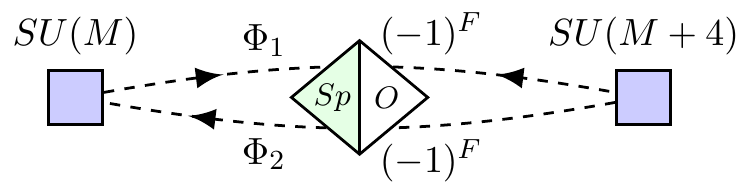}
\caption{Notation for $\quadSp$.}
\label{sfig:quad-notation-Sp}
\end{subfigure}
\hspace{1cm}
\begin{subfigure}{0.45\textwidth}
\centering
\includegraphics[width=\textwidth]{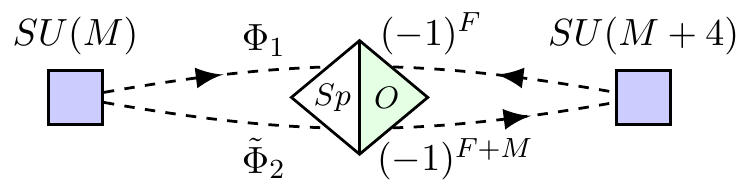}
\caption{Notation for $\quadSO$.}
\label{sfig:quad-notation-SO}
\end{subfigure}
\caption{Abstract quiver notation for $\quadSp$ and $\quadSO$. The
dashed lines indicate the mesons $\Phi_1$ and $\Phi_2$ ($\quadSp$) or $\tilde \Phi_2$ ($\quadSO$), and the attached labels are their
associated parities. The shaded half of the diamond distinguishes
between \quadSp\ and \quadSO.\\ \null\hfill {\footnotesize [Figures reproduced from~\cite{Garcia-Etxebarria:2016bpb} with modifications.]}}
\label{fig:quad-notation}
\end{figure}

In order to keep track of the SCFTs in a concise way, and also to
emphasize the fact that these sectors correspond to strongly coupled
SCFTs even near the cusps, we introduce the ``abstract quiver'' notation in
figure~\ref{sfig:quad-notation-Sp}. The dashed lines correspond to the
mesons $\Phi_1 = A_1 Y$ and $\Phi_2 = A_2 Y$, which are elementary fields in $\LSp{B}$. We also indicate the
parity $\phi = (-1)^F$ in the diagram (redundantly, next to each meson line, for reasons to be explained below). 

In addition to the mesons $\Phi_{1,2}$, which combine into the bifundamental $(\fund,\fund)$ representation of $\Sp(2M) \times \SU(M+4)$, the $\quadSp$ theory contains baryons $\mathcal{A}_k, \mathcal{S}_k$, which can be expressed in terms of the \LSp{A}\ fields as follows:\footnote{For simplicity we show the case $F>2$. For $F=1$ ($F=2$) we have $\mathcal{S}_1 = P$ ($\mathcal{S}_0 = T$).}
\begin{equation}
\begin{aligned}
  \label{eqn:quadCFTbaryons}
  \mathcal{A}_k &= A_1^k A_2^{M-k} Q^F \,, & 0&\le k \le M \,,\\
  \mathcal{S}_k &= Z^{\frac{F+k-4}{2}} R^{M+4-k} \,, & 0&\le k \le M+4\,, & (-1)^k &= (-1)^F \,.
\end{aligned}
\end{equation}
Note that the baryons $\mathcal{A}_k$ combine into the (irreducible) $M$-index antisymmetric tensor representation of $\Sp(2M)$.

\begin{figure}
\centering
\begin{subfigure}[t]{0.48\textwidth}
\centering
\includegraphics[height=6cm]{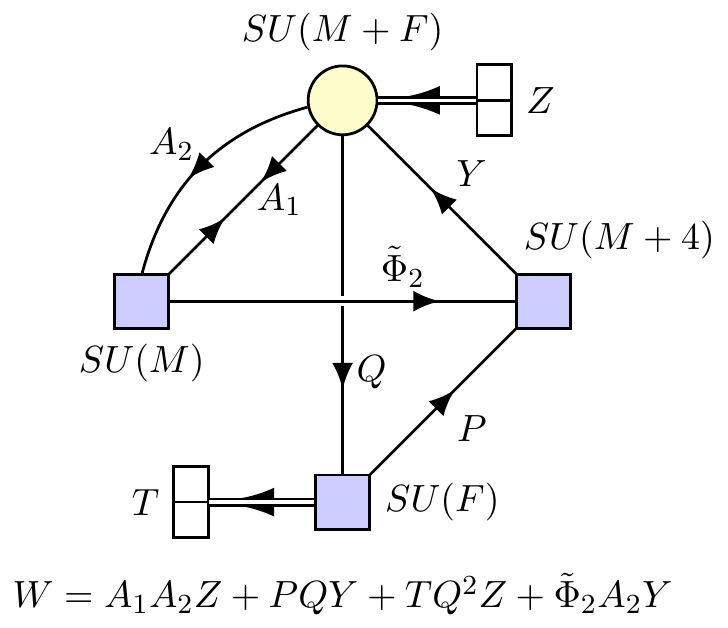}
\caption{Quiver and superpotential for \LSO{A}.}
\label{sfig:qSO-A-quiver}
\end{subfigure}
\begin{subfigure}[t]{0.48\textwidth}
\centering
\includegraphics[height=6cm]{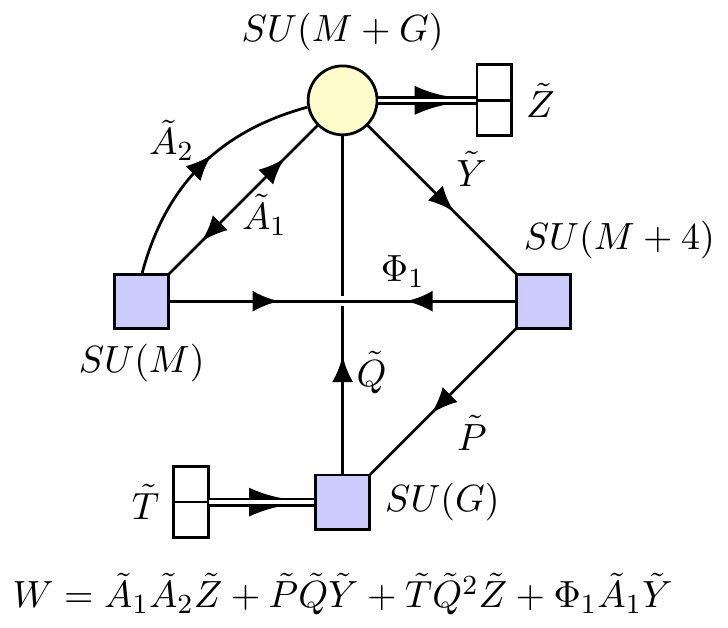}
\caption{Quiver and superpotential for \LSO{B}.}
\label{sfig:qSO-B-quiver}
\end{subfigure}
\caption{The two deconfined quivers for $\quadSO$.\\[-\baselineskip] \null\hfill {\footnotesize [Figures reproduced from~\cite{Garcia-Etxebarria:2016bpb} with modifications.]}}
\label{fig:qSO-deconfinement}
\end{figure}

\paragraph{The $\quadSO$ theories.} A second family of
theories that arise in the same context are the $\quadSO^\phi(M)$ theories. These theories have global symmetry group
$SU(M)\times \Spin(2M+8)\times U(1)^2\times U(1)_R$. Lagrangian
theories in the $\quadSO^\phi(M)$ universality class are also known, as shown in figure~\ref{fig:qSO-deconfinement} and table~\ref{table:QSO[A]-charges},
but unfortunately these Lagrangians do not preserve the full symmetry group of the theory, only a full-rank subgroup $SU(M)\times SU(M+4)\times U(1)^3\times U(1)_R$, where $SU(M+4)\times U(1)_Y$ enhances to $\Spin(2M+8)$ in the infrared under the standard embedding $U(n) \subset \SO(2n)$.\footnote{As with the $\quadSp$ theory, we are being somewhat imprecise about the global form of this manifest global symmetry group. In this case, the embedding $U(n)\subseteq \SO(2n)$ lifts to $\tilde{U}(n) \subseteq \Spin(2n)$ where $\tilde{U}(n)$ is a certain double-cover of $U(n)$, e.g., $\tilde{U}(2k) = \frac{\SU{(2k)\times\U(1)}}{\mathbb{Z}_k}$.}

\begin{table}
\begin{equation*}
\begin{array}{c|c|ccccccc}
& \SU(M+F) & \SU(M) & SU(M+4) & SU(F)\! & \U(1)_B & \!\!\!U(1)_X\!\!\! & U(1)_Y & U(1)_R\\
\hline
A_1 & \fund & \fund & \singlet & \singlet & -\frac{1}{M+F}  & 1 & \frac{M+4}{2(M+F)} & 1- \frac{M+4}{4(M+F)}\\
A_2 & \fund & \ov\fund & \singlet & \singlet & -\frac{1}{M+F}  & -1 & \frac{M+4}{2(M+F)} & 1- \frac{M+4}{4(M+F)} \\
Y & \ov\fund & \singlet & \fund & \singlet & \frac{1}{M+F} & 0 & 1-\frac{M+4}{2(M+F)} & \frac{M+4}{4(M+F)} \\
Z & \ov\asymm & \singlet & \singlet & \singlet & \frac{2}{M+F} & 0 & -\frac{M+4}{M+F} & \frac{M+4}{2(M+F)} \\
P & \singlet & \singlet & \ov\fund & \fund & \frac{1}{F} & 0 & -1+\frac{M+4}{2F} & 2+\frac{M-4}{4 F} \\
Q & \fund & \singlet & \singlet & \ov\fund & \!\!\!\!-\frac{1}{M+F} \!-\!\frac{1}{F} & 0 & \!\!\!-\!\frac{M+4}{2F}\!+\! \frac{M+4}{2(M+F)} & \,-\frac{M-4}{4F}\!-\! \frac{M+4}{4(M+F)} \\
T & \singlet & \singlet & \singlet & \asymm & \frac{2}{F} & 0 & \frac{M+4}{F} & 2+\frac{M-4}{2F} \\
\tilde \Phi_2 & \singlet & \fund & \ov\fund & \singlet & 0 & 1 & -1 & 1
\end{array} 
\end{equation*}
\caption{The charge table for \LSO{A}, which can be obtained from table~\ref{table:QSp[A]-charges} by adding the elementary meson $\tilde \Phi_2$ (flipping $\Phi_2 = A_2 Y$). (Similarly, the charge table for $\LSp{B}$ can be obtained from table~\ref{table:QSp[B]-charges} by removing the elementary meson $\Phi_2$.) Doing so breaks $\Sp(2M) \to \SU(M)\times U(1)_X$, but leads to the accidental enhancement $\SU(M+4)\times U(1)_Y \to \SO(2M+8)$ at the infrared fixed point. The $a$-maximized $R$-charge is $U(1)_R^{\text{sc}} = U(1)_R + x_M U(1)_X - \bigl( \frac{M}{4} x_M^2 + \frac{1}{3} \bigr) U(1)_B$ where $x_M$ is the middle root of $9 M x_M^3 + 9 (M +4) x_M^2 - 3(3M+8) x_M - (M+4) = 0$, 
varying between $x_0 \simeq -0.1381$ and $x_\infty \simeq -0.1018$.}
\label{table:QSO[A]-charges}
\end{table}

As before, there are two families of Lagrangians: $\LSO{A}(M,F)$ and $\LSO{B}(M,G)$, related by a deconfinement duality analogous to figure~\ref{fig:deconfinementduality} for $(-1)^F=(-1)^{G+M}$. However, unlike before the Lagrangians $\LSO{A}(M,F)$ and $\LSO{B}(M,G)$ are isomorphic after relabeling $\tilde{A}_1 \to A_2$, $\tilde{A}_2 \to A_1$, $\Phi_1 \to \tilde{\Phi}_2$, charge conjugating the gauge group and $SU(M+4)\times U(1)_Y$, and identifying $SU(G)$ with the charge conjugate of $SU(F$). For odd $M$, combining the deconfinement duality and this isomorphism, we conclude that the two parities of $F$ generate isomorphic CFTs. In fact, a more general class of deconfinement dualities shows that this remains true for even $M$~\cite{Garcia-Etxebarria:2015hua}, hence $\quadSO^+(M) \cong \quadSO^-(M)$, unlike $\quadSp^+(M) \not\cong \quadSp^-(M)$.

Note that the isomorphism between $\quadSO^+(M)$ and $\quadSO^-(M)$ involves the $\mathbb{Z}_2$ ``parity'' outer automorphism of $\Spin(2M+8)$. For instance, the baryonic operators $\mathcal{S}_k$ of the $\quadSO^\phi(M)$ CFT (see \eqref{eqn:quadCFTbaryons}) combine into a Weyl spinor representation of $\Spin(2M+8)$ whose chirality is determined by the flavor parity $\phi$.
For this reason, it is important to track $F$ parity when the $\quadSO(M)$ CFT is coupled to other sectors by a (partial) gauging of $\Spin(2M+8)$ or by superpotential couplings to $\quadSO(M)$ operators, as the $\mathbb{Z}_2$ outer automorphism acts nontrivially on these couplings.

In particular, when $\quadSO(M)$ is embedded in a larger brane tiling, typically only $SU(M+4)\times \U(1) \subset \Spin(2M+8)$ remains unbroken, 
and we need to specify
which precise subgroup this is. We do so as follows: consider the
mesons $\Phi_1 = A_1 Y$ and $\tilde \Phi_2 = \tilde A_2 \tilde Y$, which are composites in $\LSO{A}$
and $\LSO{B}$, respectively, and elementary fields in the other phase. We
attach to each of these mesons a parity given by the $F$ parity of the
phase in which this meson is \emph{composite} (this choice of
convention allows us to use the same prescription for the \quadSp\
case). That is, we assign parity $(-1)^F$ to $\Phi_1$ and
$(-1)^G$ to $\tilde \Phi_2$. The two parities are related by $(-1)^M$, so for
any fixed $M$ either parity can be specified (we will often specify
both). Specifying these parities in figure~\ref{sfig:quad-notation-SO}
tells us which precise $SU(M+4)\times U(1)$ subgroup of $\Spin(2M+8)$
is realized in the brane tiling.

\begin{figure}
	\centering
	\includegraphics[width=\textwidth]{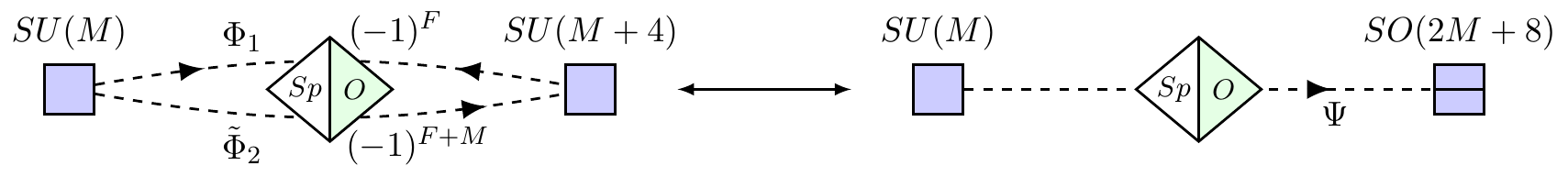}
	\caption{Comparison of different abstract quiver notations for the $\quadSO$ CFT. In the right-hand diagram the enhanced $\Spin(2M+8)$ symmetry is manifest.}
	\label{fig:Manifest-qSO}
\end{figure}

Finally, we discuss certain ``partially flipped'' versions of the $\quadSO$ theory that will appear in our construction of the $R_{2,k}$ theory for odd $k$. To simplify the discussion, we adopt an abstract quiver notation that makes the full $\Spin(2M+8)$ symmetry manifest, as shown in figure  \ref{fig:Manifest-qSO}. Next, we decompose $\Spin(2M+8) \to \Spin(2M+8-P) \times \Spin(P)$, whereupon the meson $\Phi$ in the $(\fund,\fund)$ representation of $\SU(M)\times \Spin(2M+8)$ decomposes into mesons $\Psi$ and $\Psi_P$ in the $(\fund,\fund,\singlet)$ and $(\fund,\singlet,\fund)$ representations of $\SU(M)\times \Spin(2M+8-P) \times \Spin(P)$, respectively. Next, we flip the meson $\Psi_P$ to obtain a meson $\tilde{\Psi}_P$ in the $(\ov\fund,\singlet,\fund)$ representation of $\SU(M)\times \Spin(2M+8-P) \times \Spin(P)$.

\begin{figure}
\centering
\includegraphics[width=\textwidth]{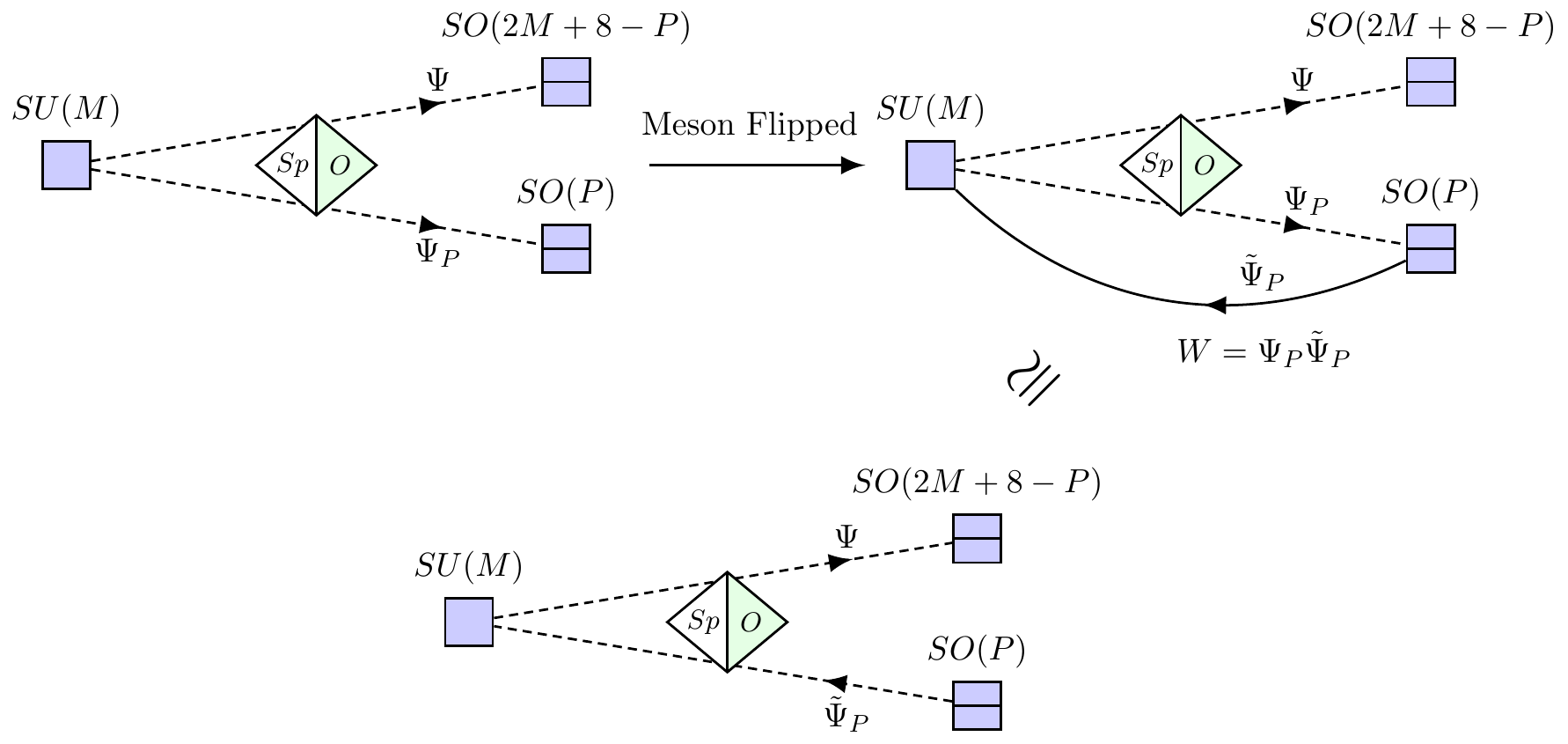}
\caption{Partially flipping the $\quadSO$ CFT in the language of the abstract quiver.}
\label{sfig:Flipped-qSO}
\end{figure}

The entire process is illustrated in figure~\ref{sfig:Flipped-qSO}. In this way, we obtain a closely related CFT with an $\SU(M)\times \Spin(2M+8-P) \times \Spin(P) \times \U(1)^2 \times U(1)_R$ symmetry. As we will see in~\S\ref{subsec:oddk}, gauging part of the global symmetry of this theory (for $P=2$ and even $M=k-1$) generates a flow to the $R_{2,k}$ (odd $k$) CFT plus a free chiral multiplet.

A similar partial flipping can applied to the $\quadSp$ theory, but as we will not make use of it in the present paper, the details are left as an exercise for the interested reader.

\subsection{Flux assignments and duality}

The map between fluxes at infinity and the choice of phase goes as
follows. Choose a basis of $H^3(X_5;\tbZ)=\bZ_2\oplus\bZ_2$ given by
two elements $\vev{a}$ and $\vev{b}$.\footnote{These basis elements
are associated to any two neighbouring non-compact divisors in the
toric diagram for $\cC_\bC(\bF_0)$. We refer to
\cite{Garcia-Etxebarria:2016bpb} for details of this construction.}
Then we can parametrise the NSNS flux in this basis by
$H_3=\alpha\vev{a}+\beta\vev{b}$. In terms of this basis, we have the
following dictionary between phases and $H_3$ flux:
\[
\begin{tabular}{r|cccc}
Phase: & \I & \II & \tII & \III \\
\hline
$H_3$ torsion & (00) & (10) & (01) & (11)
\end{tabular}
\]
Note that while the $(10)$ and $(01)$ choices of flux are different,
they are related by symmetries of the geometry, and thus lead to
isomorphic physics at the cusp.

This prescription covers the NSNS half of the dictionary between flux
and physics at the cusp. Coming back to the case of O3 planes in flat
space, this would be analogous to explaining that non-trivial NSNS
torsion takes us from $\fso$ to $\fsp$ in the $\cN=4$ theory. As in
that case, the choice of RR torsion is somewhat more subtle, and has
to do with the ranks of the gauge factors. The RR torsion can also be
parametrised as $F_3=\alpha_F\vev{a}+\beta_F\vev{b}$. Recall that a
shift of $C_0$ by 1 unit acts as $F_3\to F_3+H_3$, so for a given
nonzero choice of $H_3$ there will be multiple assignments of $F_3$
that lead to the same perturbative physics. For instance, for phase
\II, with $H_3=\vev{a}$, it is only the choice of $\beta_F$ that can
have an effect on the field theory. And indeed, for phase \II\ one
finds \cite{Garcia-Etxebarria:2016bpb}
\begin{align}
\II\colon&\qquad \beta_F \equiv N  \mod 2\, ,\\
\intertext{where $N$ refers to the rank in
figure~\ref{sfig:F03-II-quiver}.  Similarly for phase \III}
\III\colon& \qquad \alpha_F+\beta_F \equiv N \mod 2\\
\intertext{with $N$ as in figure~\ref{sfig:F03-III-quiver}.
Finally, in the case of phase
\I\ in figure~\ref{sfig:F03-I-quiver} we have two parities affecting the physics at the cusp. Let us denote $\phi_1=(-1)^{F_1}$ and $\phi_2=(-1)^{F_2}$. Then}
\I\colon& \qquad \beta_F\equiv F_1 \mod 2\\
&\qquad \alpha_F \equiv F_2 \mod 2\, .
\end{align}

With this dictionary between fluxes and parities in hand we can now
read the duality multiplets. We have defined $N$ in
figure~\ref{fig:F0-O3-phases} so that if the discrete flux agrees
between two phases with the same choice of $N$, then the theories are
in fact dual, so in what follows we will only specify the parities. In
order to keep track of this, we denote phase \I\ with parities
$\phi_1$ and $\phi_2$ as $\I^{\phi_1\phi_2}$, and we write $\II^\phi$
and $\III^\phi$ for phases \II\ and \III\ with $\phi\df (-1)^N$.

Consider for example $\I^{++}$. According to our discussion above, we
have $H_3=F_3=0$, so this phase is expected to be a $SL(2,\bZ)$
singlet. That is, all cusps in the conformal manifold induced by
changing the IIB axio-dilaton have the same effective
description. This is no longer true for $\I^{+-}$. This phase has
$H_3=0$ and $F_3=\vev{a}$. Acting with the $S$ generator of
$SL(2,\bZ)$ we obtain $H_3=\vev{a}$ and $F_3=0$, which corresponds to
$\II^+$. So in this case we have a non-trivial duality between the
ordinary cusp of type \II\ (with $N$ even) and a more exotic theory of
type \I, involving the \quadSp\ and \quadSO\ theories discussed above.

Other cases can be worked out similarly. For instance, $\I^{-+}$ is
dual to $\tII^+$ and $\I^{--}$ is dual to $\III^+$. Perhaps more
interestingly, the conformal manifold of the $\II^-$ theory involves
cusps of type $\tII^-$ and $\III^-$, as one can readily verify. Since
due to the very symmetric form of $\bF_0$ phases \II\ and \tII\ are
isomorphic, we can think of this case as a duality between $\III^-$
and $\II^-$.

\section{\alt{$\cN=1$}{N=1} Lagrangians for the \alt{$R_{2,k}$}{R2k} SCFTs}
\label{sec:Lagrangians}

In the previous section we have reviewed which kind of theories appear
in the cusps of the conformal manifold of the $\cC_\bC(\bF_0)$ SCFTs
for different choices of discrete fluxes, and in particular we have
given Lagrangians for all of them. We have also explained how all of
these theories are related by S-duality. Coming back to the
diagram~\eqref{eq:reduced-duality-square}, this provides all of the
information that we need on the left-hand side of the diagram.
Next, by Higgsing these theories we will obtain $\mathcal{N}=1$ Lagrangians flowing to the $\mathcal{N}=2$ theories that appear on the right-hand side of this diagram.

\bigskip

First, we will need to know the relation between the $\cN=2$ and
$\cN=1$ symmetry groups. The following material is standard, so we
will be brief. A longer discussion can be found in
\cite{Benini:2009mz,Tachikawa:2013kta,Tachikawa:2015bga}, for
instance. Consider an $\cN=2$ SCFT with symmetry group $G$. What we
mean by this is that the SCFT has symmetry group $G$ and an
$R$-symmetry group $SU(2)_R \times \Ur$. The $\cN=2$ theory
will have Coulomb and Higgs branches, which we explore by turning on
operators neutral under $SU(2)_R$ and $\Ur$ respectively.

The $\cN=2$ theory can be viewed as a $\cN=1$ SCFT, with an
$R$-symmetry $U(1)_R$, and an additional global symmetry $U(1)_\nR$
(so, from the $\cN=1$ point of view, our theory has non-$R$ global
symmetry $G\times U(1)_\nR$). The $\cN=1$ generators can be written in
terms of the $\cN=2$ generators as
\[
\label{eq:N=2 to N=1}
U(1)_R & = \frac{2}{3} \UsuR+\frac{1}{3} \Ur\, ,\\
U(1)_{\nR}&=\UsuR - \Ur
\]
where we have chosen a Cartan generator $\UsuR$ of $SU(2)_R$.
Our conventions for $\UsuR$ are that the $\UsuR$
charge is twice the spin (so, for instance, the spin-$1/2$
representation of $SU(2)_R$ has $\QsuR=\pm 1$).
Equivalently, we can write the $\cN=2$ generators in terms of the
$\cN=1$ generators as
\[
\UsuR &=U(1)_R+\frac{1}{3}U(1)_{\nR}\\
\Ur &=U(1)_R-\frac{2}{3}U(1)_{\nR}\, .
\]

 After establishing R charge relations between the $\cN=2$ and
 $\cN=1$ symmetry groups, we now consider the effect on the field theory of partially resolving
the $\cC_\bC(\bF_0)$ singularity to two copies of
$\bC^2/\bZ_2\times \bC$. We choose to start by studying the effect of
the partial resolution on phase \III, for reasons that will become
clear momentarily. By using the general methods of
\cite{GarciaEtxebarria:2006aq} (or simply by trial and error) it is
easy to conclude that turning on a vev for $C_1$ proportional to the
identity triggers the relevant partial resolution in the geometry,
where we are using the nomenclature in the charge
table~\eqref{eqref:F03-III-charges}. We will refer to giving such a
vev in a more gauge-invariant way as giving a vev to ``$C_1^N$'',
which is a shorthand for the gauge invariant baryon $\det(C_1)$.

This vev spontaneously breaks one linear combination of the four $U(1)$ symmetries of the parent theory, and Higgses the $\SU(N) \times \SU(N)$ gauge group
to the diagonal $\SU(N)$. After Higgsing, $C_2$ decomposes in $\adj \oplus \singlet$ and $B_{1,2}$ into two copies of $\ov\ssymm\oplus\ov\sasymm$, where the superpotential~\eqref{eq:W-phase-III} gives the antisymmetric part of $B_1$ and the symmetric part of $B_2$ a mass with $A_{11}$ and $A_{21}$, respectively.
After integrating out the massive fields and dropping the decoupled chiral field $v$ controlling the vev, an accidental $\Uz$ symmetry emerges in the infrared.
Putting all the pieces together, relabeling the fields, and choosing a new basis for the $U(1)$ symmetries that will be convenient later, we obtain:
\begin{subequations}
\begin{gather}
\renewcommand{\arraystretch}{1.4}
\begin{array}{c|c|cccc} \label{eq:flipped-Higgsed-III}
& SU(N) & \Ux & \Uz & U(1)_{\nR} & U(1)_R \\
\hline
\Phi  & \adj & 0 & 0 & -2 & \frac{2}{3} \\
S & \symm & \frac{1}{2}-\frac{1}{2 N} & \frac{1}{N} & 1 & \frac{2}{3} \\
S_b & \symmb & \frac{1}{2 N}-\frac{1}{2} & -\frac{1}{N} & 1 & \frac{2}{3} \\
A & \asymm & -\frac{1}{2 N}-\frac{1}{2} & \frac{1}{N} & 1 & \frac{2}{3} \\
A_b & \asymmb & \frac{1}{2 N}+\frac{1}{2} & -\frac{1}{N} & 1 & \frac{2}{3} \\
\phi  & 1 & 0 & 0 & -2 & \frac{2}{3}
\end{array}\\
\intertext{with superpotential}
W = \Tr(A \Phi A_b  +A \phi A_b  +S \Phi S_b  +S \phi S_b)\, . \label{eq:flipped-Higgsed-III-W}
\end{gather}
\end{subequations}
This is the $\cN=1$ description of the $\cN=2$ theory with quiver (in
$\cN=2$ notation)
\[
\label{eq:III-N=2-quiver}
\includegraphics[scale=1]{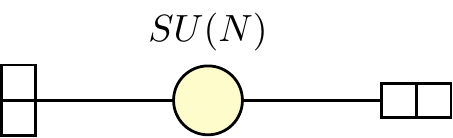}
\]
coupled to a gauge singlet chiral multiplet $\phi$.

To understand the appearance of this ``extra'' chiral field $\phi$, note that the previously-isolated singularity blows up into a $\mathbb{P}^1$ line of singularities following the partial resolution. Thus, we naturally interpret $\phi$ as the center of mass mode for the D3 brane stack moving along the $\mathbb{P}^1$. In the decompactification limit $\mathbb{P}^1 \to \mathbb{C}$, $\mathcal{N}=2$ SUSY is restored, and $\phi$ should be paired with a massless photon to complete an $\cN=2$ vector multiplet. Specifically, as dictated by $\phi$'s superpotential interactions in~\eqref{eq:flipped-Higgsed-III-W}, the massless photon in question gauges the global symmetry $\Uz$ in~\eqref{eq:flipped-Higgsed-III}. Since D-branes naturally engineer $\U(N)$ gauge theories rather than $\SU(N)$ gauge theories, the appearance of this extra photon is not a surprise.\footnote{Likewise, this is the result if we write the UV theory in~\eqref{eq:W-phase-III} as $\U(N)\times\U(N)$ gauge theory---reinstating the $\U(1)$ factors that we have ignored until now---with the caveat that one of these $U(1)$s is anomalous and so the associated photon gets a Green-Schwarz mass. Indeed, the ``accidental'' symmetry $\Uz$ can be viewed as arising from this anomalous $\U(1)$.} However, as its interactions make it infrared free, this photon (along with $\phi$) will decouple from the infrared CFT.

Thus, in order to focus on the interacting sector of the resulting $\cN=2$ SCFT, we omit this extra $U(1)$ and also remove its $\cN=2$ superpartner by adding a
new singlet field $\ov\phi$ with opposite global charges and
modifying the superpotential to
\[
W = \Tr(A \Phi A_b  +A \phi A_b  +S \Phi S_b  +S \phi S_b) + \phi\ov\phi\, .
\]
Following, e.g., \cite{Gaiotto:2015usa}, we refer to this operation (which can be applied to any gauge invariant operator $\phi$) as ``flipping $\phi$''. In this case, flipping $\phi$ gives it a mass, and after integrating it out we obtain the manifestly $\cN=2$ theory with quiver~\eqref{eq:III-N=2-quiver}.

Our general strategy now allows us to give a Lagrangian description of
the strongly coupled limit of this theory. Before doing that, we will
briefly review what is known about this case in the class $\cS$
context. As it turns out, the strong coupling behaviour depends on
whether $N$ is even or odd --- in agreement with our observation above
that the duals of $\III^+$ and $\III^-$ are rather different: the
former is dual is to $\I^{--}$ while the latter is dual to $\II^-$.

\begin{figure}
\centering
\includegraphics[width=0.5\textwidth]{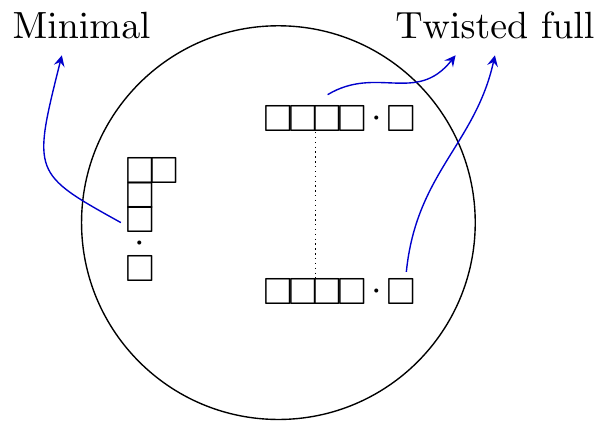}
\caption{Class $\cS$ description of the $R_{2,k}$ theories.}
\label{fig:R_{2,k}}
\end{figure}

The odd $N$ case was studied from the class $\cS$ perspective in
\cite{Chacaltana:2014nya}. Setting $N=k+1$ (where $k$ is even), we
have that the strongly coupled dual of~\eqref{eq:III-N=2-quiver} is
given by $R_{2,k} \hookleftarrow \Sp(k)$, where $R_{2,k}$ is defined
to be the SCFT arising from putting the six dimensional $A_k$
$(2,0)$ theory on a sphere with three punctures. Two of these
punctures are full twisted punctures, and the other is a minimal
untwisted puncture, see figure~\ref{fig:R_{2,k}}. We use the
notation $\cT\hookleftarrow G$ to indicate that we weakly gauge a
subgroup $G$ of the global symmetry group of the SCFT $\cT$. The
$R_{2,k}$ SCFT for $k$ even has global symmetry group
$\Sp(2k)\times U(1)$ and central charges
\begin{subequations}
\label{eq:R_{2,even}-ac}
\begin{align}
24a & = 1+\frac{19}{2}k+\frac{7}{2}k^2\, ,\\
12c & = 1+5k+2k^2\, .
\end{align}
\end{subequations}
The $\Sp(2k)$ global symmetry has level
$k_{\Sp(2k)}=k+2$. Additionally, it has a Witten (or ``global'') anomaly
\cite{Witten:1982fp,Tachikawa:2018rgw}. The Argyres-Wittig theory
\cite{Argyres:2007tq} arises in the case $N=3$, or equivalently $k=2$.

The even $N$ case was studied from the class $\cS$ perspective in
\S3.5.4 of \cite{Chacaltana:2012zy}. Again setting $N=k+1$, the S-dual is expected to be $R_{2,k}\hookleftarrow \SO(k+2)$, where the $R_{2,k}$ CFT for odd $k$ has global symmetry $\Spin(2k+4)\times U(1)$ with level $k_{\Spin(2k+4)}=2k$. 

To be precise, the S-dual depends on the global structure of the gauge group in the original quiver~\eqref{eq:III-N=2-quiver}. Since $N$ is even, there is a $\mathbb{Z}_2$ subgroup of the $\mathbb{Z}_N$ center of $\SU(N)$ under which two-index tensor reps are neutral, so we can choose the gauge group to be either $\SU(N)$, $(SU(N)/\mathbb{Z}_2)_+$ or $(SU(N)/\mathbb{Z}_2)_-$, where in the latter cases the subscript indicates the absence ($+$) or presence ($-$) of a discrete theta angle~\cite{Gaiotto:2010be,Aharony:2013hda}. In each case, there is a $\mathbb{Z}_2$ one-form symmetry~\cite{Gaiotto:2014kfa}, which is either electric, magnetic, or dyonic, respectively.

\begin{figure}
\begin{equation*}
	\begin{tikzpicture}
	\node at (0,0) {$SU(N)$};
	\node at (4,0) {$(SU(N)/\mathbb{Z}_2)_+$};
	\node at (8,0) {$ (SU(N)/\mathbb{Z}_2)_- $};
	\node at (0,-2) {$R_{2,k}\hookleftarrow \SO(N+1)_{+}$};
	\node[rotate=0,scale=1.7] at (0,.5) {$\curvearrowright $};
	\node[rotate=0,scale=1.7] at (4,.5) {$\curvearrowright $};
	\node[rotate=0,scale=1.7] at (8,.5) {$\curvearrowright $};
	\node[rotate=180,scale=1.7] at (0.1,-2.5) {$\curvearrowleft $};
	\node[rotate=180,scale=1.7] at (4.1,-2.5) {$\curvearrowleft $};
	\node[rotate=180,scale=1.7] at (8.1,-2.5) {$\curvearrowleft $};
	\node[rotate=90,scale=1.4] at (0,-1) {$\longleftrightarrow$};
	\node[rotate=90,scale=1.4] at (4,-1) {$\longleftrightarrow$};
	\node[rotate=90,scale=1.4] at (8,-1) {$\longleftrightarrow$};
	\node at (.25,-1) {$S$};\node at (4.25,-1) {$S$};\node at (8.25,-1) {$S$};
	\node at (0,1) {$T$};\node at (4,1) {$T$};\node at (8,1) {$T$};
	\node at (4,-2) {$R_{2,k}\hookleftarrow \text{Spin}(N+1) $};
	\node at (0,1) {$T$};\node at (0,-3) {$T$};
	\node at (0,1) {$T$};\node at (4,-3) {$T$};
	\node at (0,1) {$T$};\node at (8,-3) {$T$};
	\node at (8,-2) {$R_{2,k}\hookleftarrow \SO(N+1)_{-}  $};
\end{tikzpicture}
\end{equation*}
\caption{Duality orbits of the quiver \eqref{eq:III-N=2-quiver} for $N=4m$.}
\label{SUn-dual-2}
\end{figure}
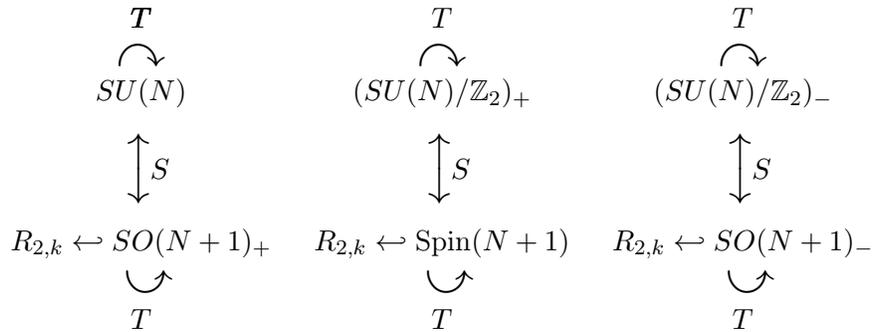

Thus, the S-dual should have a $\mathbb{Z}_2$ one-form symmetry as well. This is the case, because the gauged subgroup is embedded as $\SO(k+2) \subset \SU(k+2) \subset \Spin(2k+4)$, implying that there are no $\SO(k+2)$ spinors in the spectrum. Thus, we can gauge either $\Spin(k+2)$, $\SO(k+2)_+$ or $\SO(k+2)_-$ where the subscript again indicates presence or absence of a discrete theta angle, and there is once again an electric, magnetic, or dyonic $\mathbb{Z}_2$ one-form symmetry in each case, respectively. Examining the class $\cS$ description, we conclude that electric and magnetic lines are exchanged by the duality as usual, which implies to the duality orbits shown in figures \ref{SUn-dual-1} and \ref{SUn-dual-2} (see~\cite{Aharony:2013hda} for the action of $T$ in each case).

\begin{figure}
\begin{equation*}
\begin{tikzpicture}
	\node at (0,0) {$SU(N)$};
	\node at (4,0) {$(SU(N)/\mathbb{Z}_2)_+$};
	\node at (8,0) {$ (SU(N)/\mathbb{Z}_2)_- $};
	\node at (0,-2) {$R_{2,k}\hookleftarrow \SO(N+1)_{+}$};
	\node[rotate=0,scale=1.7] at (0,.5) {$\curvearrowright $};
	\node[rotate=180,scale=1.7] at (0.1,-2.5) {$\curvearrowleft $};
	\node[rotate=180,scale=1.7] at (4.1,-2.5) {$\curvearrowleft $};
	\node[rotate=180,scale=1.7] at (8.1,-2.5) {$\curvearrowleft $};
	\node[rotate=180,scale=1.4] at (6,0) {$\longleftrightarrow$};
	\node[rotate=90,scale=1.4] at (0,-1) {$\longleftrightarrow$};
	\node[rotate=90,scale=1.4] at (4,-1) {$\longleftrightarrow$};
	\node[rotate=90,scale=1.4] at (8,-1) {$\longleftrightarrow$};
	\node at (.25,-1) {$S$};\node at (4.25,-1) {$S$};\node at (8.25,-1) {$S$};
	\node at (0,1) {$T$};\node at (6,0.25) {$T$};
	\node at (0,1) {$T$};\node at (0,-3) {$T$};
	\node at (0,1) {$T$};\node at (4,-3) {$T$};
	\node at (0,1) {$T$};\node at (8,-3) {$T$};
	\node at (4,-2) {$R_{2,k}\hookleftarrow \text{Spin}(N+1) $};
	\node at (8,-2) {$R_{2,k}\hookleftarrow \SO(N+1)_{-}  $};
\end{tikzpicture}
\end{equation*}
\caption{Duality orbits of the quiver \eqref{eq:III-N=2-quiver} for $N=4m+2$ ($N>2$).}
\label{SUn-dual-1}
\end{figure}

For future reference, the central charges of the $R_{2,k}$ CFT with odd $k$ are
\cite{Zafrir:2019hps}
\begin{subequations}
\label{eq:R_{2,odd}-ac}
\begin{align}
24a & = -4 + \frac{9}{2}k + \frac{7}{2}k^2\, ,\\
12c & = -1 + 3k + 2k^2\, .
\end{align}
\end{subequations}
The case $N=4$ gives rise to $R_{2,3}$, which is the $E_6$
Minahan-Nemeschansky theory \cite{Chacaltana:2012zy}, or equivalently
the $T_3$ theory arising from putting the $A_2$ $(2,0)$ theory on a
sphere with three (untwisted) full punctures.

\subsection{Even \alt{$k$}{k}} \label{subsec:evenk}

We now reproduce these results from a UV $\cN=1$ Lagrangian,
constructed according to the general procedure outlined in the
introduction. We start with the even $k$ (i.e., odd $N$) case,
as it is somewhat simpler.

The starting point is to identify the operator in phase \II\ dual to
$C_1^N$. This can be done, e.g., by matching the $U(1)^4$ charges of the gauge-invariant chiral operators in the dual descriptions.
Referring to~\eqref{eq:F03-II-charges}, \eqref{eqref:F03-III-charges}, we see that
\[
\renewcommand{\arraystretch}{1.5}
\begin{array}{c|cccc}
& U(1)_B & U(1)_X & U(1)_Y & U(1)_R\\
\hline
\bigl[C_1^N\bigr]_{\text{III}} & -1 & 0 & -N & \frac{N}{2}\\
\bigl[B_1^{N-1}B_2\bigr]_{\text{II}} & -1 & 0 & -N & \frac{N}{2}
\end{array}
\]
so the phase II dual of $C_1^N$ is
$B_1^{N-1}B_2$, where we again use a condensed notation to refer to the gauge invariant baryonic operator built out of
$N-1$ copies of $B_1$ and one $B_2$. Turning on this vev triggers an
RG flow, and after integrating out the massive states we end up with
an accidental $U(1)$ in the infrared, just like in phase III. After relabeling the fields and choosing an appropriate basis for the $U(1)$ symmetries, we obtain\footnote{Although we have not indicated
it explicitly, there is naturally an extra ``$SO(1)$'' gauge factor
under which the $Y_i$ fields are charged.}
\begin{subequations}
\label{eq:recombined-phase-II}
\begin{gather}
\renewcommand{\arraystretch}{1.4}
\begin{array}{c|cc|cccc}
& SU(N) & \Sp(N-1) & \Ux & \Uz & U(1)_{\nR} & U(1)_R \\
\hline
Y_1 & \bar{\fund} & 1 & 0 & 1 & \frac{1}{N} & 1-\frac{1}{3 N} \\
Y_2 & \bar{\fund} & 1 & 0 & -1 & \frac{1}{N} & 1-\frac{1}{3 N} \\
A_1 & \asymm & 1 & 0 & 0 & -\frac{2(N+1)}{N} & \frac{2}{3 N}+\frac{2}{3} \\
A_2 & \symm & 1 & 0 & 0 & -\frac{2}{N} & \frac{2}{3 N} \\
X_1 & \bar{\fund}& \fund & \frac{1}{2} & 0 & \frac{N+1}{N} & \frac{2}{3}-\frac{1}{3 N} \\
X_2 & \bar{\fund} & \fund & -\frac{1}{2} & 0 & \frac{N+1}{N} & \frac{2}{3}-\frac{1}{3 N} \\
S & 1 & \symm & 0 & 0 & -2 & \frac{2}{3}
\end{array}
\intertext{with superpotential}
W=\Tr(A_1 X_1 X_2+A_2 S X_1 X_2+A_2 Y_1 Y_2)\, .
\end{gather}
\end{subequations}

\begin{figure}
\centering
\includegraphics[width=0.7\textwidth]{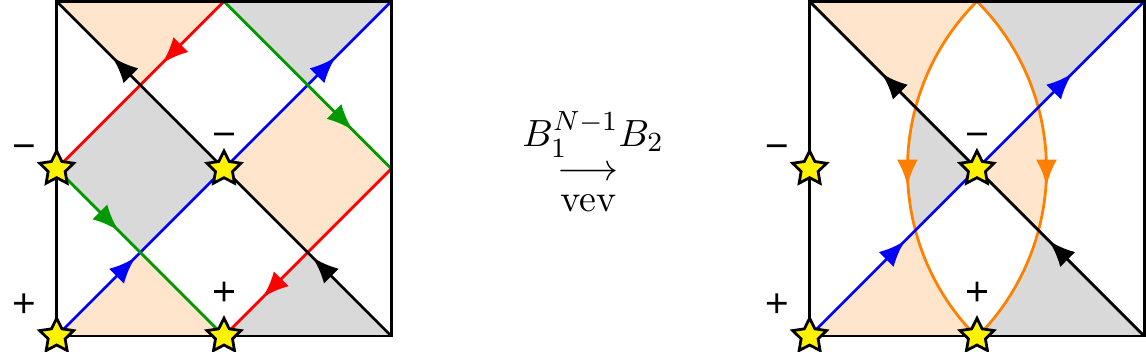}
\caption{Effect of the Higgsing on the brane tiling for phase
\II.}
\label{fig:phase-II-A2A1-Higgsing}
\end{figure}

As shown in figure~\ref{fig:phase-II-A2A1-Higgsing}, the Higgsing can also be understood from the point of view of
the brane tiling. Turning
on the vev for $B_1^{N-1}B_2$ induces a recombination of two of the
NS5 branes. The recombined system of branes will relax to a configuration with two NS5 branes on top of each other, leading to a
strongly coupled SCFT. As usual, we obtain a Lagrangian for this
SCFT by bending the recombined branes slightly, yielding the
Lagrangian description in~\eqref{eq:recombined-phase-II}. Based on the known class $\cS$ results above, we expect that the
strongly coupled sector in the brane tiling
is the $R_{2,N-1}$ theory. As an additional piece of evidence in favour
of this idea, note that the weakly-coupled part of the tiling has a gauge group $\Sp(N-1)$, in agreement with the
the class $\cS$ prediction.

This will indeed be the case, but we need to take care of one
technical point first: in order to obtain the $\mathcal{N}=2$ quiver theory~\eqref{eq:III-N=2-quiver} from partial resolution of phase \III, we had to flip the singlet $\phi$. Thus, to obtain the S-dual of theory from partial resolution of phase \II, we must flip the operator dual to $\phi$.
As before, this operator can be identified by matching its $U(1)^4$ charges with $\phi$, whose charges are shown in~\eqref{eq:flipped-Higgsed-III}.
Comparing
with~\eqref{eq:recombined-phase-II}, we identify the gauge-invariant operator
$A_2^N$ as the dual of $\phi$. The flipped theory is therefore:
\begin{subequations}
\label{eq:flipped-phase-II}
\begin{gather}
\renewcommand{\arraystretch}{1.4}
\begin{array}{c|cc|cccc}
& SU(N) & \Sp(N-1) & \Ux & \Uz & U(1)_{\nR} & U(1)_R \\
\hline
Y_1 & \bar{\fund}& 1 & 0 & 1 & \frac{1}{N} & 1-\frac{1}{3 N} \\
Y_2 & \bar{\fund} & 1 & 0 & -1 & \frac{1}{N} & 1-\frac{1}{3 N} \\
A_1 & \asymm & 1 & 0 & 0 & -\frac{2(N+1)}{N} & \frac{2}{3 N}+\frac{2}{3} \\
A_2 & \symm & 1 & 0 & 0 & -\frac{2}{N} & \frac{2}{3 N} \\
X_1 & \bar{\fund} & \fund & \frac{1}{2} & 0 & \frac{N+1}{N} & \frac{2}{3}-\frac{1}{3 N} \\
X_2 & \bar{\fund} & \fund & -\frac{1}{2} & 0 & \frac{N+1}{N} & \frac{2}{3}-\frac{1}{3 N} \\
S & 1 & \symm & 0 & 0 & -2 & \frac{2}{3} \\
\ov{\phi} & 1 & 1 & 0 & 0 & 2 & \frac{4}{3}
\end{array}\\
\intertext{with superpotential}
W=\Tr(A_1 X_1 X_2+A_2 X_1 S X_2+A_2 Y_1 Y_2) + \bar{\phi}A_2^{N}\, .
\end{gather}
\end{subequations}

Putting together the results of our analysis with the previous class
$\cS$ analysis, we conclude that this Lagrangian theory is in the same
universality class as $R_{2,N-1} \hookleftarrow \Sp(N-1)$ for odd $N$. To isolate
$R_{2,N-1}$ itself, we set the $\Sp(N-1)$ gauge coupling to zero and remove the associated vector multiplets from the theory. In fact, the $\cN =2$ adjoint vector multiplet of $\Sp(N-1)$ includes the chiral field $S$ in addition to the $\cN = 1$ adjoint vector multiplet, so to preserve $\cN=2$ supersymmetry in the infrared, we also decouple and remove $S$. 

\begin{figure}
\centering
\includegraphics[width=0.7\textwidth]{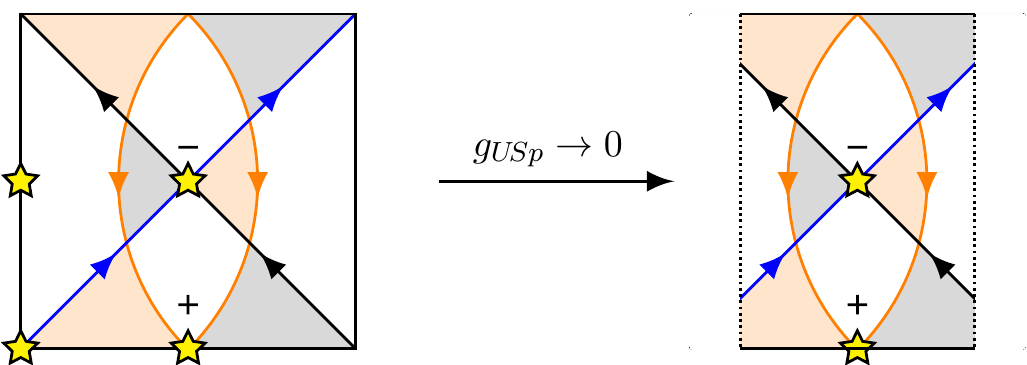}
\caption{Sending the gauge coupling of the $\Sp(N-1)$ factor to
0. In terms of the brane tiling this amounts to stretching the
torus horizontally so it becomes an infinite cylinder, or
equivalently focusing on the region close to the $SU$ factor.}
\label{fig:Phase-II-opening}
\end{figure}

In terms of the brane tiling, this corresponds to decompactifying in the horizontal direction, focusing in on the strongly coupled sector as depicted in figure~\ref{fig:Phase-II-opening}.
This sends the gauge and superpotential couplings associated to the $\Sp(N-1)$ vector multiplet to zero as the corresponding branes become infinitely large. Likewise, the vector multiplet components themselves either become non-normalizable or are pushed off to infinity, freezing these modes out of the theory.

The resulting Lagrangian theory has matter content
\begin{subequations}
\label{eq:R_{2,even}-Lagrangian}
\begin{gather}
\label{eq:R_{2,even}-charges}
\renewcommand{\arraystretch}{1.4}
\begin{array}{c|c|cccc}
& SU(N) & \Sp(2(N-1)) & \Uz & U(1)_{\nR} & U(1)_R \\
\hline
Y_1 & \bar{\fund}& 1 & 1 & \frac{1}{N} & 1-\frac{1}{3 N} \\
Y_2 & \bar{\fund} & 1 & -1 & \frac{1}{N} & 1-\frac{1}{3 N} \\
A_1 & \asymm & 1 & 0 & -\frac{2 (N+1)}{N} & \frac{2 (N+1)}{3 N} \\
A_2 & \symm & 1 & 0 & -\frac{2}{N} & \frac{2}{3 N} \\
X & \bar{\fund} & \fund & 0 & \frac{1}{N}+1 & \frac{2}{3}-\frac{1}{3 N} \\
\ov{\phi} & 1 & 1 & 0 & 2 & \frac{4}{3}
\end{array}\\
\intertext{with superpotential}
\label{eq:R_{2,even}-W}
W=\Tr(A_1 X^2 +A_2 Y_1 Y_2) + \ov{\phi}A_2^{N}\, .
\end{gather}
\end{subequations}
Note that decompactifying the brane tiling in this way manifestly enhances $\Sp(N-1) \to \SU(N-1)$ due to the disappearance of the O5 plane generating the orientifold projection in question. However, the resulting Lagrangian accidentally acquires an even larger symmetry $\Sp(2(N-1)) \supset \SU(N-1)\times \Ux$, which we have made manifest in table above.

\bigskip

We therefore conclude that the $\cN=1$ Lagrangian theory~\eqref{eq:R_{2,even}-Lagrangian} flows to the
$\cN=2$ $R_{2,N-1}$ SCFT. We now present a number of simple but
stringent tests that support this conclusion.

First, note that the manifest symmetry group of the
Lagrangian description matches with the symmetry group of the
$R_{2,N-1}$ theory for $k$ even: we find a $\Sp(2(N-1))\times \Uz$
symmetry group which we identify with the flavour group of the $\cN=2$
SCFT, and an additional $U(1)_{\nR}\times U(1)_R$ which we
identify with the manifest $\cN=1$ subgroup of the
$SU(2)_R\times \Ur$ $R$-symmetry of the $\cN=2$ SCFT. 

Second, note that the $\Sp(2(N-1))$ global symmetry has a
Witten anomaly \cite{Witten:1982fp}, coming from the $X$ fields, which
give an odd number of chiral multiplets in the fundamental of
$\Sp(2(N-1))$. This agrees with the result in \cite{Tachikawa:2018rgw}. The $a$ and $c$ central
charges are also straightforward to compute using~\cite{Anselmi:1997am,Anselmi:1997ys}
\begin{align}
a & = \frac{3}{32}(3\Tr U(1)_R^3 - \Tr U(1)_R)\,, &
c & = \frac{1}{32}(9\Tr U(1)_R^3 - 5 \Tr U(1)_R)\, . \label{eq:ac}
\end{align}
where $U(1)_R$ is the $R$-symmetry appearing in the superconformal
algebra, which can be determined by $a$-maximisation
\cite{Intriligator:2003jj}. The $U(1)_R$ given above already maximizes $a$, so the calculation is
straightforward, and we obtain
\begin{align}\label{eq:R2evenConfAnomalies}
\Tr U(1)_R^3 & = \frac{1}{27}(11k^2 + 35k - 2) \,, &
\Tr U(1)_R & = -\frac{1}{3}(k^2 + k + 2)\, .
\end{align}
Substituting into~\eqref{eq:ac} we
recover~\eqref{eq:R_{2,even}-ac} as expected. Finally, we find a mixed anomaly
\[\label{eq:R2evenMixedAnomaly}
\Tr(\Sp(2k)^2U(1)_R) = -\frac{1}{3}(k+2)
\]
which is $-\frac{1}{3}$ times the level of the $\Sp(2k)$ flavour current of the $R_{2,k}$ theory, as expected.\footnote{\label{foot:19} See, e.g., (A.4) of \cite{Benini:2009mz}. Note that we
normalize the generators of $\Sp(2k)$ so that
$\Tr \Sp(2k)^2U(1)_R=1$ for a fermion in the fundamental of $\Sp(2)$
with charge 1 under $U(1)_R$, which leads to a factor of 2
difference with the conventions in that paper.}
Written in the $\cN=2$ basis, the complete set of non-vanishing anomaly coefficients 
 are:
\begin{gather}
\renewcommand{\arraystretch}{1.6}
\begin{array}{c|c}
 \Sp(2k)^3	& 1 \bmod 2\\
 \Sp(2k)^2 U(1)_r &-(k+2)\\
 U(1)_z^2 U(1)_r &-2\\
U(1)_r^3 & -(k^2+k+2)\\
SU(2)_R^2 U(1)_r & \displaystyle\frac{1}{2}(k^2 +3 k)\\
U(1)_r &  -(k^2+k+2)\\
\end{array}\end{gather}
Besides the checks of the conformal and mixed anomalies in \eqref{eq:R2evenConfAnomalies} and \eqref{eq:R2evenMixedAnomaly}, we are not aware of a computation of the other anomaly coefficients. These are therefore a prediction of our deconfined description.

As a final check, we will focus on the rank one Argyres-Wittig theory
with symmetry $\Sp(4)\times U(1)$ \cite{Argyres:2007tq}, which is the
case $N=3$, i.e., the $R_{2,2}$ theory ($k=N-1=2$). Expressions for the Hall-Littlewood index
\cite{Romelsberger:2005eg,Kinney:2005ej,Romelsberger:2007ec,Gadde:2011uv}
of this theory were given explicitly in
\cite{Chacaltana:2014nya}. This index is an specialization of the full
superconformal index, obtained as follows. Consider the usual
variables for the superconformal index $p=tx$ and $q=t/x$, and denote
the chemical potential for the $U(1)_{\nR}$ symmetry $\nu$. Define
$\tau \df \nu (pq)^{\frac{1}{3}}$. The Hall-Littlewood index is then
\begin{equation}
\cI_{\text{HL}}(\tau) \df \cI(p,q,\tau)\bigr|_{p=q=0}
\end{equation}
where for simplicity we have turned off all chemical potentials for
non-$R$ symmetries. It is straightforward to compute this quantity
from our quiver description using the techniques in
\cite{Romelsberger:2007ec}. We obtain
\begin{equation}
\label{eq:R_{2,2}-HL}
\cI_{\text{HL}}(\tau)= 1 + 11 \tau^{2} + 10 \tau^{3} + 60 \tau^{4} + 80 \tau^{5} + 253 \tau^{6} + 350 \tau^{7} + 855 \tau^{8} + 1180 \tau^{9} + 2406 \tau^{10}+\ldots
\end{equation}
This ought to be compared with the exact form of the index, which was
found in \cite{Chacaltana:2014nya}:
\begin{equation}
\cI_{\text{HL}}(\tau) = \frac{1 + 2 \tau + 8 \tau^{2} + 20 \tau^{3} + 41 \tau^{4} + 62 \tau^{5}+ 87 \tau^{6} + 96 \tau^{7}+87 \tau^8+\cdots+\tau^{14}}{(1-\tau)^8 (1+\tau)^6 (1+\tau+\tau^2)^4}
\end{equation}
where the omitted terms in the numerator are palindromic (that is, the
coefficient of $\tau^{7-m}$ is the same as that of
$\tau^{7+m}$).  A Taylor expansion of this expression around $\tau=0$
reproduces~\eqref{eq:R_{2,2}-HL}.

\subsection{Odd \alt{$k$}{k}} \label{subsec:oddk}

We now repeat the same reasoning for even $N=k+1$. The dual phase to
$\III^+$ is $\I^{--}$. Phase \I\ has the additional complication that
it involves the strongly coupled $\quadSO$ and $\quadSp$
sectors. These can be deconfined, as reviewed in
\S\ref{sec:duality-phases}, leading to a Lagrangian theory in the same
universality class with matter content
\begin{subequations}
\label{eq:I-deconfined}
\begin{gather}
\begin{array}{c|cccc|cccc}
& SU(N-1) & SU(N-1) & SO(N+2) & \!\Sp(N-2) & U(1)_B & U(1)_X & U(1)_Y & U(1)_R \\
\hline
A^{\Rma}_{1} & 1 & \ov\fund & 1 & \fund & \frac{1}{2 (N-1)} & 1 & \frac{N+2}{2 (N-1)} & \frac{4-3 N}{4-4 N} \\
A^{\Rma}_2 & 1 & \ov\fund & 1 & \fund & \frac{1}{2 (N-1)} & -1 & \frac{N+2}{2 (N-1)} & \frac{4-3 N}{4-4 N} \\
Y^{\Rma} & 1 & \fund & \fund & 1 & \frac{1}{2-2 N} & 0 & \frac{N-4}{2 (N-1)} & \frac{N}{4 (N-1)} \\
Z^{\Rma} & 1 & \asymm & 1 & 1 & \frac{1}{1-N} & 0 & \frac{N+2}{1-N} & \frac{N}{2 (N-1)} \\
P^{\Rma} & 1 & 1 & \fund & 1 & -\frac{1}{2} & 0 & \frac{N}{2} & \frac{N}{4} \\
Q^{\Rma} & 1 & \ov\fund & 1 & 1 & \frac{N}{2 (N-1)} & 0 & \!\!\!\!\!\!-\frac{(N-2) (N+2)}{2 (N-1)} & \frac{N^2-8 N+8}{4-4 N} \\
A^{\Rmb}_1 & \ov\fund & 1 & 1 & \fund & \frac{1}{2-2 N} & \frac{N+2}{2 (N-1)} & -1 & \frac{4-3 N}{4-4 N} \\
A^{\Rmb}_2 & \ov\fund & 1 & 1 & \fund & \frac{1}{2-2 N} & \frac{N+2}{2 (N-1)} & 1 & \frac{4-3 N}{4-4 N} \\
Y^{\Rmb} & \fund & 1 & \fund & 1 & \frac{1}{2 (N-1)} & \frac{N-4}{2 (N-1)} & 0 & \frac{N}{4 (N-1)} \\
Z^{\Rmb} & \asymm & 1 & 1 & 1 & \frac{1}{N-1} & \frac{N+2}{1-N} & 0 & \frac{N}{2 (N-1)} \\
P^{\Rmb} & 1 & 1 & \fund & 1 & \frac{1}{2} & \frac{N}{2} & 0 & \frac{N}{4} \\
Q^{\Rmb} & \ov\fund & 1 & 1 & 1 & \frac{N}{2-2 N} & \!\!-\frac{(N-2) (N+2)}{2 (N-1)}\!\!\! & 0 & \frac{N^2-8 N+8}{4-4 N} \\
\Phi & 1 & 1 & \fund & \fund & 0 & -1 & -1 & 1
\end{array}\\
\intertext{and superpotential}
W=\Tr(A^{\Rma}_1 Y^{\Rma} \Phi\!+\!A^{\Rma}_2 Y^{\Rma}_1 A^{\Rmb}_1 Y^{\Rmb}\!\!+\!A^{\Rma}_1 A^{\Rma}_2 Z^{\Rma}\!+\!P^{\Rma} Q^{\Rma} Y^{\Rma}\!\!+\!A^{\Rmb}_1 A^{\Rmb}_2 Z^{\Rmb}\!+\!P^{\Rmb} Q^{\Rmb} Y^{\Rmb}\!\!+\!A^{\Rmb}_2 Y^{\Rmb} \Phi) .
\end{gather}
\end{subequations}

\begin{figure}
\centering
\includegraphics[width=0.95\textwidth]{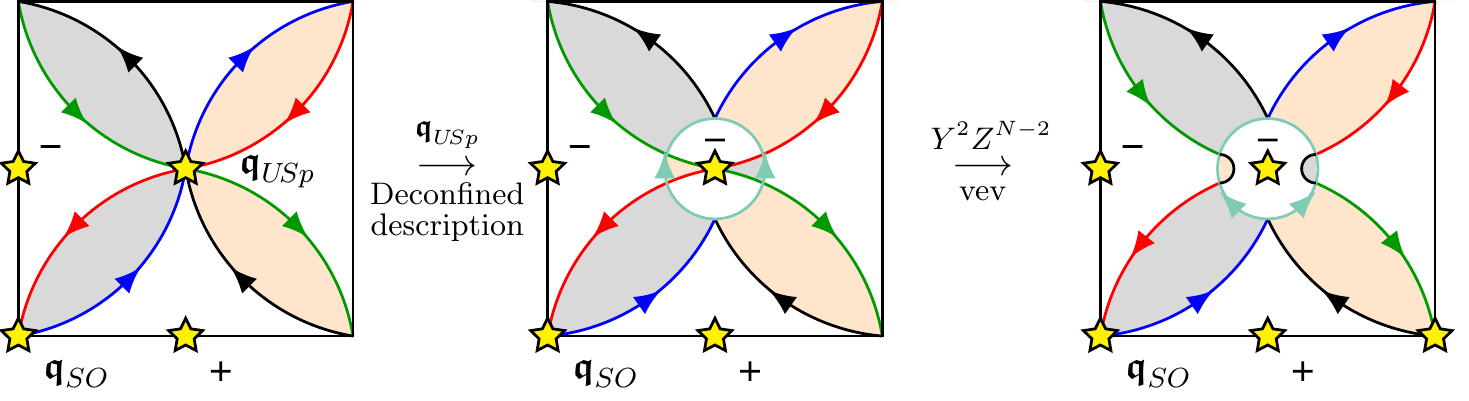}
\caption{Deconfined description of $\quadSp$ part of phase \I. We have kept $\quadSO$ confined in the tiling for clarity; deconfining it leads to an additional $SU(N-1)$ gauge
factor that we have written explicitly in~\eqref{eq:I-deconfined}. Turning on a vev for $\RmaP{Y}{2} \RmaP{Z}{N-2}$ leads to recombination of the green and red NS5 branes.}
\label{fig:I-qSp-deconfined-VEV}
\end{figure}

Now that we have this explicit Lagrangian description of phase \I\ the
analysis proceeds along the same lines as in the case of odd $N$. As a
first step, by matching $U(1)^4$ charges we determine the phase \I\ operator dual to
$C_1^N$ to be $\RmaP{Y}{2} \RmaP{Z}{N-2}$.
Tracking what happens to~\eqref{eq:I-deconfined} when we turn on
$\RmaP{Y}{2} \RmaP{Z}{N-2}$ can be done systematically, but the computation is quite technically involved. It
will be helpful to understand first the effect of the vev on the brane
tiling, as a guide to the behaviour of the field theory. Let us focus
on the effect in the $\quadSp$ sector, which we show partially
deconfined in figure~\ref{fig:I-qSp-deconfined-VEV}. In this picture, the Higgsing corresponds to a recombination of branes.

In terms of the field theory, after integrating out the fields that become
massive after Higgsing, we obtain (after some relabeling):
\begin{subequations}
\label{eq:I-Higgsed}
\begin{gather}
\setlength{\arraycolsep}{2pt}
\renewcommand{\arraystretch}{1}
{\footnotesize\begin{array}{c|cccc|cccc}
& \text{SU}(N-1) & \Sp(N-2) & SO(N+1) & \Sp(N-2) & \Ux &
\Uz & U(1)_{\nR} & U(1)_R \\
\hline
\Phi_1 & 1 & 1 & 1 & \fund & -1 &2 & 1 & 1 \\
Y^{\Rma} & 1 & \fund & \fund & 1 & 0 & 0 & 1 & 0 \\
P^{\Rma}& 1 & 1 & \fund & 1 & 0 & 0 & N & 0 \\
Q^{\Rma} & 1 & \fund & 1 & 1 & 0 & 0 & -N-1 & 2 \\
A^{\Rmb}_1 & \ov\fund & 1 & 1 & \fund & \frac{N+2}{2 (N-1)} & -\frac{1}{N-1}  &
\frac{N-2}{2-2 N} & \frac{N-2}{2 (N-1)} \\
A^{\Rmb}_2 & \ov\fund & 1 & 1 & \fund  & \frac{N+2}{2 (N-1)}& -\frac{1}{N-1}
& 1+\frac{N}{2 (N-1)} & \frac{N-2}{2 (N-1)} \\
Y^{\Rmb} & \fund & 1 & \fund & 1 & \frac{N-4}{2 (N-1)} & \frac{1}{N-1} &  \frac{N}{2-2
N} & \frac{N}{2 (N-1)} \\
Y^{\Rmb}_1 & \fund & 1 & 1 & 1 &  \frac{N-4}{2 (N-1)} &-2+\frac{1}{N-1} &
\frac{N}{2-2 N} & \frac{N}{2 (N-1)} \\
Z^{\Rmb} & \asymm & 1 & 1 & 1 &-\frac{N+2}{N-1} & \frac{2}{N-1} &  -\frac{N}{N-1}
& \frac{N}{N-1} \\
P^{\Rmb} & 1 & 1 & \fund & 1 & \frac{N}{2} & -1 & -\frac{N}{2} & \frac{N}{2} \\
P^{\Rmb}_1 & 1 & 1 & 1 & 1 & \frac{N}{2} &1 &  -\frac{N}{2} & \frac{N}{2} \\
Q^{\Rmb} & \ov\fund & 1 & 1 & 1 & \frac{(N-2) (N+2)}{2 (1-N)}
&1-\frac{1}{N-1} &  \frac{N^2}{2 (N-1)} & -\frac{(N-2)^2}{2 (N-1)} \\
\Phi & 1 & 1 & \fund & \fund & -1 & 0 & -1 & 1
\end{array}}\\
\intertext{with the superpotential}
W=\Tr(P^{\Rma} Q^{\Rma}  Y^{\Rma} \!\!+\! P^{\Rmb} Q^{\Rmb}  Y^{\Rmb}\!\!+\!\Phi_1 A^{\Rmb}_1
Y^{\Rmb}_{1}\!\!+\!P^{\Rmb}_{1} Q^{\Rmb} Y^{\Rmb}_{1}\!\!+\!A^{\Rmb}_1 Y^{\Rmb}  (Y^{\Rma})^2\Phi \!+\! A^{\Rmb}_2 Y^{\Rmb} \Phi\!+\!A^{\Rmb}_1 A^{\Rmb}_2 Z^{\Rmb})\,
\end{gather}
\end{subequations}
which agrees with the results of the brane tiling operation described above.

\begin{figure}
\centering
\includegraphics[width=0.7\textwidth]{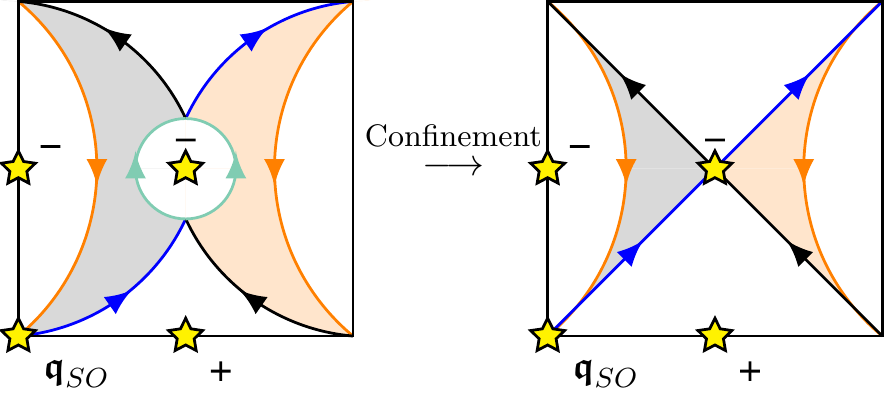}
\caption{Reconfining the result of Higgsing the deconfined $\quadSp$
sector.}
\label{fig:I-qSp-reconfine}
\end{figure}

This is a fairly imposing theory, but looking at the
brane tiling gives a clue about what to do next. As can be seen in
figure~\ref{fig:I-qSp-deconfined-VEV}, after the branes passing through the
O5$^-$ orientifold fixed point within the $\quadSp$ sector recombine
and move away, we are left with a central ``deconfinement bubble'' which will
tend to reconfine, as in figure~\ref{fig:I-qSp-reconfine}. This
suggests to that the $\Sp(N-2)$ gauge group factor in the second column of~\eqref{eq:I-Higgsed} should confine, and indeed this factor is $s$-confining
\cite{Intriligator:1995ne}, resulting in composite mesons $\cM_{\RmaP{Y}{2}} = \RmaP{Y}{2}$ and $\cM_{Y^{\Rma} Q^{\Rma}} = Y^{\Rma} Q^{\Rma}$ interacting via a superpotential. The latter meson gets a mass with $P^{\Rma}$, setting the confining superpotential to zero.

Finally, we flip the operator dual to $\phi$, which is $A_1^{\prime N-2} Q^{\Rmb}$ as is straightforward to check. We thereby obtain a Lagrangian in the same universality class as the strongly-coupled S-dual
of~\eqref{eq:III-N=2-quiver}, given by
\begin{subequations}
\begin{gather}
\renewcommand{\arraystretch}{1.2}
\begin{array}{c|ccc|cccc}
& SU(N-1) & SO(N+1) & \Sp(N-2) & \Ux & \Uz & U(1)_{\nR} & U(1)_R \\
\hline
\cM & 1 & \asymm & 1 & 0 & 0 & -2 & \frac{2}{3} \\
A_1 & \ov{\fund } & 1 & \fund  &  \frac{N+2}{2 (N-1)} & \frac{-1}{N-1} & \frac{N-2}{2 (N-1)} & \frac{N-2}{3 (N-1)} \\
A_2 & \ov{\fund } & 1 & \fund  &  \frac{N+2}{2 (N-1)} & \frac{-1}{N-1} & \!\!\!\frac{N}{2-2 N}-1 & \frac{4-3 N}{3-3 N} \\
Y & \fund  & \fund  & 1 &  \frac{N-4}{2 (N-1)} &  \frac{1}{N-1} & \frac{N}{2 (N-1)} & \frac{N}{3 (N-1)} \\
Y_{1} & \fund  & 1 & 1 & \frac{N-4}{2 (N-1)} & \frac{1}{N-1}-2 &  \frac{N}{2 (N-1)} & \frac{N}{3 (N-1)} \\
Z & \asymm & 1 & 1 & \frac{N+2}{1-N} & \frac{2}{N-1} &  \frac{N}{N-1} & \frac{2 N}{3 (N-1)} \\
P & 1 & \fund  & 1 & \frac{N}{2} & -1 &  \frac{N}{2} & \frac{N}{3} \\
P_{1} & 1 & 1 & 1 &  \frac{N}{2} & 1 & \frac{N}{2} & \frac{N}{3} \\
Q & \ov{\fund} & 1 & 1 & \frac{(N-2) (N+2)}{2(1-N)} & \frac{N-2}{N-1} &  \frac{N^2}{2-2 N} & \!\!\frac{N^2-6 N+6}{3-3 N} \\
\Phi & 1 & \fund  & \fund  &-1 & 0 &  1 & \frac{2}{3} \\
\Phi_1 & 1 & 1 & \fund  & -1 & 2 & -1 & \frac{4}{3} \\
\ov\phi & 1 & 1& 1 & 0 & 0 & 2 & \frac{4}{3}
\end{array} \\
\intertext{with superpotential}
W=\Tr(A_1 A_2 Z + A_2 Y \Phi+ A_1 Y \cM \Phi  + A_1 Y_1 \Phi_1+P Q Y+P_1 Q Y_{1})+A_{1}^{N-2} Q \ov\phi\,,
\end{gather}
\end{subequations} 
where we relabeled $\cM_{Y^2} \longrightarrow \cM$ and suppressed the primes for simplicity.

\begin{figure}
\centering
\includegraphics[width=0.7\textwidth]{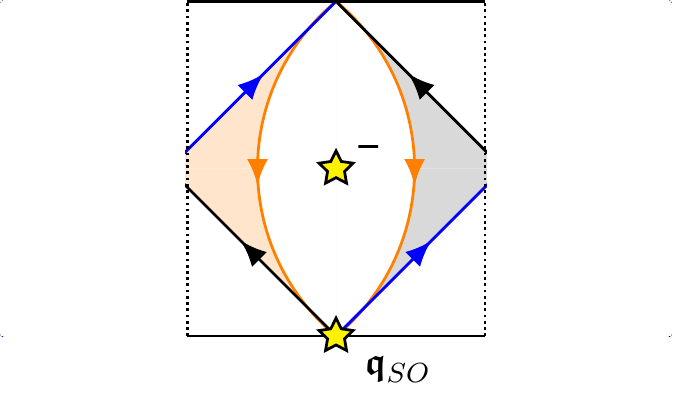}
\caption{Sending the gauge coupling of the $SO(N+1)$ factor to 0. Note that we shifted figure~\ref{fig:I-qSp-reconfine} horizontally by half a period before cutting open the torus into a cylinder, in order to focus on the strongly coupled sector.}
\label{fig:I-cutting-tiling}
\end{figure}

As discussed above, we expect this
theory to flow to $R_{2,N-1}\hookleftarrow\SO(N+1)$ per~\cite{Chacaltana:2012zy}. Therefore, we can isolate $R_{2,N-1}$ by ungauging $\SO(N+1)$ (in the $\cN=2$ sense), which also removes the chiral superpartner $\cM$.
Equivalently, in terms of the brane tiling we focus in on the
strongly coupled sector as in figure~\ref{fig:I-cutting-tiling}.
After redefining $\Uy = -(\Ux+\frac{1}{2} \Uz)$ for later convenience, we obtain the following $\cN=1$ Lagrangian theory, which is expected to flow to the $R_{2,N-1}$ theory for even $N$:
\begin{subequations}
\label{eq:R_{2,odd}-Lagrangian}
\begin{gather}
\label{eq:R_{2,odd}-charges}
\renewcommand{\arraystretch}{1.2}
\begin{array}{c|cc|ccccc}
& SU(N-1) & \Sp(N-2) & SU(N+1) & \Uy & \Uz & U(1)_{\nR} & U(1)_R \\
\hline
A_1  & \ov{\fund} & \fund & 1 & \frac{N+1}{2-2 N} & -\frac{1}{N-1} &  \frac{N-2}{2 (N-1)} & \frac{N-2}{3 (N-1)} \\
A_2 & \ov{\fund } & \fund & 1 &  \frac{N+1}{2-2 N} & -\frac{1}{N-1} & \frac{N}{2-2 N}-1 & \frac{3 N-4}{3 (N-1)} \\
Y & \fund  & 1 & \ov{\fund} &  -\frac{N-3}{2 (N-1)} & \frac{1}{N-1} & \frac{N}{2 (N-1)} & \frac{N}{3 (N-1)} \\
Y_{1} & \fund  & 1 & 1 &  \frac{N+1}{2 (N-1)} & \frac{1}{N-1}-2 & \frac{N}{2 (N-1)} & \frac{N}{3 (N-1)} \\
Z & \asymm & 1 & 1 &  \frac{N+1}{N-1} & \frac{2}{N-1} & \frac{N}{N-1} & \frac{2 N}{3 (N-1)} \\
P & 1 & 1 & \fund  &  -\frac{N-1}{2} & -1 & \frac{N}{2} & \frac{N}{3} \\
P_{1} & 1 & 1 & 1 &- \frac{1}{2} (N+1) & 1 &  \frac{N}{2} & \frac{N}{3} \\
Q & \ov{\fund } & 1 &1 &  \frac{-N^2+N+2}{2-2 N}& \frac{N-2}{N-1} & \frac{N^2}{2-2 N} & \frac{N^2-6 N+6}{3-3 N} \\
\Phi & 1 & \fund  & \fund  & 1 & 0 & 1 & \frac{2}{3} \\
\Phi_1 & 1 & \fund  & 1 & 0 & 2 & -1 & \frac{4}{3} \\
\ov\phi  & 1 & 1 & 1 & 0 & 0 & 2 & \frac{4}{3}
\end{array} \\
\intertext{with superpotential}
\label{eq:R_{2,odd}-W}
W=\Tr(A_1  A_2 Z+ A_2 Y \Phi+A_1 Y_1 \Phi_1+P Q Y+P_1 Q Y_{1})+A_{1}^{N-2} Q \ov\phi\, .
\end{gather}
\end{subequations}
Along the lines of~\cite{Garcia-Etxebarria:2015hua}, one can argue that $SU(N+1) \times \Uy$ enhances accidentally to $\Spin(2N+2)$ in the infrared, in agreement with the expected $\Spin(2N+2)\times U(1)$ flavor symmetry of the $R_{2,N-1}$ (even $N$) CFT. It is hard to imagine that one could have guessed this
$\cN=1$ Lagrangian without the aid of the brane
construction!\footnote{A different class of $\cN=1$ Lagrangians expected to flow to the
$R_{2,N-1}$ ($N$ even) CFTs and preserving a different subgroup of the full $R_{2,N-1}$ symmetry group were proposed in \cite{Zafrir:2019hps}. Unlike here, the manifest symmetry group in~\cite{Zafrir:2019hps} has a lower rank (by one) than the full $R_{2,N-1}$ symmetry group, $\U(1)_{\nR}$ being absent from the UV theory. If our proposal and that of~\cite{Zafrir:2019hps} are both correct then they should lie in the same universality class. However, we have so far been unable to relate them using known Seiberg dualities, a task made more difficult by the mismatch in the manifest symmetries. We leave this as an interesting question for future research.}

The end result can be re-expressed more simply as a partial gauging of one of the quad CFTs discussed in~\S\ref{subsec:quadCFTs}. In particular, consider the ``partially flipped'' $\quadSO$ CFT shown in figure~\ref{sfig:Flipped-qSO}, with flavor symmetry $\SU(N-2) \times \Spin(2N+2) \times \Spin(2)\times \U(1)^2 \times \U(1)_R$ (setting $M = N-2$, $P=2$). We claim that gauging $SU(N-2)\hookleftarrow \Sp(N-2)$ (with $\cN=1$ vector multiplets) generates a flow whose endpoint is the $R_{2,N-1}$ (even $N$) CFT plus a decoupled free chiral multiplet. Indeed, substituting the deconfined decription of the $\quadSO$ theory given in~\S\ref{subsec:quadCFTs}, identifying $\Spin(2) \cong \U(1)_z$, choosing an appropriate basis for the remaining $\U(1)$s (after omitting the one with a mixed $\Sp(N-2)^2 \U(1)$ anomaly), and flipping the appropriate free baryon (either $\cA_{N-2}$ and $\cA_0$, depending on which deconfined description we choose, see~\eqref{eqn:quadCFTbaryons}), we recover~\eqref{eq:R_{2,odd}-Lagrangian}.

\begin{figure}
	\centering
	\includegraphics[width=0.7\textwidth]{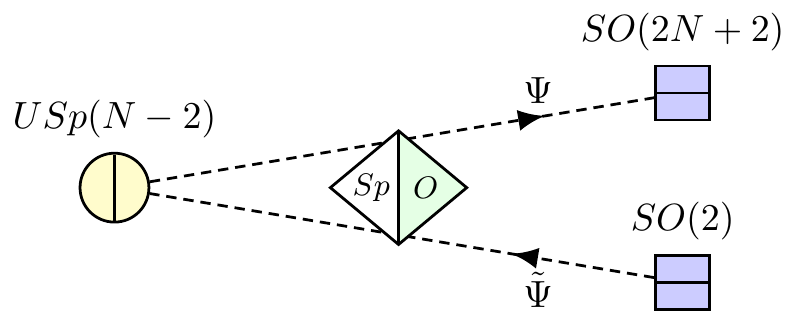}
	\caption{The abstract quiver for a partial gauging of a $\quadSO$ CFT that is expected to generate a flow to the $R_{2,N-1}$ (even $N$) CFT plus a decoupled chiral multiplet. The oppositely directed arrows on the meson lines $\Psi, \tilde{\Psi}$ indicate that the $\quadSO$ CFT is partially flipped, see figure~\ref{sfig:Flipped-qSO}.}
	\label{fig:R2k-abstract-quiver}
\end{figure}

The abstract quiver for this description of $R_{2,N-1}$ (even $N$) is shown in figure~\ref{fig:R2k-abstract-quiver}. Note that in principle we could have obtained this description directly by Higgsing the abstract quiver in figure~\ref{sfig:F03-I-quiver}; we did this calculation using the deconfined Lagrangian description only because Lagrangian methods are far more familiar. Moreover, note that the abstract quiver~\ref{fig:R2k-abstract-quiver} displays the full flavor symmetry of the $R_{2,N-1}$ CFT, lacking only the nonabelian $R$-symmetry enhancement to $\UsuR \longrightarrow \SU(2)_R$ that is never visible in $\cN = 1$ language.

The embedding of the $\U(1)$ symmetries within the $\quadSO$ description can likewise be described by the abstract charge table 
\begin{subequations}
\label{eq:R_{2,odd}-Lagrangian-B}
\begin{gather}
\label{eqn:abstractSOFlipped}
\renewcommand{\arraystretch}{1.2}
\begin{array}{c|c|cc|ccc}
		& \Sp(N-2) &  \SO(2N+2) & \SO(2)_z & \U(1)_{\nR} & \U(1)_R \\ \hline
		\quadSO  & \ast & \ast & \ast & \frac{N}{2}  & \frac{N-6}{12}  \\
		\Psi  & \fund &  \fund & 1  & 1& 2/3 \\
		\tilde{\Psi}& \ov\fund &1& \fund & -1 & 4/3 \\ \hline
		\phi & \singlet & \singlet & \singlet & 2 & 4/3
\end{array} \\
\intertext{with superpotential}
\label{eq:R_{2,odd}-abstract}
W=\mathcal{A} \phi\, .
\end{gather}
\end{subequations}
Here the second and third lines of the table indicate the charges\footnote{Note that although $\ov\fund \cong \fund$ for $\Sp(N-2)$ (the $\fund$ representation is pseudoreal), we list opposite choices for $\Psi$ and $\tilde\Psi$ as a reminder of the partial flipping.} of the $\quadSO$ mesons $\Psi$ and $\tilde{\Psi}$ and the first line indicates the admixture of the baryonic symmetry $\U(1)_B$ of the quad CFT, following the conventions of~\cite{Garcia-Etxebarria:2016bpb}. In the notation of~\eqref{eqn:quadCFTbaryons}, the baryon $\mathcal{A}$ is either $\mathcal{A}_{N-2}$ or $\mathcal{A}_0$  depending on which deconfined description we pick, the right one being fixed by the $\U(1)$ charges.

Let us provide some evidence that this Lagrangian theory and its partially-gauged-$\quadSO$ cousin are in the
same universality class as the $R_{2,N-1}$ theory. The global symmetry
of the $R_{2,N-1}$ theory is $\Spin(2N+2)\times U(1)$ times
$SU(2)_R\times \Ur$. The $\cN=1$ description will break the
$R$-symmetry factor to $U(1)_{\nR}\times U(1)_R$, so ideally we would
like to have an $\cN=1$ description with global symmetry group
$\Spin(2N+2)\times U(1) \times U(1)_{\nR} \times U(1)_R$. This is precisely the manifest symmetry of the abstract quiver shown in figure~\ref{fig:R2k-abstract-quiver}, although in our explicit Lagrangian description~\eqref{eq:R_{2,odd}-Lagrangian} only a maximal
subgroup $SU(N+1)\times U(1)$ of $\Spin(2N+2)$ (of equal rank) is manifest.

We next compare the central charges. We have, setting $N=k+1$ as above,
\[
\Tr U(1)_R^3 & = \frac{1}{27} (11 k^{2} + 9 k - 28)\, \\
\Tr U(1)_R & = -\frac{1}{3}(k^{2} + 3 k + 4)\, .
\]
When substituted into~\eqref{eq:ac} these values lead to the expected
$a$ and $c$ central charges given in~\eqref{eq:R_{2,odd}-ac}. Similarly, we find
\[
\Tr \Spin(2k+4)^2 U(1)_R = -\frac{1}{3}(2k)
\]
in agreement with the expected central charge for the flavour
symmetry (up to the same factor of $-\frac{1}{3}$ we found in the
even $k$ case).
As before, the complete list of non vanishing anomalies in the $\cN=2$ basis is given by:
\begin{gather}
\renewcommand{\arraystretch}{1.6}
\begin{array}{c|c}
 \Spin(2 k + 4)^2 U(1)_r & - 2 k\\
   U(1)_z^2 U(1)_r & -8\\
   U(1)_r^3 &  -(k^2 + 3 k+4)\\
   SU(2)_R^2 U(1)_r  & \displaystyle\frac{1}{2}(k^2+ k -2) \\
  U(1)_r &  -(k^2 + 3 k+4)\\
\end{array}\end{gather}
These anomalies match the ones computed in \cite{Zafrir:2019hps} (for the anomaly computation $\Spin(2k+4)\simeq SO(2k+4)$). Note, however,
that the ones involving $\Spin(2k+4)$ differ by a
factor of 2 with respect to the conventions in that paper, due to the
different normalization of the $\Spin(2k+4)$ generators, see the discussion in
footnote~\ref{foot:19}.

Finally, we can compute the superconformal index and compare with
known results. We will do it for $R_{2,3}$, which is a well studied
example, as it is the rank one $E_6$ Minahan-Nemeschansky (also known
as $T_3$). In this case we have the following matter content:
\begin{subequations}
\label{eq:E6-quiver}
\begin{gather}
\renewcommand{\arraystretch}{1.2}
\begin{array}{c|cc|ccccc}
& SU(3) & SU(2) & SU(5) & \Uy & \Uz & U(1)_{\nR} & U(1)_R \\
\hline
A_1 & \ov\fund & \fund & 1 & -\frac{5}{6} & -\frac{1}{3} & \frac{1}{3} & \frac{2}{9} \\
A_2 & \ov\fund & \fund & 1 & -\frac{5}{6} & -\frac{1}{3} & -\frac{5}{3} & \frac{8}{9} \\
Y & \fund & 1 & \ov\fund & -\frac{1}{6} & \frac{1}{3} & \frac{2}{3} & \frac{4}{9} \\
Y_{1} & \fund & 1 & 1 & \frac{5}{6} & -\frac{5}{3} & \frac{2}{3} & \frac{4}{9} \\
Z & \ov\fund & 1 & 1 & \frac{5}{3} & \frac{2}{3} & \frac{4}{3} & \frac{8}{9} \\
P & 1 & 1 & \fund & -\frac{3}{2} & -1 & 2 & \frac{4}{3} \\
P_{1} & 1 & 1 & 1 & -\frac{5}{2} & 1 & 2 & \frac{4}{3} \\
Q & \ov\fund & 1 & 1 & \frac{5}{3} & \frac{2}{3} & -\frac{8}{3} & \frac{2}{9} \\
\Phi & 1 & \fund & \fund & 1 & 0 & 1 & \frac{2}{3} \\
\Phi_1 & 1 & \fund & 1 & 0 & 2 & -1 & \frac{4}{3} \\
\ov\phi  & 1 & 1 & 1 & 0 & 0 & 2 & \frac{4}{3}
\end{array} \\
\intertext{with superpotential}
\label{eq:E6-W}
W=\Tr(A_1 A_2 Z+ A_2 Y \Phi+ A_1 Y_1 \Phi_1+P Q Y+P_1 Q Y_{1})+A_{1}^{2} Q \ov\phi\, .
\end{gather}
\end{subequations}

With this field content at hand it is easy to compute the superconformal index (using the prescription in
\cite{Romelsberger:2007ec}) to a
fairly high order and compare it with known results
\cite{Gadde:2010te}. We find for the first few orders, in the
conventions of \cite{Garcia-Etxebarria:2015hua},
\[
1&+\nR^2 t^{4/3} \biggl[1+X_{0,1,0,0,0}+z X_{0,0,0,1,0}+\frac{X_{0,0,0,0,1}}{z}\biggr] \\
&+t^2 \biggl[\frac{1}{\nR^6}-\biggl(2+X_{0,1,0,0,0} +z X_{0,0,0,1,0}+\frac{X_{0,0,0,0,1}}{z}\biggr)\biggr]\\
&-t^{7/3} J_1 \biggl[\frac{1}{\nR^4}-\nR^2 \biggl( 2+X_{0,1,0,0,0}+z  X_{0,0,0,1,0}+\frac{X_{0,0,0,0,1}}{z}\biggr)\biggr]+\ldots
\]
where we have grouped the $SU(5)\times \Uy$ characters into
$\Spin(10)$ characters $X_{a,b,c,d,e}$, denoted by $z$ the $\Uz$
chemical potential, and by $\nR$ the $U(1)_{\nR}$ chemical
potential. This index agrees perfectly with the index computed in
\cite{Gadde:2010te},\footnote{This index was given an alternative
  $\cN=1$ Lagrangian interpretation in \cite{Gadde:2015xta}.} if we embed $\Spin(10)\times \Uz\to E_6$. We
have verified that the agreement persists to at least the order $t^{11/3}$, but the
resulting expressions are a bit too unwieldy to display here.

Finally, let us briefly note that the theory
in~\eqref{eq:E6-quiver} has no 1-form symmetries (i.e., the gauge group is simply connected and the spectrum is complete). This matches
the expected answer \cite{Tachikawa:2013hya}, which can also
be deduced from a BPS quiver analysis along the lines of
\cite{DelZotto:2020esg} and the general class-$\cS$ analysis
in \cite{Bhardwaj:2021pfz}.

\section{Higgs branch deformations}

\label{sec:Higgs}

We have constructed candidate Lagrangian descriptions for the
$R_{2,k}$ theories, for all $k$. So far, our evidence for the validity
of these descriptions is that the global symmetries, central charges
and superconformal indices match between our $\cN=1$ theories and the
ones expected for the $R_{2,k}$ theories (whenever those are known).

In the rest of this paper we will provide further evidence for these
proposals by rederiving known results about the moduli space and
deformations of these theories. We start in this section by studying
what happens as we move on the Higgs branch.

\subsection{Dimension of the Higgs branch for \alt{$R_{2,\text{k}}$}{R2k}}

The quaternionic dimension of the Higgs branch of $\cN=2$ theories is
$\dim_{\mathbb{H}}(\cH) = 24(c-a)$~\cite{Gaiotto:2008nz}.
The values of $a$ and $c$ for the $R_{2,k}$ theories were given
in~\eqref{eq:R_{2,even}-ac} and \eqref{eq:R_{2,odd}-ac} above, from
which we obtain\footnote{This formula assumes that on a generic point on a Higgs branch the IR
theory is a theory of free hypers. This is something that is indeed
true for our $\cN=1$ theory, and true for $R_{2,k}$ for rank up to two
\cite{Martone:2021ixp}, but we are not aware of a computation showing
that it is true for $k>2$, which would be a prediction of our analysis.
We thank the referee for highlighting this assumption.}
\[\label{eq:dim(H)-expected-even}
\dim_{\mathbb{H}}(\cH) = 1 + \frac{k (k+1)}{2}
\]
in the even $k$ case and
\[
\label{eq:dim(H)-expected-odd}
\dim_{\mathbb{H}}(\cH) = 1 + \frac{(k+1)(k+2)}{2}
\]
in the odd $k$ case. We now reproduce these formulae from our
$\cN=1$ Lagrangian descriptions. 

The computation is simpler for even $k$, so we discuss this case first. The UV Lagrangian constructed in~\S\ref{subsec:evenk} is shown in~\eqref{eq:R_{2,even}-Lagrangian}. We turn on a vev for the baryonic Higgs branch operator $X^{N-1} Y_1$, which completely Higgses the gauge group. Note that the $A_1$ F-term forces the $X$ vevs to span an isotropic subspace of $\Sp(2N-2)$, breaking $\Sp(2N-2) \to \U(N-1)$. In particular, the field $X$ in the $\SU(N)\times\Sp(2N-2)$ irrep $(\ov\fund,\fund)$ decomposes into two fields $X_1$ and $X_2$ in $\SU(N)\times\U(N-1)$ irreps $(\ov\fund,\fund)$ and $(\ov\fund,\ov\fund)$, respectively. Choosing the vev to be in the component $X_1^{N-1} Y_1$ for definiteness, the end result is
\[
\renewcommand{\arraystretch}{1.2}
\begin{array}{c|ccc}
	& \text{U}(N-1)' & \text{U}(1)_{\nR}' & \text{U(1})_R'   \\
	\hline
        A_2 & \symm & 2 & \frac{4}{3} \\
        X_2 & \symmb & 0 & 0 \\
	\ov\phi  & 1 & 2 & \frac{4}{3} \\
	v & 1 & 0 & 0
\end{array} \label{eqn:evenk}
\]
with vanishing superpotential, where the primes indicate that we have mixed the (center of) the indicated symmetry groups with $\U(1)_z$ from~\eqref{eq:R_{2,even}-Lagrangian} (under which $X^{N-1} Y_1$ carries unit charge) to isolate the subgroups preserved by the vev.\footnote{While these symmetries are further enhanced in the IR, this is not very important in the present argument.} Thus, we obtain
\[
\dim_\bH(\cH) = \frac{1}{2}\left(1+1+\frac{1}{2}N(N-1)+\frac{1}{2}N(N-1)\right) = 1+\frac{1}{2}N(N-1)
\]
free hypermultiplets, in agreement with~\eqref{eq:dim(H)-expected-even} for $N=k+1$. 

Next, we consider the odd $k$ case, for which the UV Lagrangian theory is shown in~\eqref{eq:R_{2,odd}-Lagrangian}. Turning on a vev for the gauge-singlet Higgs branch operator $P_1$ breaks one linear combination of the $U(1)$ global symmetries and gives a mass to $Q$ and $Y_1$.\footnote{Note that since it has nonzero $\Uy$ charge, $P_1$ sits inside a nontrivial (spinor) representation of $\Spin(2N+2)$, and the vev will break $\Spin(2N+2) \to \SU(N+1) \times \Uy$. This fact is obscured in the UV Lagrangian, where only the subgroup $\SU(N+1) \times \Uy$ was manifest to begin with.} Integrating these out, we obtain
\begin{subequations}
\begin{gather}
\renewcommand{\arraystretch}{1.2}
\begin{array}{c|cc|cccc}
& SU(N-1) & \Sp(N-2) & SU(N+1) & \Uy' & U(1)_{\nR}' & U(1)_R'   \\
\hline
A_1 & \ov{\fund} & \fund & 1 & \frac{1+N}{1-N} & 1 & \frac{2}{3} \\
A_2 & \ov{\fund} & \fund & 1 & \frac{1+N}{1-N} & -1 & \frac{4}{3} \\
Y & \fund& 1 & \ov{\fund} & \frac{2}{-1+N} & 0 & 0 \\
Z & \asymm& 1 & 1 & \frac{2 (1+N)}{-1+N} & 0 & 0 \\
P & 1 & 1 & \fund & -N & N & \frac{2 N}{3} \\
\Phi  & 1 & \fund & \fund & 1 & 1 & \frac{2}{3} \\
\Phi_1 & 1 & \fund& 1 & 1+N & -1-N & -\frac{2}{3}(N-2) \\
\ov\phi  & 1 & 1 & 1 & 0 & 2 & \frac{4}{3}\\
v & 1 & 1 & 1 & 0 & 0 & 0
\end{array}\\
\intertext{with superpotential}
W=\Tr\biggl(A_1 A_2 Z +A_2 \Phi   Y+\frac{1}{v} A_1 P Y \Phi_1 \biggr)\, .
\end{gather}
\end{subequations}
Here we explicitly include $P_{1}$---now denoted by $v$ as a reminder that it has a vev---in our set of light fields since we are interested in
counting flat directions in the IR. To further simplify the result, we
deconfine the antisymmetric tensor field $Z$ and Seiberg dualize 
$SU(N-1)$ to obtain 
\begin{subequations}
\begin{gather}
\renewcommand{\arraystretch}{1.3}
\begin{array}{c|ccc|cccc}
& SU(N-2) & \Sp(N-2) & \Sp(N-4) & SU(N+1) & \Uy' & U(1)_{\nR}' & U(1)_R'   \\
\hline
\widetilde{A_{1}} & \fund & \fund & 1 & 1 & 0 & -1 & \frac{4}{3} \\
\widetilde{A_2} & \fund & \fund & 1 & 1 & 0 & 1 & \frac{2}{3} \\
\widetilde{Y} & \ov{\fund} & 1 & 1 & \fund & 1 & 0 & 0 \\
\widetilde{H_{z}}& \ov{\fund} & 1 & \fund & 1 & 0 & 0 & 0 \\
\widetilde{P_{z}} & \fund & 1 & 1 & 1 & -(1+N) & 0 & 2 \\
\cM_{A_1 Y} & 1 & \fund & 1 & \ov{\fund} & -1 & 1 & \frac{2}{3} \\
\cM_{Y P_{z}} & 1 & 1 & 1 & \ov{\fund} & N & 0 & 0\\
P & 1 & 1 & 1 & \fund & -N & N & \frac{2 N}{3} \\
\Phi_1 & 1 & \fund & 1 & 1 & 1+N & -(1+N) & \frac{2(2-N)}{3}  \\
\ov\phi  & 1 & 1 & 1 & 1 & 0 & 2 & \frac{4}{3} \\
v & 1 & 1 & 1 & 1 & 0 & 0 & 0
\end{array}
\intertext{with superpotential}
\label{eq:R_{2,odd}-intermediate-W}
W=\Tr\biggl(\widetilde{A_1}  \widetilde{A_2} \widetilde{H_{z}}^2+ \frac{1}{v} \Phi_1 \cM_{A_1 Y} P  +\cM_{A_1 Y} \widetilde{A_1} \widetilde{Y}+\widetilde{P_{z}}  \widetilde{Y}  \cM_{Y P_{z}}\biggr)\, .
\end{gather}
\end{subequations}
Here the fields with tildes on top are the Seiberg dual quarks and $\cM_{A_1 Y}$,
$\cM_{Y P_z}$ are two of the composite mesons, whereas the remaining mesons acquire masses via the superpotential and have been integrated out.\footnote{Note that $\widetilde{H_{z}}$ and $\widetilde{P_{z}}$ are the Seiberg dual quarks of fields $H_z$ and $P_z$ arising from deconfining $Z$; likewise $\cM_{Y P_z}$ is a composite involving one of these fields.}

Since the $\Sp(N-4)$ gauge factor now has $N-2$ flavors, it confines with a quantum-deformed moduli space~\cite{Intriligator:1995ne}, forcing the composite $\cM_{\widetilde{H_z} \widetilde{H_z}}$ (in the $\asymmb$ irrep of $\SU(N-2)$) to get a vev. This Higgses $SU(N-2)$ to $\Sp(N-2)$ as well as giving a mass to $\widetilde{A_1}$ and $\widetilde{A_2}$, but breaks no global symmetries (since $\widetilde{H_z}$ is already neutral). After Higgsing and integrating out the massive matter, we are left with
\begin{subequations}
\begin{gather}
\renewcommand{\arraystretch}{1.2}
\begin{array}{c|cc|cccc}
& \Sp(N-2) & \Sp(N-2) & SU(N+1) & \Uy & U(1)_{\nR}' & U(1)_R'   \\
\hline
\widetilde{Y} & \fund& 1 & \fund & 1 & 0 & 0 \\
\widetilde{P_{z}} & \fund & 1 & 1 & -(1+N) & 0 & 2 \\
\cM_{A_1 Y} & 1 & \fund & \ov{\fund} & -1 & 1 & \frac{2}{3} \\
\cM_{Y P_{z}} & 1 & 1 & \ov{\fund} & N & 0 & 0\\
P & 1 & 1 & \fund & -N & N & \frac{2 N}{3} \\
\Phi_1 & 1 & \fund & 1 & 1+N & -(1+N) & \frac{2(2-N)}{3} \\
\ov\phi  & 1 & 1 & 1 & 0 & 2 & \frac{4}{3} \\
v & 1 & 1 & 1 & 0 & 0 & 0
\end{array}\\
\intertext{with superpotential}
\label{eq:Higgs-odd-deconfined-W}
W = \Tr\biggl(\widetilde{P_{z}} \cM_{Y P_{z}}\widetilde{Y} + \frac{1}{v} \Phi_1 \cM_{A_1 Y } P\biggr)\, .
\end{gather}
\end{subequations}
Now both $\Sp(N-2)$ gauge group factors are s-confining. After confinement, the superpotential~\eqref{eq:Higgs-odd-deconfined-W} gives masses to a number of the fields.

The end result is
\[
\renewcommand{\arraystretch}{1.2}
\begin{array}{c|cccc}
& \text{SU}(N+1) & \text{U}(1)_{y} & \text{U}(1)_{\nR}' & \text{U(1})_R'   \\
\hline
(\widetilde{Y})^2 & \asymm & 2 & 0 & 0 \\
(\cM_{A_1 Y})^2 & \asymmb & -2 & 2 & \frac{4}{3} \\
\ov\phi  & 1 & 0 & 2 & \frac{4}{3} \\
v & 1 & 0 & 0 & 0
\end{array}
\]
with vanishing superpotential, where $(\widetilde{Y})^2$ and $(\cM_{A_1 Y})^2$ denote two of the composite mesons resulting from s-confinement. This is a free theory (note the similarity to~\eqref{eqn:evenk}), so we can now trivially compute the dimension of the Higgs branch by counting the number of free hypermultiplets:
\[
\dim_\bH(\cH) = \frac{1}{2}\left(1+1+\frac{1}{2}N(N+1)+\frac{1}{2}N(N+1)\right) = 1+\frac{1}{2}N(N+1)
\]
in agreement with~\eqref{eq:dim(H)-expected-odd} since $N=k+1$.

\subsection{\alt{$(A_1,D_4)$}{(A1,D4)} Argyres-Douglas from partially closing \alt{$R_{2,2}$}{R22} punctures}
\label{sec:(A_1,D_4)}

As a further check of the Higgs branch for
$R_{2,\text{even}}$, we verify the proposal in
\cite{Beem:2020pry} that there is a Higgsing of the $R_{2,2}$ theory
leading to the $(A_1,D_4)$ Argyres-Douglas theory.\footnote{There are
a number of other properties of the $R_{2,2}$ theory that we could
compare to the results in \S3.2 of \cite{Beem:2020pry}. We
leave these checks to the interested reader.}

\begin{figure}
\centering
\includegraphics[height=6.5cm]{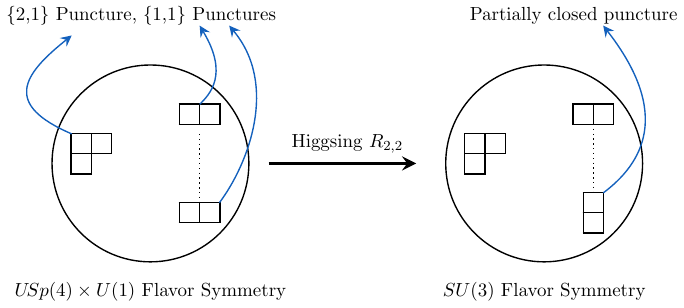}
\caption{The class $\cS$ description of Higgsing $R_{2,2}$ to the $(A_1,D_4)$ Argyres-Douglas theory}
\label{sfig:Partially Closing Puncture}
\end{figure}

In the class-$\cS$ description (see the left hand side of
figure~\ref{sfig:Partially Closing Puncture}), $R_{2,2}$ comes from an
$A_{2}$ theory on a sphere with an untwisted puncture and two twisted
punctures.  An $SU(2) \times SU(2)$ symmetry enhancing to $\Sp(4)$ is
associated with the twisted punctures, and $U(1)$ is associated with
the untwisted puncture.
It was argued in \cite{Beem:2020pry} that partially closing one of the
twisted punctures, as in the right hand side of
figure~\ref{sfig:Partially Closing Puncture}, we get the $(A_1,D_4)$
Argyres-Douglas theory and one decoupled hypermultiplet.

\begin{subequations} \label{eq:R_{2,2}}
Specializing~\eqref{eq:R_{2,even}-Lagrangian} to the case $N=3$, the Lagrangian whose infrared fixed is expected to be $R_{2,2}$ has the matter content
\begin{gather}
\renewcommand{\arraystretch}{1.3}
\label{eq:R_{2,2}-matter}
\begin{array}{c|c|cccc}
& \SU(3) & \Sp(4) & \Uz & U(1)_{\nR} & U(1)_R \\
\hline
Y_1 & \ov{\fund} & 1 & 1 & \frac{1}{3} & \frac{8}{9} \\
Y_2 & \ov{\fund} & 1 & -1 & \frac{1}{3} & \frac{8}{9} \\
A_1 & \ov\fund & 1 & 0 & -\frac{8}{3} & \frac{8}{9} \\
A_2 & \symm & 1 & 0 & -\frac{2}{3} & \frac{2}{9} \\
X & \ov{\fund} & \fund & 0 & \frac{4}{3} & \frac{5}{9} \\
\bar{\phi}
& 1 & 1 & 0 & 2 & \frac{4}{3}
\end{array}\\
\intertext{and superpotential:}
W=\Tr(A_2 Y_1 Y_2 +A_1 X X)+\bar{\phi} A_2^3
\end{gather}
\end{subequations}
Partially closing a puncture corresponds to turning on a nilpotent vev for the moment map operator of the flavour symmetry associated to the puncture. Moment maps have spin one under the $\cN=2$
$SU(2)_R$ symmetry and are neutral under the $\cN=2$ $\Ur$
symmetry. Thus per~\eqref{eq:N=2 to N=1} they have $U(1)_{\nR}\times
U(1)_R$ charges $(2, 4/3)$. It is straightforward to show that the only chiral operators of this form in~\eqref{eq:R_{2,2}} are $A_2 X^2$ and $\ov\phi$; since these transform in the adjoint representations of $\Sp(4)$ and $\Uz$, respectively, they are the moment maps in question. In particular, in terms of the manifest $\SU(2)_1\times\SU(2)_2$ flavor symmetries of the twisted punctures, $X$ decomposes into $X_1$ and $X_2$ in the $\SU(2)_1\times\SU(2)_2$ irreps $(\fund,\singlet)$ and $(\singlet,\fund)$ respectively. Thus, to partially close the second twisted puncture, we give a nilpotent vev to the $\SU(2)_2$ moment map $A_2 X_2^2$.

In the special case of $\su(2)$, there is a unique non-trivial nilpotent orbit. For instance,
upon decomposing the matter content in terms of a maximal torus
$U(1)_2 \subset \SU(2)_2$
\[
\renewcommand{\arraystretch}{1.3}
\begin{array}{c|c|ccccc}
& \SU(3) & \SU(2)_1 & U(1)_{2} & \Uz & U(1)_{\nR} & U(1)_R \\
\hline
Y_1 & \ov{\fund} & 1 & 0 & 1 & \frac{1}{3} & \frac{8}{9} \\
Y_2 & \ov{\fund} & 1 & 0 & -1 & \frac{1}{3} & \frac{8}{9} \\
A_1 & \ov\fund & 1 & 0 & 0 & -\frac{8}{3} & \frac{8}{9} \\
A_2 & \symm & 1 & 0 & 0 & -\frac{2}{3} & \frac{2}{9} \\
X_1 & \ov{\fund}& \fund & 0 & 0 & \frac{4}{3} & \frac{5}{9} \\
X_{2 a} & \ov{\fund} & 1 & 1 & 0 & \frac{4}{3} & \frac{5}{9} \\
X_{2 b} & \ov{\fund} &1 & -1 & 0 & \frac{4}{3} & \frac{5}{9} \\
\ov\phi  & 1 & 1 & 0 & 0 & 2 & \frac{4}{3}
\end{array}
\]
the operators $A_2 X_{2a}^2$ and $A_2 X_{2b}^2$ both sit on the nontrivial nilpotent orbit.
Giving a vev (denoted by $v$) to $A_2 X_{2b}^2$ Higgses
$SU(3)$ down to $SU(2)$, and the superpotential gives masses to several fields. Integrating out the massive matter results in a Lagrangian field content
\begin{subequations}
\begin{gather}
\renewcommand{\arraystretch}{1.2}
\begin{array}{c|c|cccc}
& \SU(2) & \SU(2)_1 & \Uz & U(1)_{\nR} & U(1)_R \\
\hline
Y_{1} & \fund & 1 & 1 & \frac{1}{2} & \frac{5}{6} \\
Y_{2} & \fund & 1 & -1 & \frac{1}{2} & \frac{5}{6} \\
A_{1} & 1 & 1 & 0 & -3 & 1 \\
A_{2} & \symm & 1 & 0 & -1 & \frac{1}{3} \\
X_{11} & \fund & \fund & 0 & \frac{3}{2} & \frac{1}{2} \\
X_{12} & 1 & \fund & 0 & 1 & \frac{2}{3} \\
X_{2a} & 1 & 1 & 0 & 2 & \frac{4}{3} \\
\bar{\phi} & 1 & 1 & 0 & 2 & \frac{4}{3} \\
v & 1 & 1 & 0 & 0 & 0
\end{array}\\
\intertext{with superpotential:}
W=\Tr(Y_{1} Y_{2} A_{2}+A_{1} X_{11}^2) + v \bar{\phi} A_{2}^2 \, .
\end{gather}
\end{subequations}
Since $X_{12}$ is decoupled,
we identify it with the expected free hyper. Likewise, we identify the
decoupled chiral field $X_{2a}$ as the $\cN=2$
partner of the flat direction $v$.

Removing these fields, we end up with the Lagrangian
\begin{subequations} \label{eqn:A1D4Lagrangian}
\begin{gather}
\renewcommand{\arraystretch}{1.2}
\label{eq:(A_1,D_2)}
\begin{array}{c|c|cccc}
& \SU(2) & \SU(2)_1 & \Uz & U(1)_{\nR} & U(1)_R \\
\hline
Y_{1} & \fund & 1 & 1 & \frac{1}{2} & \frac{5}{6} \\
Y_{2} & \fund & 1 & -1 & \frac{1}{2} & \frac{5}{6} \\
A_{1} & 1 & 1 & 0 & -3 & 1 \\
A_{2} & \symm & 1 & 0 & -1 & \frac{1}{3} \\
X_{1} & \fund & \fund & 0 & \frac{3}{2} & \frac{1}{2} \\
\bar{\phi} & 1 & 1 & 0 & 2 & \frac{4}{3}
\end{array}\\
\intertext{with superpotential:}
W=\Tr(Y_{1} Y_{2} A_{2}+A_{1} X_{1}^2) + \bar{\phi} A_{2}^2\, . \label{eqn:A1D4W}
\end{gather}
\end{subequations}
This is in perfect agreement with the matter content of Lagrangian description of the $(A_1,D_4)$ theory proposed in \cite{Agarwal:2016pjo} up to a change of basis for the flavor symmetries.\footnote{Note that the decoupled Coulomb branch operator $A_2^2$ is not explicitly flipped in~\cite{Agarwal:2016pjo} as is done here, accounting for the absence of $\ov\phi$ from their table. Although the superpotential~\eqref{eqn:A1D4W} was not given in~\cite{Agarwal:2016pjo}, it can in principle be inferred from the matter content and symmetries.}

At this point we can appeal to the convincing arguments of
\cite{Agarwal:2016pjo} that the theory with matter
content~\eqref{eq:(A_1,D_2)} flows to the $(A_1,D_4)$
theory. Nevertheless, for the convenience of the reader, we collect
here the results of some simple checks that can be performed
on~\eqref{eqn:A1D4Lagrangian}. First, it is straightforward to compute the central charges
\[
(a,c)=\left(\frac{7}{12},\; \frac{2}{3}\right)
\]
which agree with the expected values for the $(A_1,D_4)$ theory. Likewise, the
superconformal index of the theory is straightforward to
compute (as in~\cite{Agarwal:2016pjo})
\[
1&+\frac{t}{\nR^3}+t^{4/3} \biggl(\nR^2 X_{1,1}-\frac{J_1}{\nR}\biggr)+\nR t^{5/3}+t^2 \biggl(-1-X_{1,1}+\frac{J_1}{\nR^3}+\frac{1}{\nR^6}\biggr)  \\ 
&+t^{7/3}\biggl(\nR^2 J_1 (1+X_{1,1}) -\frac{1+J_2}{\nR}-\frac{J_1}{\nR^4}\biggr) 
+t^{8/3} \biggl(\nR^4 X_{2,2}+\nR J_1+\frac{2}{\nR^2}\biggr)   \\
&+t^3\biggl(-J_1 (2+X_{1,1})+\frac{-1+J_2}{\nR^3}+\frac{J_1}{\nR^6}+\frac{1}{\nR^9}\biggr) + \ldots \,.
\]
where the result organizes into complete $\SU(3) \supset \SU(2)_1 \times \Uz$ characters $X_{m,n}$, consistent with the expected accidental enhancement $\SU(2)_1 \times \Uz \to \SU(3)$ in the infrared.

\section{Coulomb branch deformations}

\label{sec:Coulomb}

We now examine the Coulomb branch of our proposed Lagrangians and compare with what is known about the Coulomb branch of the $R_{2,k}$ theories. As the Coulomb branch is parameterized by operators that are neutral under $SU(2)_R$ and all flavour symmetries, from the $\cN=1$ point of view Coulomb branch operators satisfy $U(1)_{\nR}=-3U(1)_R$ (see~\eqref{eq:N=2 to N=1}) and are neutral under the other flavour symmetries. After classifying such operators below in both even and odd $k$ cases, we consider the effect of giving them vevs.

\subsection{Even \alt{$k$}{k} Coulomb branch}

The Lagrangian~\eqref{eq:R_{2,even}-Lagrangian} expected expected to flow to the $R_{2,k}$ theory for even $k$ is reproduced below for
convenience:
\begin{gather}
\renewcommand{\arraystretch}{1.3}
\begin{array}{c|c|cccc}
& SU(N) & \Sp(2(N-1)) & \Uz & U(1)_{\nR} & U(1)_R \\
\hline
Y_1 & \bar{\fund}& 1 & 1 & \frac{1}{N} & 1-\frac{1}{3 N} \\
Y_2 & \bar{\fund} & 1 & -1 & \frac{1}{N} & 1-\frac{1}{3 N} \\
A_1 & \asymm & 1 & 0 & -\frac{2 (N+1)}{N} & \frac{2 (N+1)}{3 N} \\
A_2 & \symm & 1 & 0 & -\frac{2}{N} & \frac{2}{3 N} \\
X & \bar{\fund} & \fund & 0 & \frac{1}{N}+1 & \frac{2}{3}-\frac{1}{3 N} \\
\ov{\phi} & 1 & 1 & 0 & 2 & \frac{4}{3}
\end{array}
\tag{\ref{eq:R_{2,even}-charges} (bis)} \\
\intertext{with superpotential}
W=\Tr(A_1 X X+A_2 Y_1 Y_2) + \ov{\phi} A_2^N \, .
\tag{\ref{eq:R_{2,even}-W} (bis)}
\end{gather}

Since all the chiral fields have non-negative charge under $\UsuR =U(1)_R+\frac{1}{3}U(1)_{\nR}$, only those with vanishing charge, i.e., $A_1$ and $A_2$, can appear in Coulomb branch operators. Thus, the complete set of Coulomb branch operators is given by the $\SU(N)$ baryons $\cO_p=A_1^{2 p} A_2^{N-2 p}$ for $p=1,\ldots,k/2$ (the decoupled baryon $\mathcal{O}_0 = A_2^N$ having been set to zero by the $\ov\phi$ F-term).
These operators have conformal dimension
\[
\Delta \cO_p = \frac{3}{2} Q_{U(1)_R}[\cO_p] = 2p+1
\]
so we find a set of Coulomb branch operators with dimensions
$3,5,\ldots,k+1$, as expected \cite{Chacaltana:2014nya}.

Giving a vev to a single such operator $\cO_p$ will break
$SU(N)\to \Sp(2p)\times \SO(N-2p)$ and initiate a flow. After integrating
out massive matter in the IR this leads to two decoupled sectors. The
first one is manifestly $\cN=2$ supersymmetric, having matter content
\begin{subequations}
\begin{gather}
\begin{array}{c|c|cc}
& \SO(N-2 p) & \Sp(2N-2) & \UsuR \\
\hline
A_{1} & \asymm & 1 & 0 \\
X & \fund & \fund & 1 \\
\end{array} \\
\intertext{and superpotential}
W=\Tr(A_{1} X^2)\, .
\end{gather}
\end{subequations}
This is an $\cN=2$ $SO(N-2p)$ gauge theory with $N-1$ hypermultiplets in the $\fund$ representation. (This theory is infrared free for all $p$).

The second factor has matter content
\begin{subequations} \label{eqn:USpNotN2}
\begin{gather}
\begin{array}{c|c|cc}
& \Sp(2 p) & \SO(2) & \UsuR \\
\hline
Y & \fund & \fund & 1 \\
A_{2} & \symm & 1 & 0 \\
\bar{\phi} & 1 & 1 & 2 \\
v & 1 & 1 & 0 \\
\end{array}\\
\intertext{and superpotential}
W=v\bar{\phi} A_{2}^{2 p}+\Tr(A_2 Y^2)\,,
\end{gather}
\end{subequations}
where $v$ is the chiral superfield with a vev.

This theory \emph{would} be manifestly $\cN=2$ supersymmetric, were it not for the $v \ov\phi A_{2}^{2p}$ superpotential term. To understand the effect of this extra term, consider the case $p=1$. Without the $v \ov\phi A_{2}^{2p}$ superpotential term, we obtain a manifestly $\cN=2$ supersymmetric $\SU(2)$ gauge theory with $\cN_f = 1$ flavor, together with a free hypermultiplet made out of $v$ and $\ov\phi$. The exact Coulomb branch solution to
the interacting part of this theory is well known
\cite{Seiberg:1994rs,Seiberg:1994aj}: at low energies the theory flows 
to an $\cN=2$ Maxwell theory, except at three points on the Coulomb
branch where additional magnetic / dyonic hypermultiplets become massless. These three
points are symmetrically distributed around the origin. In particular,
there is no extra hypermultiplet at the origin of the Coulomb
branch. Turning the $v\bar{\phi} A_{2}^{2}$ coupling back on gives a mass
to the Coulomb branch operator $A_{2}^2$ of the $SU(2)$ theory, together
with $\ov\phi$, so the Coulomb branch of the $SU(2)$ theory is
lifted. We end up with a free $\cN=2$ $U(1)$ vector multiplet made of
the $\cN=1$ $U(1)$ vector multiplet and $v$. 

Thus, in $\cN=2$ language, upon giving the dimension-three Coulomb branch operator $\cO_1 = A_1^2 A_2^{N-2}$ a vev, the $R_{2,k}$ (even $k$) theory flows to an $SO(N-2)$ gauge theory with $N-1 = k$ hypers in the $\fund$ representation together with a pure-glue $\U(1)$ gauge theory. Note that since $\tau = e^{\pi i/3}$ at the origin of the Coulomb branch of the $N_f=1$ $\SU(2)$ Seiberg-Witten theory~\cite{Seiberg:1994aj}, the $\U(1)$ holomorphic gauge coupling is frozen at this value along this portion of the Coulomb branch.\footnote{In fact, this is required for the discrete symmetries to match between the UV and IR. A careful analysis of~\eqref{eqn:USpNotN2} reveals the presence of an additional $\mathbb{Z}_{2p+1}$ discrete symmetry. For $p=1$, the only way that this $\mathbb{Z}_3$ symmetry can act on the low-energy effective theory is as an electromagnetic duality symmetry $\mathbb{Z}_3 \subset \SL(2,\mathbb{Z})$, which is broken unless $\tau = e^{\pi i/3}$.}

We expect that a similar analysis will yield an explicitly $\cN=2$ description of~\eqref{eqn:USpNotN2} for all $p$, but we not attempt it here.

\subsection{Odd \alt{$k$}{k} Coulomb branch}

The odd $k$ case behaves very similarly, but the technical analysis is
fairly cumbersome, so we will be very brief, and just describe the
results. Recall from~\eqref{eq:R_{2,odd}-Lagrangian} the matter
content for this theory. We can identify a set of Coulomb
branch operators of the form
\[
\cO_p \df A_1^{N-2-2p} A_{2}^{2p} Q
\]
with $p=1,\ldots, (k-1)/2$ and
\[
\Delta \cO_p = 2p+1\, .
\]
That is, we have a set of Coulomb branch operators of dimensions
$3,5,\ldots,k$, which is again the expected answer
\cite{Chacaltana:2010ks}.

Turning on a vev for $\cO_p$ we again flow in the infrared to two
decoupled sectors. The first is manifestly $\cN=2$, with matter content
\begin{subequations} \label{eqn:R2oddkCoulomb1}
\begin{gather}
\begin{array}{c|c|cc}
& \Sp(N-2-2p) & \SO(2N+2) & \UsuR \\
\hline
B & \symm & 1 & 0 \\
Y & \fund & \fund & 1 \\
\end{array} \\
\intertext{and superpotential}
W=\Tr(B Y^2)\, .
\end{gather}
\end{subequations}
The second sector is precisely as in the even $k$ case:
\begin{subequations} \label{eqn:R2oddkCoulomb2}
\begin{gather}
\begin{array}{c|c|cc}
& \Sp(2p) & \SO(2) & \UsuR \\
\hline
X & \fund & \fund & 1 \\
A & \symm & 1 & 0 \\
\bar{\phi} & 1 & 1 & 2 \\
v & 1 & 1 & 0
\end{array}\\
\intertext{with superpotential}
W=v\bar{\phi} A^{2p}+\Tr(A X^2)\, .
\end{gather}
\end{subequations}
As before, this theory flows to an $\cN=2$ pure-glue $U(1)$ gauge theory with $\tau = e^{\pi i/3}$ when $p=1$. We leave an analysis of the case $p>1$ to future work.

As a non-trivial consistency check, consider $R_{2,3}$, for which the only Coulomb branch operator is $\cO_1$. Putting $N=4$ and $p=1$, we find that the sector~\eqref{eqn:R2oddkCoulomb1} disappears whereas \eqref{eqn:R2oddkCoulomb2} becomes the $\cN=2$ pure-glue $\U(1)$ theory with $\tau = e^{\pi i /3}$, as argued above. As $R_{2,3}$ is the $E_6$ Minahan-Nemeschansky theory~\cite{Minahan:1996fg}, this is the expected result.

\section{Mass deformation}

\label{sec:mass}

Finally, let us comment briefly on the effect of turning on mass
deformations, and how to see that they reproduce the expected results
in a simple but interesting example: the mass deformation of the rank
one $E_6$ Minahan-Nemeschansky theory to $\cN=2$ $SU(2)$ with 5
flavours. This deformation is natural from the point of view of the
F-theory realisation of the Minahan-Nemeschansky theories: what we are
doing is taking one of the $C$ 7-branes on the $A^5BC^2$ $E_6$
stack to infinity, leaving $A^5BC$, namely an eight dimensional
$SO(10)$ theory.\footnote{We refer the reader unfamiliar with the relevant
F-theory constructions to \cite{DeWolfe:1998zf} for background and
notation.} The worldvolume theory for a D3 probe on this stack
is precisely $SU(2)$ with five flavours.

We are therefore interested in identifying a relevant gauge invariant
operator leaving a $SO(10)$ subgroup of the flavour symmetry
unbroken. A quick look to~\eqref{eq:E6-quiver} suggests a natural
candidate: $\ov\phi$.
Adding the mass deformation term $ m \ov\phi$ to the
superpotential~\eqref{eq:E6-W} leads to
\[
W_{\text{$E_6$-deform}}=A_1  A_2 Z+A_2 Y \Phi+ A_1 Y_{1} \Phi_1+P Q Y+P_{1} Q Y_{1}+(A_{1}^2 Q+m) \ov\phi \,,
\]
which will force the $\SU(3)$ baryon $A_{1}^2 Q$ to get a vev. This breaks $\SU(3)\times \SU(2) \to \SU(2)$ (embedded as the diagonal subgroup of $\SU(2)\times\SU(2) \subset \SU(3)\times\SU(2)$). After integrating out the resulting massive fields, we obtain the matter content
\begin{subequations}
\begin{gather}
\renewcommand{\arraystretch}{1.3}
\begin{array}{c|c|ccc}
& \SU(2) & \SU(5) & \Uy  & \UsuR \\
\hline
S & \symm & 1 & 0  & 0 \\
A & \fund & \ov\fund & -1  & 1 \\
B & \fund & \fund & 1 & 1
\end{array}\\
\intertext{with superpotential}
W= \Tr(B S A) \,,
\end{gather}
\end{subequations}
which is indeed the $\cN=2$ $SU(2)$ theory with
five flavours.\footnote{Note that the flavor symmetry is actually $\SO(10)$. Here we only display the $\SU(5)\times \Uy$ subgroup that was manifest in the UV Lagrangian we started with.}

Although this theory is infrared free, we can reach more interesting theories by further mass deformation. First, giving a mass to a single hypermultiplet yields the superconformal $\SU(2)$ theory with four flavors. From there, it is possible to reach $(A_1, D_4)$ by a further mass deformation~\cite{Agarwal:2016pjo}, as well as the $(A_1,A_3)$ and $(A_1,A_2)$ theories~\cite{Maruyoshi:2016aim}. As these mass deformations are thoroughly explored in existing literature, we will not discuss them any further here.

\section{Conclusions and further directions}

\label{sec:conclusions}

In this paper we have introduced a new approach for systematically
constructing $\cN=1$ Lagrangians for the $R_{2,k}$ $\cN=2$
SCFTs. These Lagrangians pass a multitude of very nontrivial checks:
symmetries, anomalies, central charges and superconformal indices all
match with the expected $\cN=2$ fixed points in the IR, and various
properties of their moduli spaces and mass deformations all agree with
the expected results.

The appearance of the $R_{2,k}$ theories is ultimately due to the fact
that our parent $\cN=1$ theory is the complex cone over $\bF_0$. The
methods developed in \cite{Garcia-Etxebarria:2016bpb} are nevertheless
much more general, so a natural question is which other $\cN=2$
theories can be reached by applying the same methods to other classes
of singularities. A natural class of spaces to consider is the
Calabi-Yau cones over $Y^{2n,0}$
\cite{Gauntlett:2004zh,Gauntlett:2004yd,Gauntlett:2004hh,Gauntlett:2004hs,Martelli:2004wu},
which generalises the $n=1$ case studied here.

Along similar lines, it would be interesting to drop some of the
assumptions in \cite{Garcia-Etxebarria:2016bpb}, for example by
allowing for the presence of flavour branes and non-compact
orientifolds. This is again likely to lead to new $\cN=1$ Lagrangians
for interesting $\cN=2$ theories.

More generally, we would like to develop a more direct method of
deriving our results. Our approach is certainly roundabout: we are
using $\cN=1$ dualities to understand $\cN=2$ dualities! This is very
surprising, and contrary to the usual expectation that having more
supersymmetry makes analysis of duality simpler. While the fundamental new ideas in our analysis of ``brane bending'' and ``deconfinement'' --- introduced
in \cite{Garcia-Etxebarria:2015hua,Garcia-Etxebarria:2016bpb} to
understanding interacting $\cN=1$ SCFTs --- require us to deviate from purely $\cN=2$ supersymmetric language, there is no obvious reason why they cannot be applied more directly to the class-$\cS$ construction.  Understanding whether this is possible --- and if so how to do so systematically ---
is a natural challenge raised by our results.

\acknowledgments

We thank F.~Albertini, P.~Argyres, F.~Carta, S.~Cremonesi, M.~Lemos and
S.~Razamat for helpful discussions. I.G.E.\ is partially supported by
STFC grant ST/T000708/1 and by the Simons Foundation collaboration
grant on Global Categorical Symmetries (award number 888990). The
research of B.H.\ and A.K.S.\ was supported by National Science
Foundation grant PHY-1914934.

\appendix

\section{Superconformal index for \alt{$R_{2,2k}$}{R2,2k}} \label{R2kSCI}

We compute the superconformal index for $R_{2,2k}$ from the matter content given in~\eqref{eq:R_{2,even}-charges} with $N=2k+1$.  We are using conventions for the fugacities $t$ and $J_n$ similar to \cite{Garcia-Etxebarria:2015hua}.  We display only a small number of terms in the index and computations to different ranks and higher order in $t$ can be done using the computer program of \cite{Garcia-Etxebarria:2015hua}.

\subsubsection*{Index for $R_{2,2}$:}

\begin{multline*}
1+{\nR}^2 t^{4/3} \bigl(1+X_{2,0}\bigr)
+\frac{t^{5/3} X_{1,0}}{{\nR}^2} 
+t^2
\Bigg(-2-X_{2,0}+\frac{1}{{\nR}^6}+{\nR}^3 \Bigl(\frac{X_{0,1}}{z}+z X_{0,1}\Bigr)-J_{1}
X_{1,0}\Bigg) \\
+t^{7/3}
\Bigg(-\frac{J_1+X_{1,0}}{{\nR}^4}-{\nR}^2 \bigl(-X_{1,0}-J_{1}
(2+X_{2,0})\bigr)\Bigg) 
+t^{8/3} \Bigg({\nR}
\Bigl(-\frac{X_{0,1}}{z}-z X_{0,1}\Bigr)\\
+\frac{2+2 J_{1}
X_{1,0}}{{\nR}^2}+{\nR}^4 (1+X_{0,2}+X_{2,0}+X_{4,0})\Bigg)
+t^3 \Bigg(\frac{J_{1}}{{\nR}^6}+{\nR}^3 \Bigl(\frac{J_{1}
X_{0,1}}{z}+z J_{1} X_{0,1}\Bigr)\\
-(2+J_{2}) X_{1,0}-J_{1} (3+X_{2,0})+X_{3,0}\Bigg) 
+t^{10/3}
\Bigg(\frac{-\frac{1}{z}-z}{{\nR}}+\frac{-1-J_{2}-J_{1}
X_{1,0}+X_{2,0}}{{\nR}^4}\\
+{\nR}^5 \Bigl(\frac{X_{0,1}+X_{2,1}}{z}+z
(X_{0,1}+X_{2,1})\Bigr)
+{\nR}^2 \Bigl(-1-X_{0,2}+J_{1} X_{1,0}\\
-4 X_{2,0}+J_{2}
(2+X_{2,0})-X_{2,1}-J_{1} X_{3,0}-X_{4,0}\Bigr)\Bigg)+\ldots
\end{multline*}

\subsubsection*{Index for $R_{2,4}$:}

\begin{multline*}
1+ t^{4/3} {\nR}^2 \bigl(1+X_{2,0,0,0}\bigr)\\
+t^2 \left(-2-X_{2,0,0,0}+\frac{1}{{\nR}^6}\right) 
+t^{7/3}
\left(-\frac{J_{1}-X_{1,0,0,0}}{{\nR}^4}+{\nR}^2 J_{1} (2+X_{2,0,0,0})\right) \\
+t^{8/3} \left(\frac{2-J_{1}
X_{1,0,0,0}}{{\nR}^2}+{\nR}^4 (1+X_{0,2,0,0}+X_{2,0,0,0}+X_{4,0,0,0})\right) \\
+t^3
\left(X_{0,0,1,0}+\frac{J_{1}-X_{1,0,0,0}}{{\nR}^6}+X_{1,0,0,0}-J_{1} (3+X_{2,0,0,0})\right)\\
+t^{10/3} \Bigl(\frac{1}{{\nR}^{10}}+{\nR}^5
\left(\frac{X_{0,0,0,1}}{z}+z X_{0,0,0,1}\right)+\frac{-1-J_{2}+2 J_{1} X_{1,0,0,0}+X_{2,0,0,0}}{{\nR}^4}\\
+{\nR}^2 \bigl(-1-J_{1}
X_{0,0,1,0}-X_{0,2,0,0}-4 X_{2,0,0,0}+J_{2} (2+X_{2,0,0,0})\bigr)\\
-{\nR}^2 \bigl(X_{2,1,0,0}-X_{4,0,0,0}\bigr)\Bigr) +\ldots 
\end{multline*}

\subsubsection*{Index for $R_{2,6}$:}

\begin{multline*}
1 +{\nR}^2 t^{4/3} \bigl(1+X_{2,0,0,0,0,0}\bigr)\\
+t^2 \left(-2-X_{2,0,0,0,0,0}+\frac{1}{{\nR}^6}\right) +t^{7/3}
\left(-\frac{J_{1}}{{\nR}^4}+{\nR}^2 J_{1} (2+X_{2,0,0,0,0,0})\right)\\
+t^{8/3} \left(\frac{2}{{\nR}^2} +{\nR}^4 (1+X_{0,2,0,0,0,0}+X_{2,0,0,0,0,0}+X_{4,0,0,0,0,0})\right)\\
+t^3 \left(\frac{J_{1}+X_{1,0,0,0,0,0}}{{\nR}^6}-J_{1}
(3+X_{2,0,0,0,0,0})\right)+\ldots
\end{multline*}

\bibliographystyle{JHEP}
\bibliography{refs}

\end{document}